\documentclass[twocolumn]{aastex631}
\usepackage{amsmath}
\usepackage[english]{babel}
\usepackage{natbib}

\shorttitle{eDisk-XXIV }

\begin{document}

\title{Early Planet Formation in Embedded Disks (eDisk). XXIV: Systematic Investigation of Disk Structures based on Visibility Analysis}

\author[0000-0002-0554-1151]{Mayank Narang}
\affiliation{Academia Sinica Institute of Astronomy \& Astrophysics, 11F of Astronomy-Mathematics Building, AS/NTU, No. 1, Sec. 4, Roosevelt Rd., Taipei 106319, Taiwan, R.O.C.}

\author{Jerry Xu}
\affiliation{Department of Astronomy, University of Illinois, 1002 West Green St, Urbana, IL 61801, USA}

\author[0000-0002-4540-6587,sname='Leslie Looney']{Leslie W. Looney}
\affiliation{Department of Astronomy, University of Illinois, 1002 West Green St, Urbana, IL 61801, USA}

\author[0000-0003-0998-5064]{Nagayoshi Ohashi}
\affiliation{Academia Sinica Institute of Astronomy \& Astrophysics, 11F of Astronomy-Mathematics Building, AS/NTU, No. 1, Sec. 4, Roosevelt Rd., Taipei 106319, Taiwan, R.O.C.}

\author{Anika Khandavalli}
\affiliation{Department of Astronomy, University of Illinois, 1002 West Green St, Urbana, IL 61801, USA}

\author[0000-0002-9209-8708]{Patrick Sheehan}
\affiliation{National Radio Astronomy Observatory, 520 Edgemont Rd., Charlottesville, VA 22903 USA}

\author[0000-0001-5058-695X]{Jonathan P. Williams}
\affiliation{Institute for Astronomy, University of Hawai‘i at Mānoa, 2680 Woodlawn Dr., Honolulu, HI 96822, USA}

\author[0000-0003-0845-128X]{Shigehisa Takakuwa}
\affiliation{Department of Physics and Astronomy, Graduate School of Science and Engineering, Kagoshima University, 1-21-35 Korimoto, Kagoshima, Kagoshima 890-0065,
Japan}
\affiliation{Academia Sinica Institute of Astronomy \& Astrophysics, 11F of Astronomy-Mathematics Building, AS/NTU, No. 1, Sec. 4, Roosevelt Rd., Taipei 106319, Taiwan, R.O.C.}

\author[0000-0001-9133-8047]{Jes K. J{\o}rgensen}
\affiliation{Niels Bohr Institute, University of Copenhagen,
{\O}ster Voldgade 5-7, 1350, Copenhagen K, Denmark}

\author[0000-0002-9143-1433]{Ilseung Han}
\affiliation{Institut de Ci\`encies de l'Espai (ICE-CSIC), Campus UAB, Can Magrans S/N, E-08193 Cerdanyola del Vall\`es, Catalonia, Spain}

\author{Woojin Kwon}
\affiliation{Department of Earth Science Education, Seoul National University, 1 Gwanak-ro, Gwanak-gu, Seoul 08826, Republic of Korea}
\affiliation{SNU Astronomy Research Center, Seoul National University, 1 Gwanak-ro, Gwanak-gu, Seoul 08826, Republic of Korea}

\author[0000-0002-7402-6487]{Zhi-Yun Li}
\affiliation{University of Virginia, 530 McCormick Rd., Charlottesville, Virginia 22904, USA}

\author[0000-0002-4372-5509]{Nguyen Thi Phuong}
\affiliation{Vietnam National Space Center, Vietnam Academy of Science and Technology, 18 Hoang Quoc Viet, Nghia Do, Hanoi, Vietnam}

\author[0000-0002-6195-0152]{John J. Tobin}
\affiliation{National Radio Astronomy Observatory, 
520 Edgemont Rd., Charlottesville, VA 22903 USA}

\begin{abstract}
The dust continuum emission from young protostellar disks encodes key information about their mass distribution and early evolution, yet uniform high-resolution comparative studies remain limited. We present a systematic uv-plane analysis of parametric intensity models applied to ALMA Band-6 (1.3 mm) observations of 23 disks (19 protostellar systems with 4 being in binary) from the eDisk sample, spanning Gaussian profiles to power-law cores with exponential tails (PLCT), including asymmetric extensions. Gaussian models generally fail to reproduce the centrally peaked emission and extended outer structure observed in most disks, whereas the PLCT framework provides a significantly improved description of radial brightness profiles. Incorporating azimuthal asymmetries further reduces residuals in 15 of 17 inclined disks, indicating that departures from axisymmetry are common at early stages. Only two disks, L1489 IRS and Oph IRS63, exhibit clear gap and ring substructures, while most appear smooth at the spatial resolution and sensitivity of our observations. These systems are among the most evolved in the sample, and the absence of flat-spectrum sources limits the evolutionary range probed, {suggesting that the detection of prominent gaps and rings is not common} in the earliest phases of disk evolution. Using a uniform definition of disk radius based on the 95\% enclosed flux, we find a positive correlation with stellar mass, $R_{\rm disk} \propto M_{\star}^{1.5 \pm 0.1}$, with disks in binary systems systematically smaller than those around isolated protostars. While the models capture overall morphology and large-scale asymmetries, distinguishing intrinsic structures from radiative transfer effects in optically thick regions remains challenging.

\end{abstract}

\section{Introduction} \label{Intro}

The ubiquity of exoplanetary systems has transformed our understanding of planet formation. Statistical surveys from missions like Kepler and TESS reveal that planetary systems are common, but their architectures vary widely, often reflecting underlying trends with stellar mass, metallicity, and age \citep[e.g.,][]{2012Natur.486..375B,2015ApJ...809....8B,2016AJ....152..187M,2018AJ....156..221N,2019AJ....158..109H,2022AJ....164..181U,2023AJ....165..265M,2024AJ....168....7B}. Since planets form within the circumstellar disks that surround young stars, the initial conditions and evolution of these disks play a crucial role in shaping the final planetary system.

With the advent of the Atacama Large Millimeter/submillimeter Array (ALMA), it has become increasingly clear that substructures in protoplanetary disks are very common in older disks \citep[e.g.,][]{2015ApJ...808L...3A, 2018ApJ...869L..41A, 2018ApJ...869...17L, 2021MNRAS.501.2934C,2024PASJ...76..437Y,2025A&A...696A.232G}. These substructures include a variety of features such as concentric bright and dark rings \citep[e.g.,][]{2018ApJ...869L..42H,2024PASJ...76..437Y,2025A&A...696A.232G}, which are often interpreted as the signatures of pressure bumps or dust traps. Large inner holes or gaps, frequently associated with transitional disks, represent another common morphology \citep[e.g.,][]{2014prpl.conf..497E, 2023EPJP..138..225V}. In addition to radial structures, azimuthal asymmetries such as spirals and vortices have been observed \citep{2013A&A...556A..76V, 2016Sci...353.1519P}. Among the observed features, wide gaps and ring-like structures appear to be the most common. Understanding how these substructures arise, and how they correlate with stellar and disk properties, is essential to linking the early stages of disk evolution to the wide variety of planetary systems observed today.

Several mechanisms have been proposed to explain these substructures in disks such as envelope infall (e.g., \citealt[e.g.,][]{2015ApJ...805...15B}; \citealt{2022ApJ...928...92K}), zonal flows \citep[e.g.,][]{2009ApJ...697.1269J,2014ApJ...796...31B}, formation of dead zones \citep[e.g.,][]{2013ApJ...765..114D}, snow lines \citep[e.g.,][]{2017A&A...600A.140S, 2020MNRAS.495.3160O}, photoevaporation \citep[e.g.,][]{2017RSOS....470114E, 2022EPJP..137.1357E}, gravitational instabilities \citep{1964ApJ...139.1217T, 2016ARA&A..54..271K}, gravitational perturbations from a companion \citep{2019MNRAS.483.4114C}, and active planet formation \citep[e.g.,][]{2015ApJ...809...93D, 2018ApJ...869L..47Z}. Despite numerous proposed explanations, the origin of these features remains uncertain with disk-planet interactions as the leading hypothesis.

The ubiquity of substructures such as rings, gaps, and asymmetries in Class II disks strongly suggests that significant disk evolution has already occurred by the Class II. This observation points to the likelihood that the processes leading to planet formation are initiated much earlier, during the protostellar phase. Numerical simulations indicate that the large reservoirs of gas and dust present in protostellar disks surrounding Class 0/I protostars may create more favorable conditions for the rapid growth of solids and the formation of planetesimals, compared to the more evolved, mass-depleted disks of Class II objects \citep[e.g.,][]{2017ApJ...838..151T, 2019MNRAS.484.1574T}.

Additional support for the early formation of planets, particularly gas giants like Jupiter, comes from geochemical studies of primitive meteorites. The distribution of volatiles in non-carbonaceous (NC) and carbonaceous chondrites (CC) provides important clues about the conditions in the early solar system \citep[see][]{2020SSRv..216..133B}. These studies suggest a distinct separation between NC materials, which are believed to have originated in the inner solar system, and CC materials, which are associated with the outer regions of the protoplanetary disk. This compositional dichotomy may have been driven by the early formation of a proto-Jupiter, whose presence could have acted as a barrier to material transport between the inner and outer disk. Alternatively, this separation might reflect the development of a pressure maximum or a dust trap within the disk, which could have efficiently segregated solids based on their size and composition. Together, these findings underscore the importance of the protostellar phase in shaping the architecture and composition of planetary systems.

Recent detections of substructures in protostellar disks provide further evidence for the early onset of planet formation \citep[e.g.,][]{2020Natur.586..228S, 2017ApJ...840L..12S, 2018ApJ...857...18S, 2020ApJ...902..141S, 2021MNRAS.501.2934C, 2023ApJ...951....8O,2023ApJ...951...11Y,2023ApJ...958...98F, 2025PASJ...77..572S,2025ApJ...993..120H, 2025A&A...700A.235H}. High-resolution ALMA observations ($\leq$ 0.1'') reveal that ringlets and large gaps, ranging from tens to hundreds of au in radius, have formed in some disks prior to the T Tauri (or Class II) phase. Most observed substructures are found in Class I and flat-spectrum protostars, with only two Class 0 protostars showing substructures so far \citep{2020ApJ...902..141S}. 

If these substructures in protostellar disks are indeed the result of disk-(proto)planet interactions, it suggests that planet formation begins as early as the protostellar phase. By studying protostars where substructures have already formed in their disks, we can gain valuable insights into the initial physical and chemical conditions that set the stage for planet formation. Such studies are critical for understanding the timing, mechanisms, and environments in which planets begin to form, providing a window into the earliest phases of planetary system evolution.

An important question remains: how common are substructures in protostellar disks? Most of the substructures identified in these disks have been detected through surveys that differ significantly in their sensitivity, angular resolution, and sample selection criteria. This variation makes it difficult to assess the ubiquity of substructures in a statistically meaningful way. To address this challenge systematically, the ALMA Large Program, Early Planet Formation in Embedded Disks (eDisk) \citep{2023ApJ...951....8O}, was conducted.

\begin{figure*}
\centering
\includegraphics[width=0.8\linewidth]{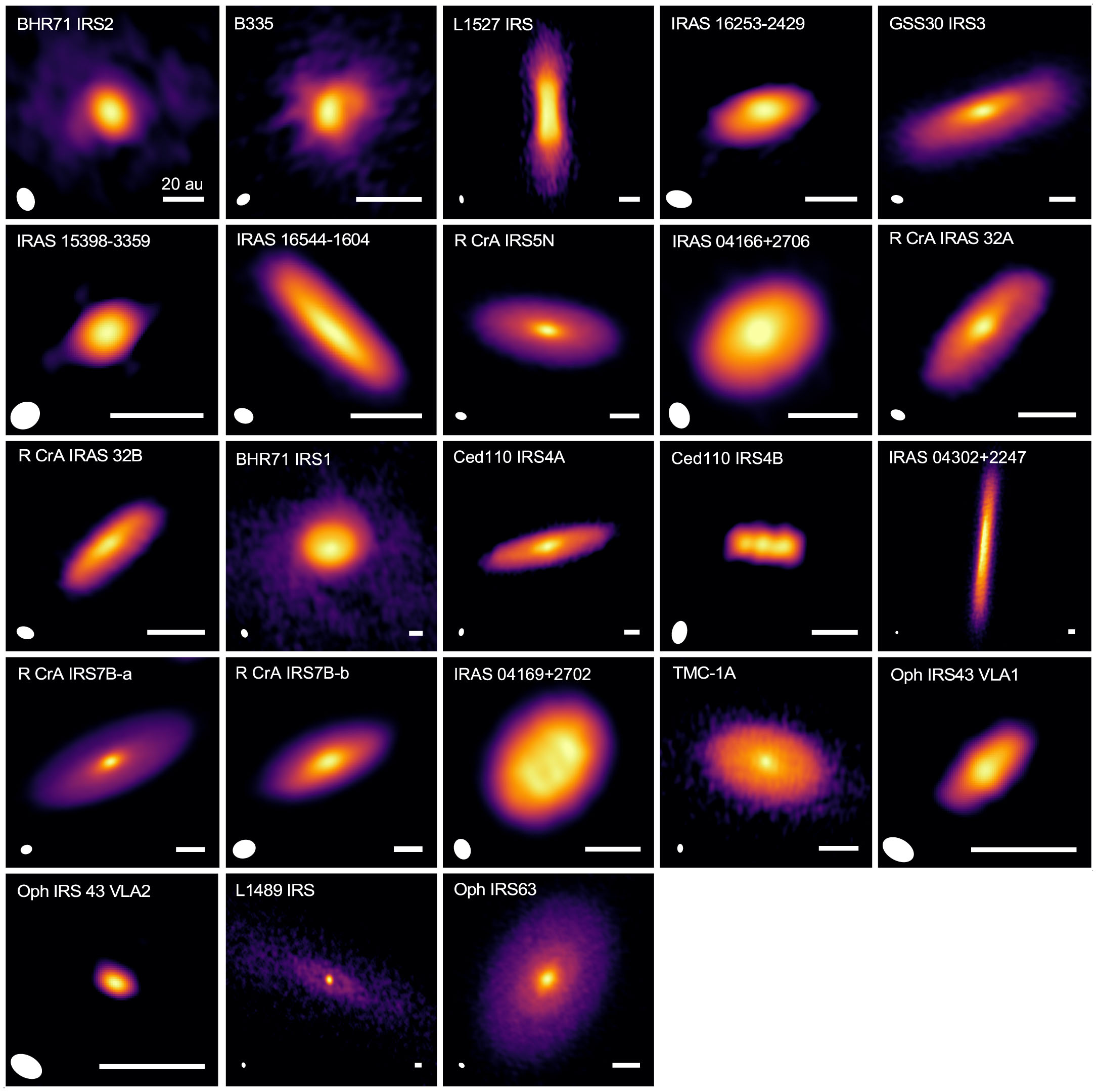}
\caption{The gallery of protostellar disks detected in the eDisk sample (adapted from \citealt{2023ApJ...951....8O}). All the maps are shown in individual asinh-stretch intensity scales, except for IRAS 04169+2702, which is
shown in a linear intensity scale. The protostars are ordered by bolometric temperature, T${\rm bol}$. The
beam size and the scale of 20 au are shown in the lower left and lower right corners of each panel, respectively. }
\label{Fig1}
\end{figure*}

The eDisk program conducted a comprehensive survey of 19 protostellar systems, including 12 Class 0 and 7 Class I sources (see Table \ref{tab:obs} and Figure \ref{Fig1}), with the goal of characterizing their substructures. Although the sample was intentionally selected to exclude binaries, four of the 19 systems were later identified as binary sources with the high resolution eDisk observations. Observations were carried out at a high angular resolution of 0.\arcsec04 in continuum emission at a wavelength of 1.3 mm, providing unprecedented detail on the morphology of these young disks \citep{2023ApJ...951....8O}. The eDisk sample (Table \ref{tab:obs}) spans a wide range of bolometric temperatures (T$_\mathrm{bol}$) and bolometric luminosities (L$_\mathrm{bol}$), allowing us to explore disks across different evolutionary stages and physical conditions. Furthermore, all the targets are located within 200 pc, ensuring that the resolution is sufficient to resolve small-scale structures within the disks.  Another important and unique aspect of the eDisk program is that dynamical masses are estimated uniformly for the entire sample.

In Figure~\ref{Fig1}, we present the ALMA continuum images of 23 protostellar disks from the eDisk sample, including four disks associated with secondary components in binary systems. All of the eDisk individual source papers (see references in Table \ref{tab:obs}) fit disk intensities in the image plane using simple Gaussian models or, in a few cases, more complex structure models \citep[e.g.,][]{2023ApJ...954...67S}. In general, there was little evidence for substructure in most sources, with the exception of a few notable cases (e.g., Oph IRS 63 and L1489 IRS). However, there has not yet been a systematic comparison of model fits to the source intensity distributions across the full sample.

To investigate the structural properties of protostellar disks in the eDisk sample, we systematically fit simple analytical intensity models directly to the interferometric visibilities in the uv plane using pdspy \cite{2017ApJ...851...45S,2019ApJ...874..136S,2022ApJ...934...95S}. This approach represents a compromise between overly simplistic image-plane modeling (which neglects uv-sampling effects and envelope emission) and computationally expensive radiative transfer modeling. The primary objective of this analysis is to capture the fundamental geometric characteristics of the disks while maintaining a balance between model complexity and interpretability.

We utilized a range of analytical profiles to model key disk parameters, including disk inclination, position angle, and radial intensity distribution. Our modeling approach was informed by insights from previous studies, as well as our own experience with disk structures, allowing us to refine the fitting process effectively. Working in the visibility domain offers several advantages, including avoiding artifacts produced by image deconvolution algorithms such as \texttt{CLEAN}, which can smear or distort disk structures, and enabling the simultaneous characterization of both compact and extended emission components. This approach provides a more reliable representation of the intrinsic disk structure while also allowing disk radii to be measured in a uniform manner across the sample.

The paper is organized as follows. Section~2 describes the observations and data reduction procedures. In Section~3, we outline the methodology used to fit the $uv$ visibilities with simple two-dimensional analytical models. The resulting fits and inferred disk properties are presented in Section~4, followed by a discussion of the implications of our results in Section~5. Finally, we summarize our conclusions in Section~6.

\section{Observations and data reduction}
\subsection{Observations}
The observations for the eDisk survey were carried out over multiple sessions between April 2021 and July 2022 as part of the ALMA Large Program eDisk (Project ID: 2019.1.00261.L, PI: N. Ohashi). These observations were conducted in ALMA Band 6 at a wavelength of 1.3 mm. Additionally, supplementary data were acquired through ALMA Director’s Discretionary Time (DDT) observations (Project ID: 2019.A.00034.S, PI: J. Tobin). 

The projected baselines for the observations spanned a range from 15 meters to $\sim$12600 meters, allowing for a spatial resolution of $\sim$ 7 au at a distance of 140~pc in the continuum (0.04\arcsec). This was achieved using the C43-8 and C43-5 antenna configurations. 

The ALMA correlator was set up to observe both 1.3 mm continuum emission and several molecular lines, including CO ($J=2-1$) and its isotopologues, alongside other molecular tracers. These spectral lines provided critical insights into the physical and chemical conditions of the gas within the circumstellar environments of the embedded protostars \citep[e.g.,][]{2025arXiv251221454F,2025A&A...704A.133S,2025AJ....169..192N,2025ApJ...985..166K,2026arXiv260209837N}. Such data offer valuable information on the dynamics, composition, and evolution of the disks. For an in-depth description of the observational setup and correlator configuration, please refer to \cite{2023ApJ...951....8O}.

For two sources in our sample, TMC-1A and B335, comparable ALMA observations were previously conducted and were retrieved from the ALMA archive for inclusion in our eDisk study. Additionally, due to scheduling constraints, only short baseline data were obtained for Oph IRS 63 as part of the eDisk program. To supplement this, we incorporated long baseline data from archival ALMA observations of Oph IRS 63 (ALMA Program ID 2015.1.01512.S PI: D. Segura-Cox).

\subsection{Data Reduction}

{The initial calibration was restored using the standard ALMA pipeline packaged with CASA version 6.2.1 \citep[CASA;][]{casa2022} following a custom script developed specifically for the eDisk Large Program \citep{Tobin2023}. For QA0 “semi-pass” sources affected by severe phase decorrelation the raw data was manually reprocessed to assess their potential for self-calibration. All subsequent operations were performed on individual Measurement Sets (MSs) rather than concatenated files, preventing interpolation artifacts, spectral window offsetting, and metadata corruption. Spectrally averaged continuum datasets were constructed by flagging expected lines and line-free ranges identified in the pipeline cont.dat files, translating LSRK frequency ranges into channel ranges, and averaging windows down to $\sim 30\text{ MHz}$ channels. Relative flux density scales between Execution Blocks (EBs) were normalized by comparing azimuthally averaged visibility amplitude profiles across overlapping baselines out to $800\text{ k}\lambda$, resolving typical $10\text{--}15\%$ inter-EB calibration offsets. For targets severely degraded by decorrelation, a two-pass method was applied where self-calibration was completed first to derive accurate relative scale factors prior to re-running the reduction pipeline.}

{Self-calibration was performed on the compact configuration data prior to combining them with extended configuration observations. Calibration solution intervals were structured to evenly divide scans, ranging from full observation durations (inf\_EB) to isolate systematic antenna position offsets down to single integrations (int) to mitigate rapid atmospheric phase fluctuations. Deconvolution during self-calibration utilized CASA’s 'tclean' with auto-multithresh masking, paired with clean thresholds that decreased logarithmically from Peak (S/N)/15 (i.e., begin with the assumption that we would trust features with intensities $>$ (Peak S/N)/15, which means that for a Peak S/N = 150, we would trust features $>$ 10$\sigma$.) down to 3$\sigma$.  Incremental gain solutions were calculated across all available bandwidth using combine='spw' and linear phase frequency scaling (interp='linearPD'). Iterations were retained only if the re-imaged S/N improved using the starting model, with phase-only self-calibration performed prior to attempting amplitude self-calibration. }

{Final continuum images were synthesized across a comprehensive parameter space to evaluate the trade-off between angular resolution and surface brightness sensitivity. Deconvolution was executed across Briggs robust parameters ranging from -2 to +2, as well as with $uv$-tapering applied starting at 1000 k$\lambda$, 2000 k$\lambda$, and 3000 k$\lambda$ (for robust values of 1 and 2). Standard image grids comprised 6000 $\times$ 6000 pixels with a pixel scale of 0.003\arcsec, expanding to 14000 $\times$ 14000 pixels for wide-field targets, achieving a typical synthesized beam size of ~0.05\arcsec at robust = 0.5. For further details see \cite{2023ApJ...951....8O}. }

\begin{deluxetable*}{ccccccccccc}
\tablecaption{Target information \label{tab:obs}}
\tablehead{
Name & RA & Dec & Class & Distance & T$\rm _{bol}$ & L$\rm _{bol}$ & $M_{\star}$ & rms &   ref \\
 & (hms) & (dms) & & (pc) & (K)& (L$_\odot$) & (M$_\odot$) & ($\mu$Jy / beam) &  \\
}
\startdata
BHR71 IRS2 & 12:01:34.008 & -65:08:48.08 & 0 & 176 & 39 & 1.1 & $0.20_{-0.04}^{+0.10}$ &30 &1 \\
B335 & 19:37:00.900 & +07:34:09.81 & 0 & 165 & 41 & 1.4 & $0.08_{-0.01}^{+0.11}$& 25  &2 \\
L1527 IRS & 04:39:53.878 & +26:03:09.43 & 0 & 140 & 41 & 1.3 & $0.42_{-0.07}^{+0.04}$ &18 &3 \\
IRAS 16253-2429 & 16:28:21.615 & -24:36:24.33 & 0 & 139 & 42 & 0.16 & $0.11_{-0.04}^{+0.05}$ & 23  &4 \\
IRAS 16544-1604 (CB 68) & 16:57:19.643 & -16:09:24.02 & 0 & 151 & 50 & 0.89 & $0.19_{-0.05}^{+0.09}$ &20  &5 \\
GSS30 IRS3 & 16:26:21.715 & -24:22:51.09 & 0 & 138 & 50 & 1.7 & $0.36_{-0.06}^{+0.10}$ &19  &6 \\
IRAS 15398-3359 & 15:43:02.232 & -34:09:06.96 & 0 & 155 & 50 & 1.4 & $0.09_{-0.03}^{+0.03}$ &34  &7 \\
R CrA IRS5N & 19:01:48.480 & -36:57:15.39 & 0 & 147 & 59 & 1.4 & $0.36_{-0.05}^{+0.1}$&14  &8 \\
IRAS 04166+2706 & 04:19:42.505 & +27:13:35.83 & 0 & 156 & 61 & 0.4 & $0.28_{-0.09}^{+0.19}$ &19 &9 \\
R CrA IRAS 32A & 19:02:58.722 & -37:07:37.39 & 0 & 150 & 64 & 1.6 & $0.60_{-0.14}^{+0.22}$&24  &10 \\
R CrA IRAS 32B & 19:02:58.642 & -37:07:36.39 & 0 & 150 & --- & --- & $0.29_{-0.07}^{+0.12}$&24  &10 \\
BHR71 IRS1 & 12:01:36.476 & -65:08:49.37 & 0 & 176 & 66 & 10 & $0.39_{-0.10}^{+0.09}$  &48 &1 \\
Ced110 IRS4A & 11:06:46.369 & -77:22:32.88 & 0 & 189 & 68 & 1 & $1.61_{-0.24}^{+0.22}$ &15 &11 \\
Ced110 IRS4B & 11:06:46.772 & -77:22:32.76 & 0 & 189 & 68 & 10 & $0.04_{-0.02}^{+0.02}$ &15 &11 \\
IRAS 04302+2247 & 04:33:16.499 & +22:53:20.23 & I & 160 & 88 & 0.43 & $1.91_{-0.27}^{+0.12}$ &15 &12 \\
R CrA IRS7B-a & 19:01:56.420 & -36:57:28.66 & I & 152 & 88 & 5.1 & $2.71_{-0.26}^{+0.44}$&32  &13 \\
R CrA IRS7B-b & 19:01:56.385 & -36:57:28.11 & I & 152 & --- & --- & --- &32 &13 \\
IRAS 04169+2702 & 04:19:58.477 & +27:09:56.82 & I & 156 & 163 & 1.5 & $0.69_{-0.14}^{+0.57}$ & 19 &14 \\
TMC-1A & 04:39:35.202 & +25:41:44.22 & I & 137 & 183 & 2.3 & $0.87_{-0.06}^{+0.08}$ &28 &2 \\
Oph IRS43 VLA1 & 16:27:26.906 & -24:40:50.81 & I & 137 & 193 & 4.1 & $5.00_{-3.00}^{+1.00}$ &53 &15 \\
Oph IRS43 VLA2 & 16:27:26.911 & -24:40:51.40 & I & 137 & --- & --- & $1.00_{-0.3}^{+0.3}$& 53&15 \\
L1489 IRS & 04:04:43.080 & +26:18:56.12 & I & 146 & 213 & 3.4 & $1.84_{-0.27}^{+0.22}$ &17 &16 \\
Oph IRS63 & 16:31:35.654 & -24:01:30.08 & I & 132 & 348 & 1.3 & $0.62_{-0.14}^{+0.12}$ &16  &17 \\
\enddata
\tablecomments{
References: (1) \citet{2024ApJ...974...21G}, (2) \citet{2023ApJ...951....8O}, (3) \citet{2023ApJ...951...10V}, (4) \citet{2023ApJ...954..101A}, (5) \citet{2023ApJ...953..190K}, (6) \citet{2024AA...690A..46S}, (7) \citet{2023ApJ...958...60T}, (8) \citet{2023ApJ...954...69S}, (9) \citet{2025ApJ...992...18P}, (10) \citet{2024ApJ...966...32E}, (11) \citet{2023ApJ...954...67S}, (12) \citet{2023ApJ...951....9L}, (13) \citet{2024ApJ...969..125Y}, (14) \citet{2025ApJ...993..120H}, (15) \citet{2023ApJ...958...20N}, (16) \citet{2023ApJ...951...11Y}, (17) \citet{2023ApJ...958...98F}. The protostellar properties (L$_\mathrm{bol}$, T$_\mathrm{bol}$, Distance) are from \citet{2023ApJ...951....8O} and \citet{2024ApJ...969..125Y} (but also see \citealt{2023ApJS..266...32P}, \citealt{2024ApJ...962L..16N}), while the protostellar masses ($M_{\star}$) are from Aso et al. (2026 submitted) when available. 
 }
\end{deluxetable*}

\section{UV visibility analysis}

To streamline the analysis and improve consistency, we categorized the eDisk sample into three distinct model classes: Inclined disks, binary, and edge-on disks, based on their morphological features. This classification served as a framework to guide the fitting process, enabling us to tailor the models to the specific structural characteristics observed within each disk.

\subsection{Inclined disks} \label{sec:Inclined math}

For inclined disks, we start with two primary models: a standard Gaussian profile and a modified viscous accretion disk model from \cite{1974MNRAS.168..603L}, commonly referred to as the power-law core with an exponential tail (PLCT) profile. The Gaussian profile is the simplest model and has been widely used to describe a range of astrophysical targets \citep[e.g.,][]{2016ApJ...828...46A,2023AJ....166..184M}. Below, we provide detailed descriptions of the standard models applied.

The Gaussian profile is given as 

\begin{equation}
 I_{\nu, g} (r) = I_{\nu,0} \exp{\left(-\frac{r^2}{2 \, {r_{w}}^2}\right)}\\
\end{equation}
where $I_{\nu, g}(r)$ is the intensity as a function of radius $r$, $r_{w}$ is the standard deviation width of the disk. For a disk with inclination ($i$) and position angle ($p.a.$), we can express the coordinate system and $r$ of the disk as 

\begin{gather} \label{eq:pa&i}
 xp = (x-x_0) \cos(p.a.) + (y-y_0) \sin(p.a.)\\ 
 yp = -(x-x_0) \sin(p.a.) + (y-y_0) \cos(p.a.)\\ 
 r = \sqrt{{xp}^2 + \frac{{yp}^2}{\cos^2 i}}
\end{gather}
where $x$ and $y$ are the sky-plane coordinates along the major and minor axis of the disk, while $x_0$ and $y_0$ are the centers of the disk in the $xy$ coordinate frame. 

The PLCT profile is a modification of the standard viscous accretion disk profile with two-components, a power law that represents the inner protostellar disk, and the exponential {tail} that represents the edge of the outer disk and the inner envelope component 
as

\begin{equation}
 I_{\nu, PLCT} (r) = I_{\nu,0} \, \left(\frac{r}{{r_{w}}}\right)^{-\gamma_1} \, \exp{\left[-\left(\frac{r}{{r_{w}}}\right)^{2-\gamma_2}\right]} \,.
 \label{eq:PLCT}
\end{equation}

If $\gamma_1$ = $\gamma_2$ we get the standard PLCT profile from \cite{1974MNRAS.168..603L} and if $\gamma_1$ = $\gamma_2$ = 0 we retrieve the standard Gaussian profile. 
Allowing the two power-law indices to vary independently provides additional flexibility and yields improved fits to the observed intensity distributions for many sources. One of the key advantages of the Gaussian and the PLCT models is their widespread use in the literature, which can allow for direct comparisons between our findings and those from previous studies. Additionally, the 2D Gaussian model is particularly notable as it is commonly used for determining disk sizes. This familiarity and established use in the community make it a reliable baseline for our analysis.

{In practice, to avoid a singularity at the disk center, the radial coordinate $r$ in all power-law terms is replaced by $\sqrt{r^2+\mathrm{smooth}^2}$, where parameter $\mathrm{smooth}$ is fixed to a negligibly small value (0.1 pixel).}

We can further add more complexities to the model by adding rings, gaps, and azimuthally asymmetric components. These components capture the substructures and the brightness asymmetries often seen in protoplanetary disks. 

The brightness distribution of a ring is given by,
\begin{equation}
 F_{\nu, ring}(r) = F_{\nu,r_0} \exp{\left(-\frac{(r - r_{c,r})^4}{2 \, {r_{w,r}}^4}\right)} \label{eq:exp_gap2}
\end{equation} 
where $r_{c,r}$ is the radius of the center of the ring, and $r_{w,r}$ is the half-width of the ring. We used the Gaussian to the power 4 as it  changes the profile from a classic bell shape to a flat-top, sharp-shouldered profile, which is likely suitable for rings. If $F_{\nu,r_0}$ is positive then the disk model has a bright ring but if negative values the model has a gap. This is a multiplicative component rather than an additive component to the PLCT that modulates the brightness intensity. 

The azimuthally asymmetric component can either be described as,

\begin{equation}
 F_{\nu,a,}(r,\phi) = F_{\nu,a_0} \exp{\left(-\frac{(r - r_{c,a})^4}{2 \, {r_{w,a}}^4}\right)} \, \exp{\left(-\frac{(\phi - \phi_{c,a})^4}{2 \, {\phi_{w,a}}^4}\right)}
 \label{eq7}
\end{equation}
where $r_{c,a}$ is the radial center of the brightness asymmetry and $\phi_{c,a}$ is the azimuthal center of the asymmetry, $r_{w,a}$ and $\phi_{w,a}$ are the characteristic radial and azimuthal widths \citep{2013Sci...340.1199V}. Again depending on the sign of $F_{\nu,a_0}$ we can add a bright spot  or a dip in the flux. These rings and brightness asymmetries can be directly multiplied to the PLCT profile to produce the disk intensity profile $I_{\nu, disk}(r)$:

\begin{equation}
\begin{aligned}
I_{\nu,\mathrm{disk}}(r,\phi) &= I_{\nu,\mathrm{PLCT}}(r) \times \Big(1 + F_{\nu,\mathrm{ring}}(r) \\
&\qquad + F_{\nu,a}(r,\phi) + F_{\nu,a2}(r,\phi)\Big)
\end{aligned}
 \label{eq9}
\end{equation}

The multiplicative and additive nature of our models enables us to account for multiple gaps and radial asymmetries.  One of the key advantages of our model definition and implementation is its inherent flexibility, which allows for the incorporation of additional complexities into the base model (see e.g., Equation \ref{eq9}). This modular approach enables us to iteratively refine and expand upon previous best-fit results. Given the diversity observed in disk structures, no single model can adequately reproduce all disks with high fidelity. To address this, we conducted an extensive exploration of various asymmetry models, systematically building upon the foundational PLCT profile. We can add the azimuthally asymmetric component as described in equation \ref{eq7}, which adds an asymmetry at a radius $r_{c,a}$ and angle $\phi_{c,a}$. The sign of $I_{\nu, a_0}$ indicates if its a brightness enhancement or a dark spot. This gives us the PLCT with asymmetry model (see Section 4.1.3). By including asymmetry into the PLCT model, we aim to explore more complex structures and physical properties of the disks that might not be fully captured by the simpler, symmetric models. This combination of models provides a robust framework for evaluating and interpreting the disk structures in our sample. We also made our best fit models incorporating additional components beyond the standard models (see Section \ref{sec:best-fit} and Appendix \ref{advance_models}).

\subsection{Edge-on disks} \label{sec:edge-on math}

Simple Gaussian and modified PLCT-based disk intensity models fail to reproduce the observed profiles of edge-on disks (L1527 and IRAS 04302+2247). For these sources, we adopted the modeling approach described by \citet{2022ApJ...934...95S}. Specifically, we used an exponentially tapered rectangular profile to avoid sharp edges and abrupt truncation, which can introduce artifacts such as ringing in the uv-domain.

\begin{equation}
 \label{equation:rectangle}
 I_{\nu,rec} = I_{\nu,0} \, \exp\left\{ -\frac{\left|x - x_0\right|^{\gamma_x}}{2 \, x_w^{\gamma_x}} - \frac{\left|y - y_0\right|^{\gamma_y}}{2 \, y_w^{\gamma_y}}\right\}
\end{equation}
where $x_w$ and $y_w$ represent the width of the rectangle along the major and minor axes of the disk, $\gamma_x$ and $\gamma_y$ characterize how smoothly/sharply the rectangle is tapered beyond $x_w$ and $y_w$. {Following \cite{2022ApJ...934...95S}, we fix $\gamma_x=\gamma_y=4$ in the baseline rectangle models. These exponents are treated as free parameters only in the models that include the flaring term, which will be described later. As can be seen in Figure \ref{Fig1} and from \cite{2022ApJ...934...95S} we see that the edge-on disks can vary in brightness along the major axis ($x$).} We can add a power-law modulation term to the rectangle model $I_{\nu,rec}$ to account for it that is offset from the disk center by $\delta$ along the major axis which is given as: 

\begin{equation}
 I_{pl} = I_{\nu,rec} \left(\frac{\left|x'\right|}{x_w}\right)^{-\gamma}
\end{equation}
where $x'$ is $x+\delta$, and $\gamma$ is the power-law index. We note that the disk width $y_w$ changes as we move away from the disk center. To account for the disk flaring along the vertical axis we introduce a flaring term to $y_w$. 

\begin{equation}
 y_w = y_{w,0} \, \left[1 + A \left(\frac{\left|x'\right|}{x_w}\right)\right],
\end{equation}
where $A$ controls how much wider the disk is at $x' = x_w$ compared to $x' = 0$ and can be turned on/off depending on the models. When the flaring term is included, we also allow $\gamma_x$ and $\gamma_y$ in Equation~\ref{equation:rectangle} to vary freely. We further observed that the disk width remains relatively flat near the center, while the flaring-induced widening becomes more apparent at larger radii. To account for this behavior and to smooth the sharp curvature at the center caused by the direct connection of flaring components from both sides, we introduced a sigmoid function to regulate the flaring term:
\begin{equation}
 f_{\mathrm{sig}}(x') = \frac{1}{1+\exp{\left(-\frac{\left|x'\right|-x_s}{w}\right)}}.
\end{equation}
The flaring term then becomes:
\begin{equation}
 y_w = y_{w,0} \, \left[1 + A \left(\frac{\left|x'\right|}{x_w}\right) \cdot f_{\mathrm{sig}}(x')\right] \label{eq:sigmoid_flaring}
\end{equation}
where $x_s$ controls the turn-on radius of the flaring, and $w$ determines the width of this transition. This formulation ensures that the disk remains flat and smooth near the center and gradually flares out at larger radii.

In practice, to avoid a singularity at the disk center, $\left|x'\right|$ in the power-law terms is replaced by 

\begin{equation}
 \mathbf {\left|x'\right| \mapsto  \sqrt{x'^2+\mathrm{smooth}^2} }
\end{equation}

where $\mathrm{smooth}$ is fixed to a negligibly small value (0.1 pixel). In the flared models, we allow $\mathrm{smooth}$ to vary freely, so that it no longer serves as a mere softening, but controls the shape of the central brightness profile. Variable $\mathrm{smooth}$ sets the scale interior to which the power-law modulation saturates to a constant. Since the rectangle-based models are constructed symmetrically about the disk midline, $\mathrm{smooth}$ determines the width over which the two sides of the profile are smoothly joined, which becomes particularly important when the flaring term allows the two sides to deform independently.

We can further modify the simple power-law component, which can account for a more peaked central brightness. We used a smooth broken power-law $I_{\nu, bpl}$ with two power-law indexes $\gamma_{in}$ and $\gamma_{out}$ representing the inner and outer power-law slopes

\begin{equation}
 \label{equation:broken_powerlaw}
 I_{\nu, bpl} = I_{\nu,rec} \left(\frac{\left|x'\right|}{x_b}\right)^{-\gamma_{in}} \, \left\{\frac{1}{2}\left[1 + \left(\frac{\left|x'\right|}{x_b}\right)^{1/\Delta}\right]\right\}^{(\gamma_{in} - \gamma_{out})\Delta}
\end{equation}
where $x_b$ is the point where the transition
from inner to outer slope occurs and $\Delta$ controls the smoothness of the transition from the inner to outer power-law. 

To account for asymmetric brightness profiles on the two sides of the disk's major axis, we allow the model to adopt two distinct outer slope indices, $\gamma_{out,1}$ and $\gamma_{out,2}$, corresponding to the regions on each side of the major axis, respectively. To ensure a smooth transition between two slopes across the disk mid-plane, we introduce a weighting function based on a hyperbolic tangent:
\begin{gather}
 f_{w}(x') = \frac{1}{2}\left[1 + \tanh\left(\frac{x'}{\epsilon}\right)\right] \\
 \gamma_{\mathrm{out}}(x') = f_{w}(x') \gamma_{\mathrm{out},1} + \left[1 - f_{w}(x')\right] \gamma_{\mathrm{out},2} \label{eq:smooth_gamma_transition}
\end{gather}
where $\epsilon$ is a transition scale parameter that controls how rapidly the model switches between $\gamma_{out,1}$ and $\gamma_{out,2}$. This formulation guarantees a smooth continuous slope transition near the disk center and ensures that at larger radii each of the two outer slopes dominates its respective side of the disk.

Even though for our edge-on systems we do not see any such gaps in Figure \ref{Fig1}, we opted to also include a gap model for our fitting of the edge-on disks. \cite{2022ApJ...934...95S} for the L1527 IRS disk showed that models with gaps did produce the highest Bayesian evidence, but it was not statistically significantly better than models without a gap. To prevent ringing in the model from sharply truncating the intensity in the gap, we use a smooth gap model parameterized as a Gaussian subtracted from the gap-free model.
\begin{equation}
 I_{\nu, gapped} = I_{\nu, bpl} \, \left\{1 - (1 - \Delta_{gap}) \, \exp\left[-\frac{(\left|x'\right| - x_{gap})^2}{2 \, {w_{gap}}^2}\right]\right\} \label{eq:edge-on-bpl_sym_gapped}
\end{equation}
Here $x_{gap}$ represents the center of the gap, $\Delta_{gap}$ the multiplicative factor by which the intensity is reduced at the center of the gap, and $w_{gap}$ the width of the gap.

For IRAS 04302, the disk exhibits an asymmetric structure primarily along the minor-axis direction, while the structure itself is extended along the major axis. To reproduce this morphology, we introduce an additional asymmetric component, which is a localized enhancement aligned with the disk midplane. This component is given by:
\begin{gather}
 S(x,y) = S_x(x) S_y(y)\\
 S_x(x) = \exp{\left[-\left(\max{\left(0,\frac{|x-x_g|}{x_{w,g}}-1\right)}\right)^4\right]}\\
 S_y(y) = \exp\left[-\left(\frac{y-y_g}{y_{w,g}}\right)^4\right]
\end{gather}
Here, $x_g$ and $y_g$ define the location of the structure, while $x_{w,g}$ and $y_{w,g}$ control its extent along the major and minor axes, respectively. This structural component is implemented as a multiplicative modulation to the base model:
\begin{equation}
 I_{\nu, disk}(x,y) = I_{\nu, bpl}(x,y)[1+ a \cdot S(x,y)] \label{eq:edge-on-bpl-ystructure}
\end{equation}
This formulation allows the asymmetric structure to extend along the major axis with a flat-top 4th-order Gaussian-like profile, while dropping rapidly by 4th-order Gaussian profile along the minor axis.

\subsection{Binary sources}
Three of our sources are close binaries (Ced 110 IRS4, R CrA IRS7B, and Oph IRS43 VLA1). In these cases, we modeled the primary component using the models described in Section~3.1, while the much weaker and more compact secondary component was fitted using either a simple Gaussian or a PLCT model. By fitting the positional offsets, $x_0$ and $y_0$, the code is able to automatically determine the center of the secondary component.  Then by simply adding the two models we can fit the uv visibilities of the system such as 

\begin{equation}
 I_{\nu, binary} = I_{\nu, primary} + I_{\nu, secondary}
\end{equation}

\subsection{Final fit} \label{sec:final_fit}

Using the above models we can fit the disk emission, but there might be some contribution from the envelope especially at shorter baselines. To account for envelope emission we also add a large Gaussian $I_{\nu, E}$: 

\begin{equation} \label{eq:genv}
 I_{\nu, E} (r) = I_{\nu,E_0} \exp{\left(-\frac{r^2}{2 \, {r_{En}}^2}\right)}\\
\end{equation}
where $r_{En}$ is the envelope radius.

Similar to \cite{2020ApJ...902..141S, 2022ApJ...934...95S} rather than fitting for the peak surface brightness we fit the total integrated flux. The total integrated flux $I_{\nu}$ is given as: 

\begin{equation}
 I_{\nu} = I_{\nu, E} + I_{\nu,disk}\\
\end{equation}
where $I_{\nu,disk}$ can be either one of our inclined model or edge-on model or even our binary model. \\

\subsection{Model Fitting}

{To fit the visibilities we} utilized the Python package pdspy\footnote{\url{https://pdspy.readthedocs.io/en/latest/index.html}} \citep{2017ApJ...851...45S,2019ApJ...874..136S,2022ApJ...934...95S}, which provides tools for reading the observed measurement sets and extracts the visibilities into Python data structures. {{The visibilities were first averaged onto a two-dimensional uv grid with a cell size of 4 k$\lambda$ using the pdspy uv.average routine, with multi-frequency synthesis (mfs) enabled, which  combines visibilities observed at different frequencies by scaling their $u,v$ coordinates to a common reference frequency. The resulting real and imaginary visibilities are shown as a function of deprojected baseline length in Figure \ref{fig:vis}.}}

Model visibilities were generated by Fourier-transforming a synthetic model intensity image and sampling it with the same baselines as the observed data in the u-v plane. This process was performed using the \textit{GALARIO} library \citep{2018MNRAS.476.4527T}, which enables efficient computation of model visibilities. The resulting model visibilities were then subtracted from the observed data to evaluate the fit.

We employed the \textit{dynesty} package \citep{2020MNRAS.493.3132S}, a nested sampling algorithm, to explore the parameter space and sample the posterior distributions of the model parameters. Nested sampling is particularly advantageous in this context, as it provides robust estimates of the Bayesian evidence, or Log$Z$, for comparing different models. This metric allowed us to assess the relative statistical significance of different models in replicating the observed features. The primary goal of this modeling effort is to develop a simplified framework capable of capturing the key features in the observed data. This approach provides a quantitative basis for evaluating the significance of the observed features. 

{Model parameter space was sampled using the dynamic nested sampling algorithm implemented in \texttt{dynesty} \citep{2020MNRAS.493.3132S}. Uniform priors were assigned to all free parameters across bounded physical intervals. Sampler convergence was dictated by the remaining Bayesian evidence, terminating when the estimated remaining evidence dropped below $\Delta \ln \mathcal{Z} < 0.05$ for inclined disks or $\Delta \ln \mathcal{Z} < 1.0$ for edge-on geometries. Post-run convergence was confirmed by manually inspecting diagnostic sampling plots to ensure smooth parameter evolution and convergence, and by verifying that all posterior probability distributions were well-resolved (ideally Gaussian) without railing against prior boundaries.}

\begin{figure*}
\centering
\includegraphics[width=0.85\linewidth]{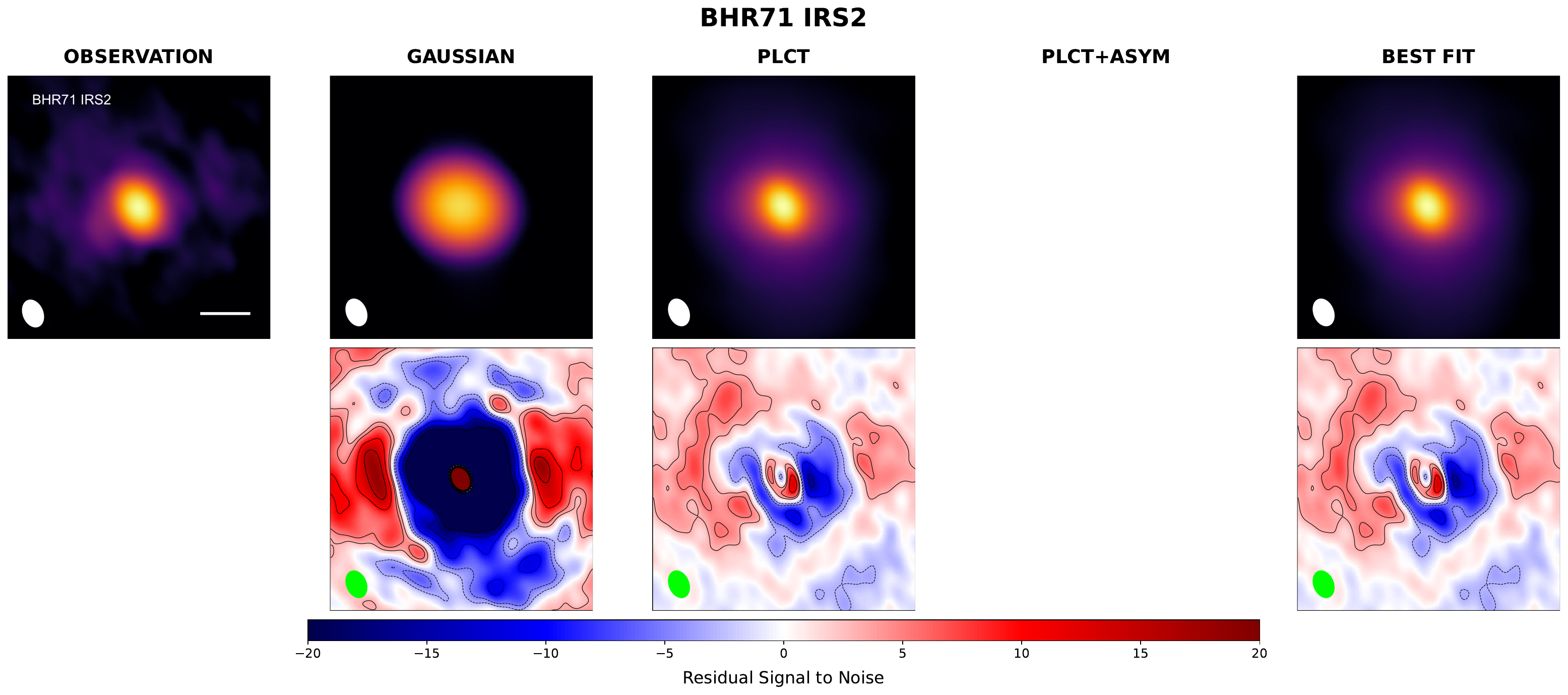}
\includegraphics[width=0.85\linewidth]{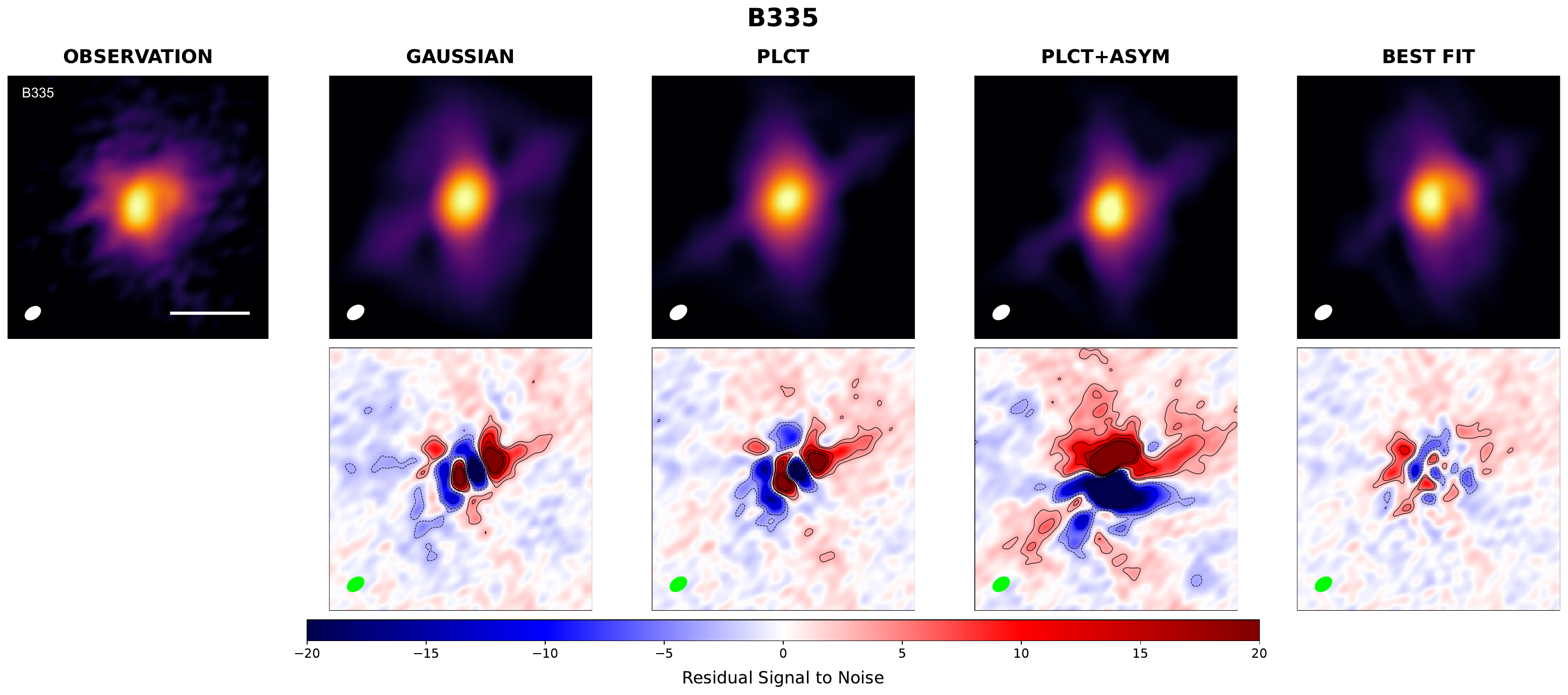}
\includegraphics[width=0.85\linewidth]{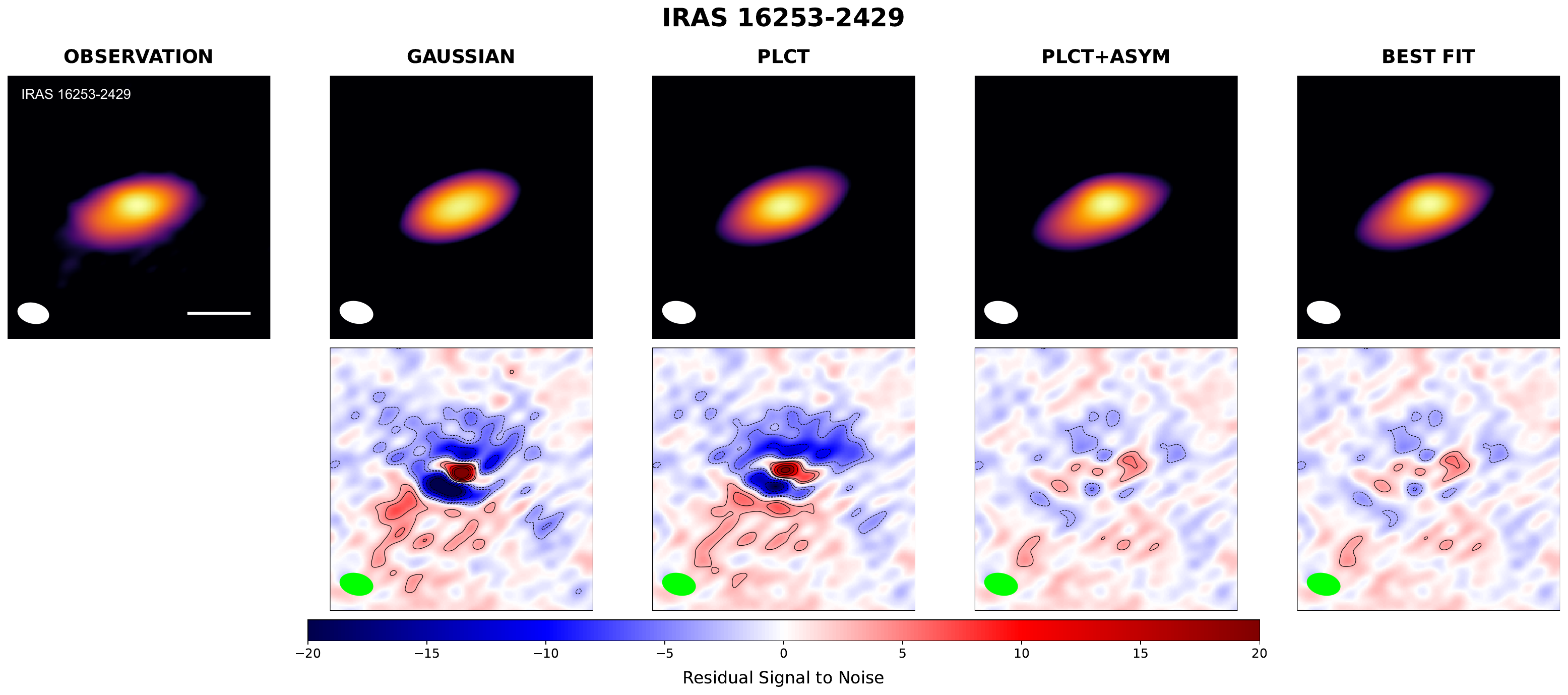}
\caption{The various fits for the eDisk targets are shown. For each source, the first column in the top row presents the ALMA eDisk observations, the second column shows the Gaussian fit, the third column displays the PLCT profile, the fourth column presents the asymmetric PLCT profile, and the fifth column shows the best-fit model. The color scale for the observations and models are same as Figure \ref{Fig1} and \cite{2023ApJ...951....8O}. The beam size and the scale of 20 au are shown in the lower left and lower right corners of each panel. The residuals are shown in the bottom row. The residual contours represent signal-to-noise ratios of [-20, -15, -10, -5, -3, 3, 10, 15, 20], based on the background rms values specific to each source (see Table~\ref{tab:obs}).}
\label{Fig2}
\end{figure*}

\figsetstart
\figsetnum{2}
\figsettitle{Model Comparisons for eDisk Targets}

\figsetgrpstart
\figsetgrpnum{2.1}
\figsetgrptitle{BHR71 IRS2}
\figsetplot{BHR71IRS2_model_comparison.pdf}
\figsetgrpnote{The various fits for the eDisk targets are shown. For each source, the first column in the top row presents the ALMA eDisk observations, the second column shows the Gaussian fit, the third column displays the PLCT profile, the fourth column presents the asymmetric PLCT profile, and the fifth column shows the best-fit model. The color scale for the observations and models are same as Figure \ref{Fig1} and \cite{2023ApJ...951....8O}. The beam size and the scale of 20 au are shown in the lower left and lower right corners of each panel. The residuals are shown in the bottom row. The residual contours represent signal-to-noise ratios of [-20, -15, -10, -5, -3, 3, 10, 15, 20], based on the background rms values specific to each source (see Table~\ref{tab:obs}).}
\figsetgrpend

\figsetgrpstart
\figsetgrpnum{2.2}
\figsetgrptitle{B335}
\figsetplot{B335_model_comparison.pdf}
\figsetgrpnote{The various fits for the eDisk targets are shown. For each source, the first column in the top row presents the ALMA eDisk observations, the second column shows the Gaussian fit, the third column displays the PLCT profile, the fourth column presents the asymmetric PLCT profile, and the fifth column shows the best-fit model. The color scale for the observations and models are same as Figure \ref{Fig1} and \cite{2023ApJ...951....8O}. The beam size and the scale of 20 au are shown in the lower left and lower right corners of each panel. The residuals are shown in the bottom row. The residual contours represent signal-to-noise ratios of [-20, -15, -10, -5, -3, 3, 10, 15, 20], based on the background rms values specific to each source (see Table~\ref{tab:obs}).}
\figsetgrpend

\figsetgrpstart
\figsetgrpnum{2.3}
\figsetgrptitle{IRAS 16253}
\figsetplot{IRAS16253_model_comparison.pdf}
\figsetgrpnote{The various fits for the eDisk targets are shown. For each source, the first column in the top row presents the ALMA eDisk observations, the second column shows the Gaussian fit, the third column displays the PLCT profile, the fourth column presents the asymmetric PLCT profile, and the fifth column shows the best-fit model. The color scale for the observations and models are same as Figure \ref{Fig1} and \cite{2023ApJ...951....8O}. The beam size and the scale of 20 au are shown in the lower left and lower right corners of each panel. The residuals are shown in the bottom row. The residual contours represent signal-to-noise ratios of [-20, -15, -10, -5, -3, 3, 10, 15, 20], based on the background rms values specific to each source (see Table~\ref{tab:obs}).}
\figsetgrpend

\figsetgrpstart
\figsetgrpnum{2.4}
\figsetgrptitle{GSS30 IRS3}
\figsetplot{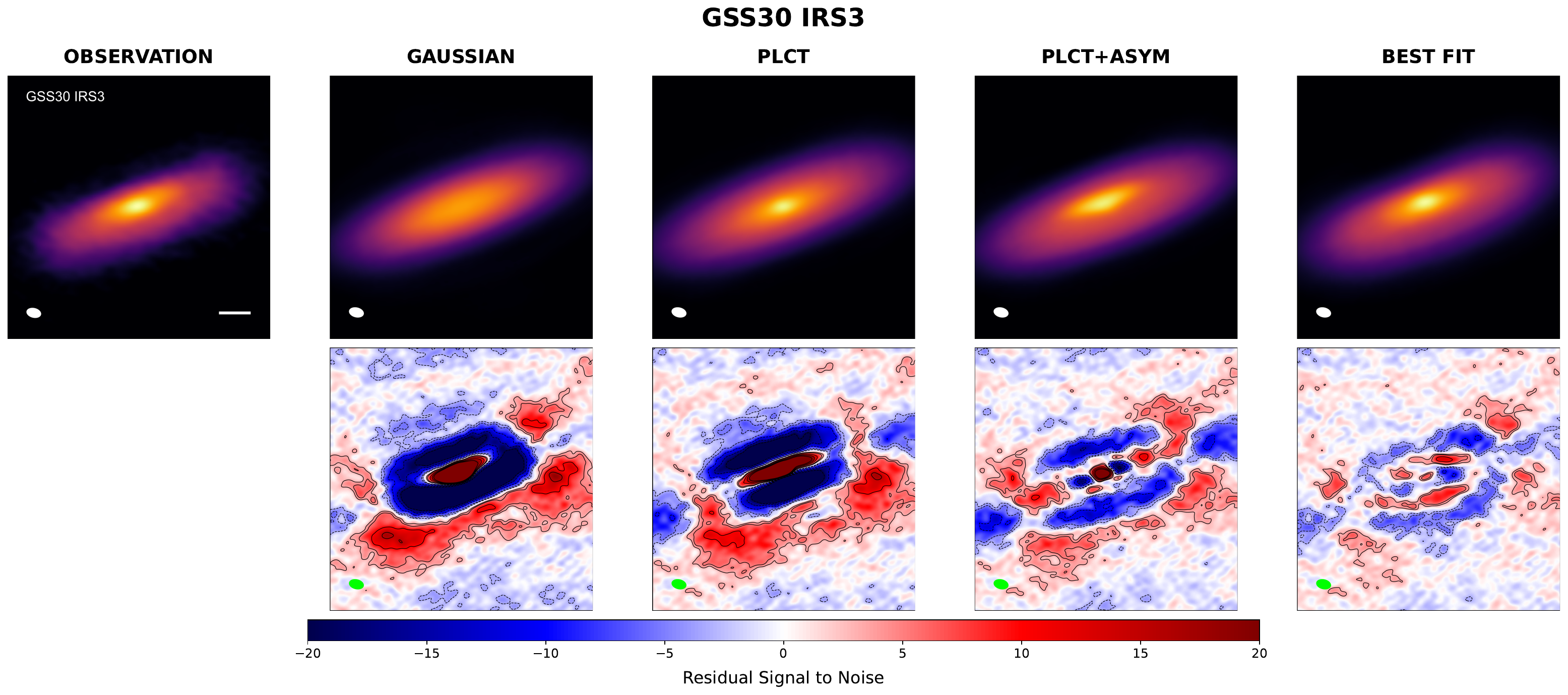}
\figsetgrpnote{The various fits for the eDisk targets are shown. For each source, the first column in the top row presents the ALMA eDisk observations, the second column shows the Gaussian fit, the third column displays the PLCT profile, the fourth column presents the asymmetric PLCT profile, and the fifth column shows the best-fit model. The color scale for the observations and models are same as Figure \ref{Fig1} and \cite{2023ApJ...951....8O}. The beam size and the scale of 20 au are shown in the lower left and lower right corners of each panel. The residuals are shown in the bottom row. The residual contours represent signal-to-noise ratios of [-20, -15, -10, -5, -3, 3, 10, 15, 20], based on the background rms values specific to each source (see Table~\ref{tab:obs}).}
\figsetgrpend

\figsetgrpstart
\figsetgrpnum{2.5}
\figsetgrptitle{IRAS 15398}
\figsetplot{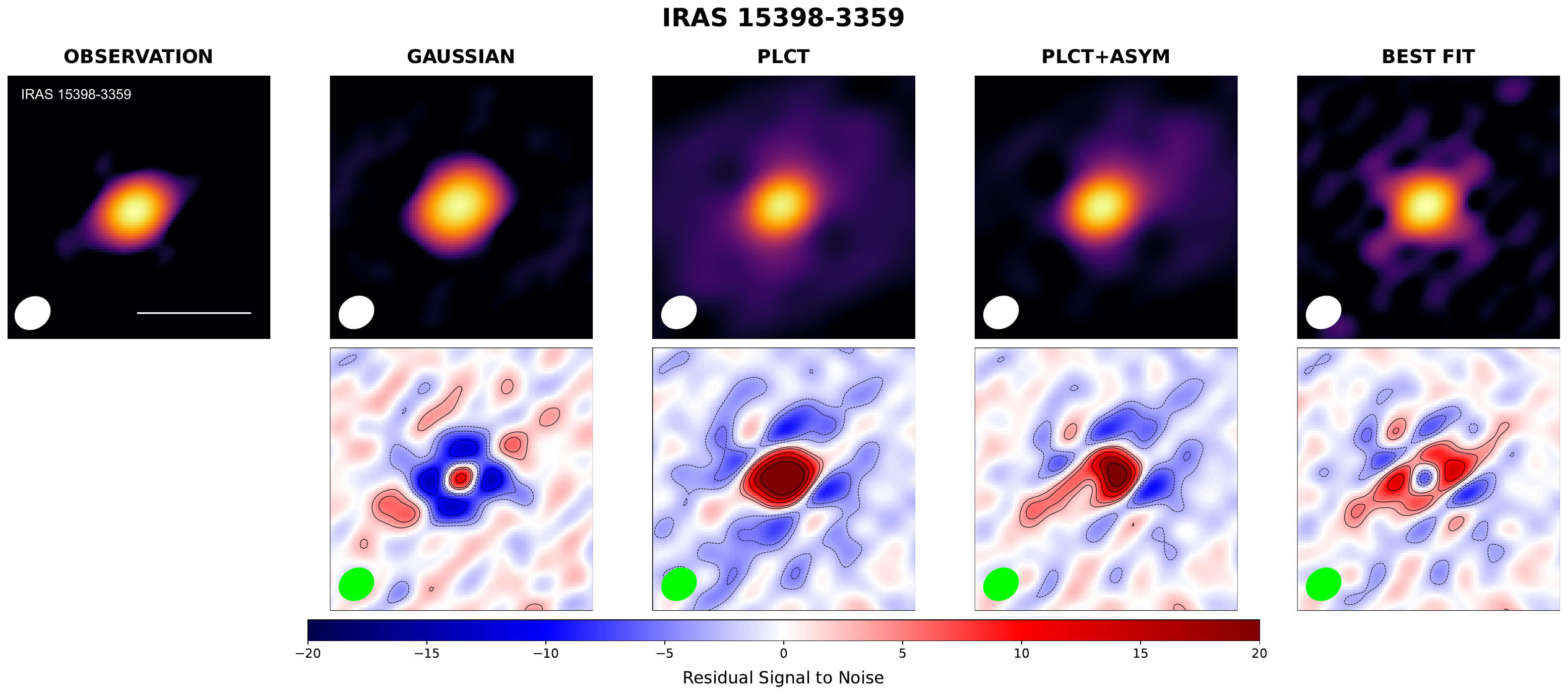}
\figsetgrpnote{The various fits for the eDisk targets are shown. For each source, the first column in the top row presents the ALMA eDisk observations, the second column shows the Gaussian fit, the third column displays the PLCT profile, the fourth column presents the asymmetric PLCT profile, and the fifth column shows the best-fit model. The color scale for the observations and models are same as Figure \ref{Fig1} and \cite{2023ApJ...951....8O}. The beam size and the scale of 20 au are shown in the lower left and lower right corners of each panel. The residuals are shown in the bottom row. The residual contours represent signal-to-noise ratios of [-20, -15, -10, -5, -3, 3, 10, 15, 20], based on the background rms values specific to each source (see Table~\ref{tab:obs}).}
\figsetgrpend

\figsetgrpstart
\figsetgrpnum{2.6}
\figsetgrptitle{CB68}
\figsetplot{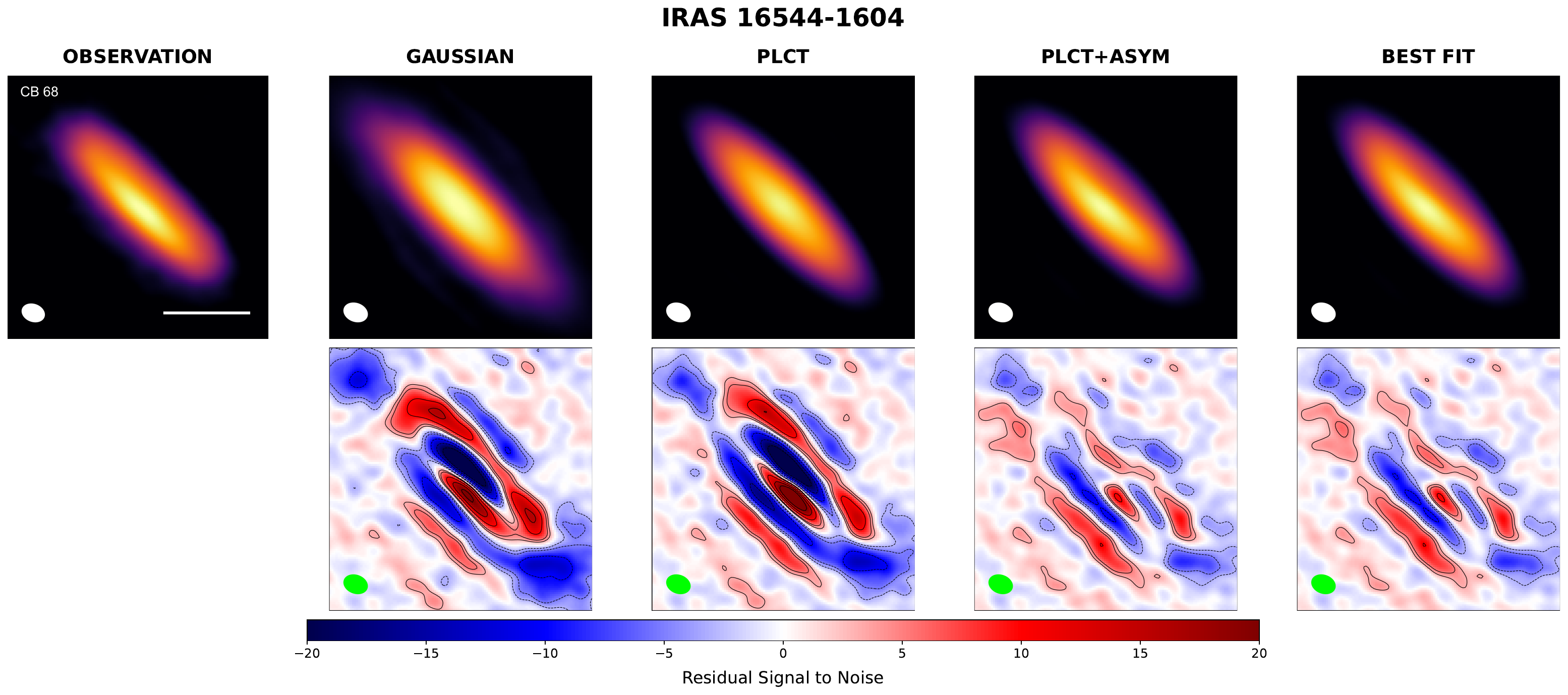}
\figsetgrpnote{The various fits for the eDisk targets are shown. For each source, the first column in the top row presents the ALMA eDisk observations, the second column shows the Gaussian fit, the third column displays the PLCT profile, the fourth column presents the asymmetric PLCT profile, and the fifth column shows the best-fit model. The color scale for the observations and models are same as Figure \ref{Fig1} and \cite{2023ApJ...951....8O}. The beam size and the scale of 20 au are shown in the lower left and lower right corners of each panel. The residuals are shown in the bottom row. The residual contours represent signal-to-noise ratios of [-20, -15, -10, -5, -3, 3, 10, 15, 20], based on the background rms values specific to each source (see Table~\ref{tab:obs}).}
\figsetgrpend

\figsetgrpstart
\figsetgrpnum{2.7}
\figsetgrptitle{IRS 5N}
\figsetplot{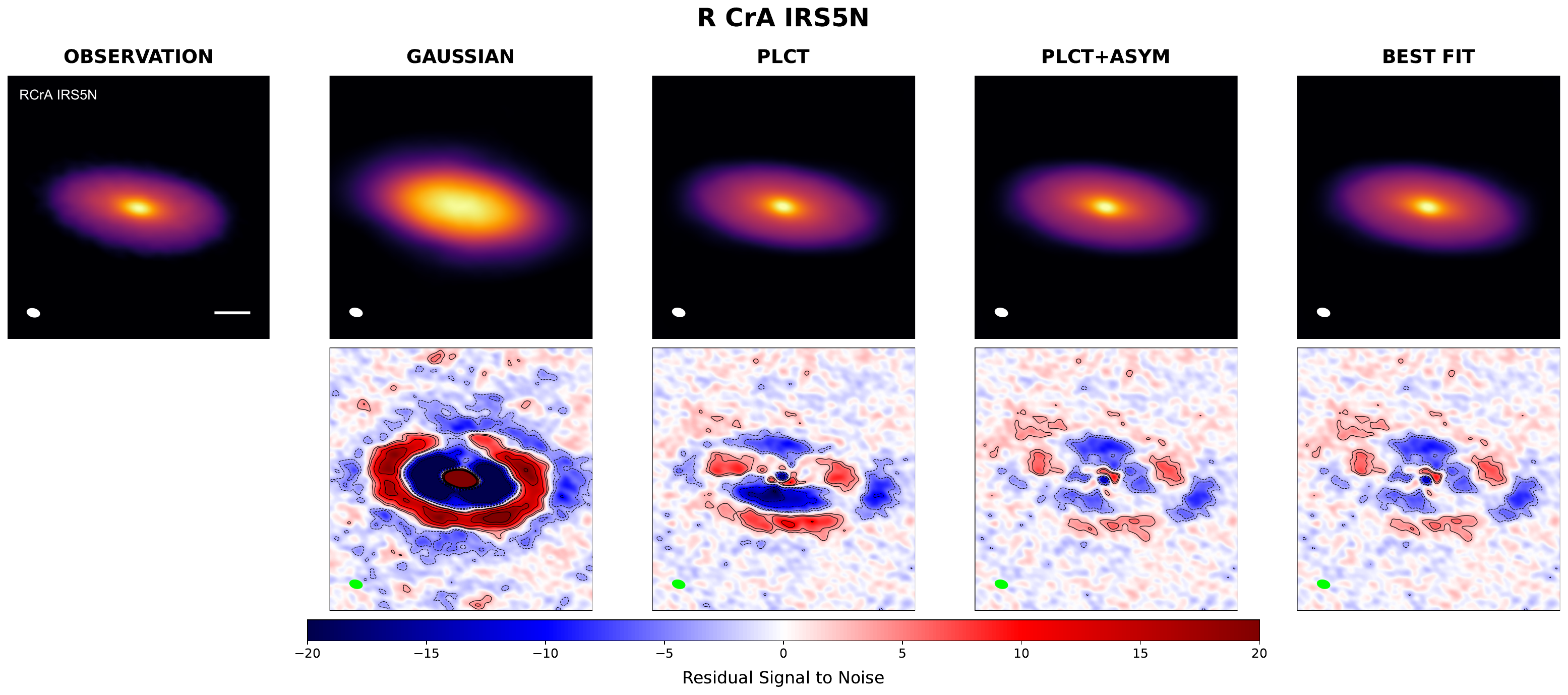}
\figsetgrpnote{The various fits for the eDisk targets are shown. For each source, the first column in the top row presents the ALMA eDisk observations, the second column shows the Gaussian fit, the third column displays the PLCT profile, the fourth column presents the asymmetric PLCT profile, and the fifth column shows the best-fit model. The color scale for the observations and models are same as Figure \ref{Fig1} and \cite{2023ApJ...951....8O}. The beam size and the scale of 20 au are shown in the lower left and lower right corners of each panel. The residuals are shown in the bottom row. The residual contours represent signal-to-noise ratios of [-20, -15, -10, -5, -3, 3, 10, 15, 20], based on the background rms values specific to each source (see Table~\ref{tab:obs}).}
\figsetgrpend

\figsetgrpstart
\figsetgrpnum{2.8}
\figsetgrptitle{IRAS 04166}
\figsetplot{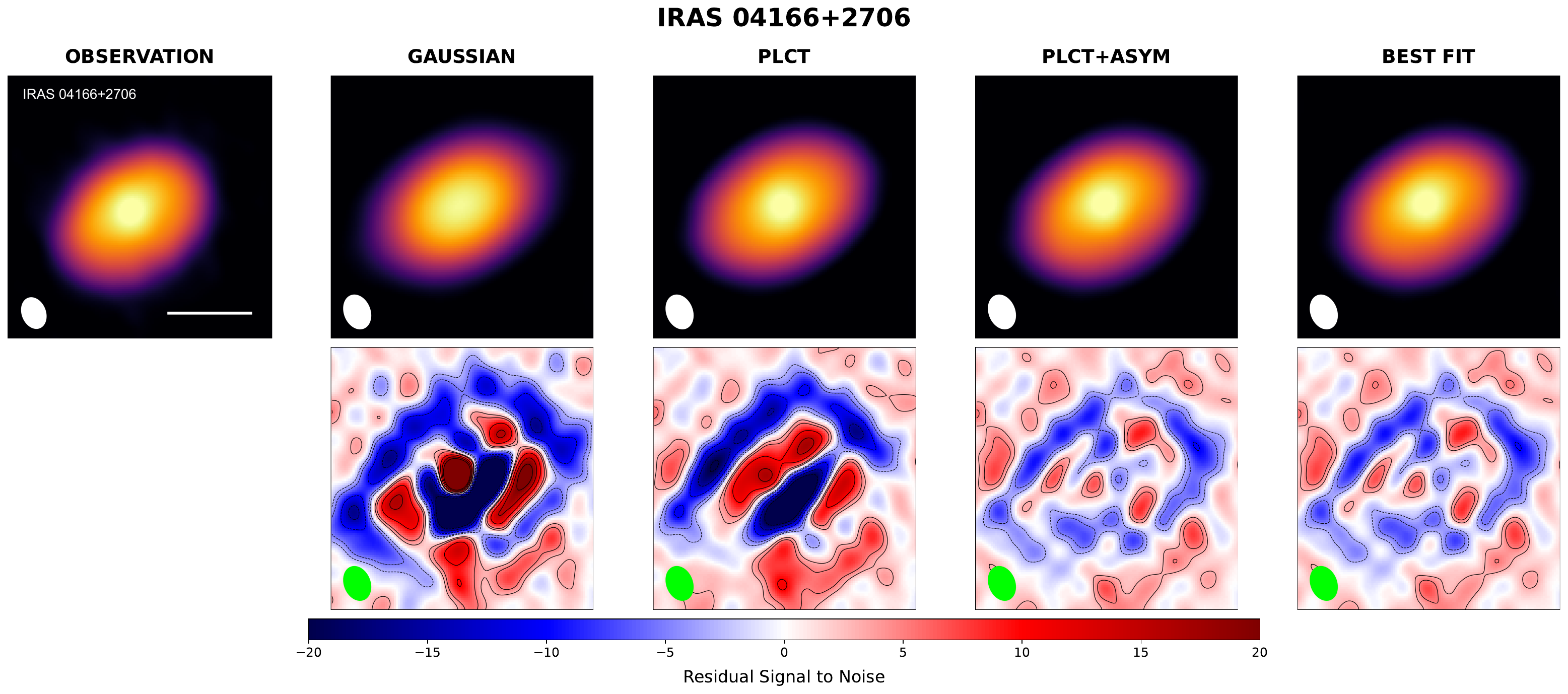}
\figsetgrpnote{The various fits for the eDisk targets are shown. For each source, the first column in the top row presents the ALMA eDisk observations, the second column shows the Gaussian fit, the third column displays the PLCT profile, the fourth column presents the asymmetric PLCT profile, and the fifth column shows the best-fit model. The color scale for the observations and models are same as Figure \ref{Fig1} and \cite{2023ApJ...951....8O}. The beam size and the scale of 20 au are shown in the lower left and lower right corners of each panel. The residuals are shown in the bottom row. The residual contours represent signal-to-noise ratios of [-20, -15, -10, -5, -3, 3, 10, 15, 20], based on the background rms values specific to each source (see Table~\ref{tab:obs}).}
\figsetgrpend

\figsetgrpstart
\figsetgrpnum{2.9}
\figsetgrptitle{IRAS 32}
\figsetplot{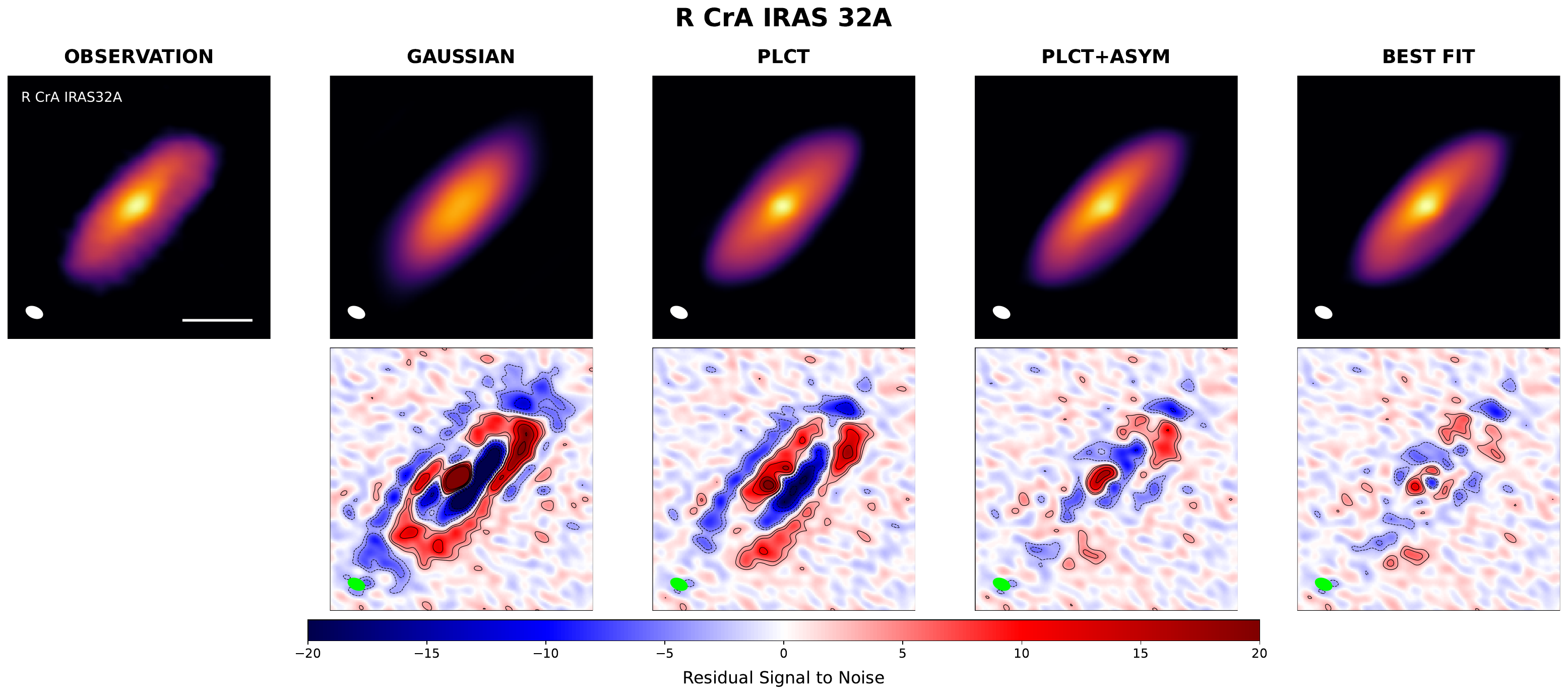}
\figsetgrpnote{The various fits for the eDisk targets are shown. For each source, the first column in the top row presents the ALMA eDisk observations, the second column shows the Gaussian fit, the third column displays the PLCT profile, the fourth column presents the asymmetric PLCT profile, and the fifth column shows the best-fit model. The color scale for the observations and models are same as Figure \ref{Fig1} and \cite{2023ApJ...951....8O}. The beam size and the scale of 20 au are shown in the lower left and lower right corners of each panel. The residuals are shown in the bottom row. The residual contours represent signal-to-noise ratios of [-20, -15, -10, -5, -3, 3, 10, 15, 20], based on the background rms values specific to each source (see Table~\ref{tab:obs}).}
\figsetgrpend

\figsetgrpstart
\figsetgrpnum{2.10}
\figsetgrptitle{IRAS 32-b}
\figsetplot{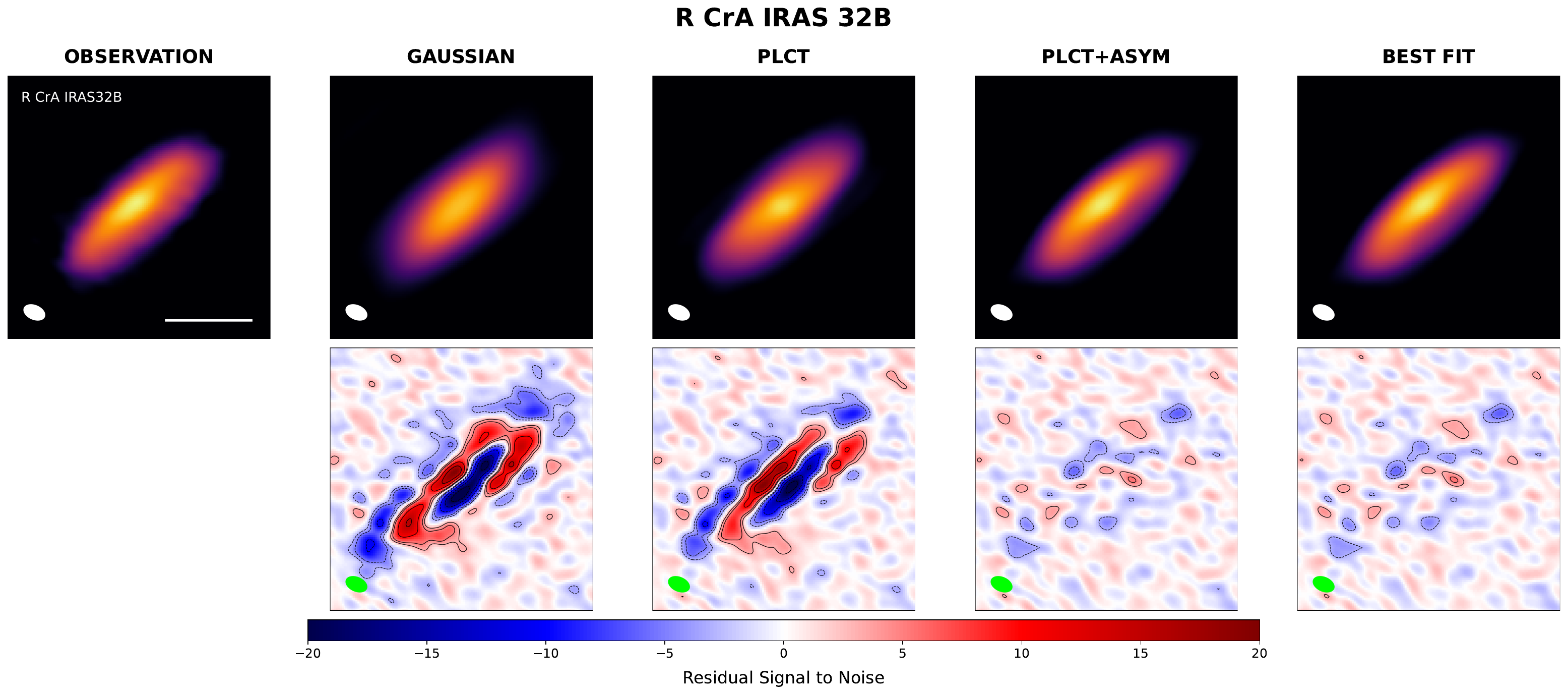}
\figsetgrpnote{The various fits for the eDisk targets are shown. For each source, the first column in the top row presents the ALMA eDisk observations, the second column shows the Gaussian fit, the third column displays the PLCT profile, the fourth column presents the asymmetric PLCT profile, and the fifth column shows the best-fit model. The color scale for the observations and models are same as Figure \ref{Fig1} and \cite{2023ApJ...951....8O}. The beam size and the scale of 20 au are shown in the lower left and lower right corners of each panel. The residuals are shown in the bottom row. The residual contours represent signal-to-noise ratios of [-20, -15, -10, -5, -3, 3, 10, 15, 20], based on the background rms values specific to each source (see Table~\ref{tab:obs}).}
\figsetgrpend

\figsetgrpstart
\figsetgrpnum{2.11}
\figsetgrptitle{BHR71 IRS1}
\figsetplot{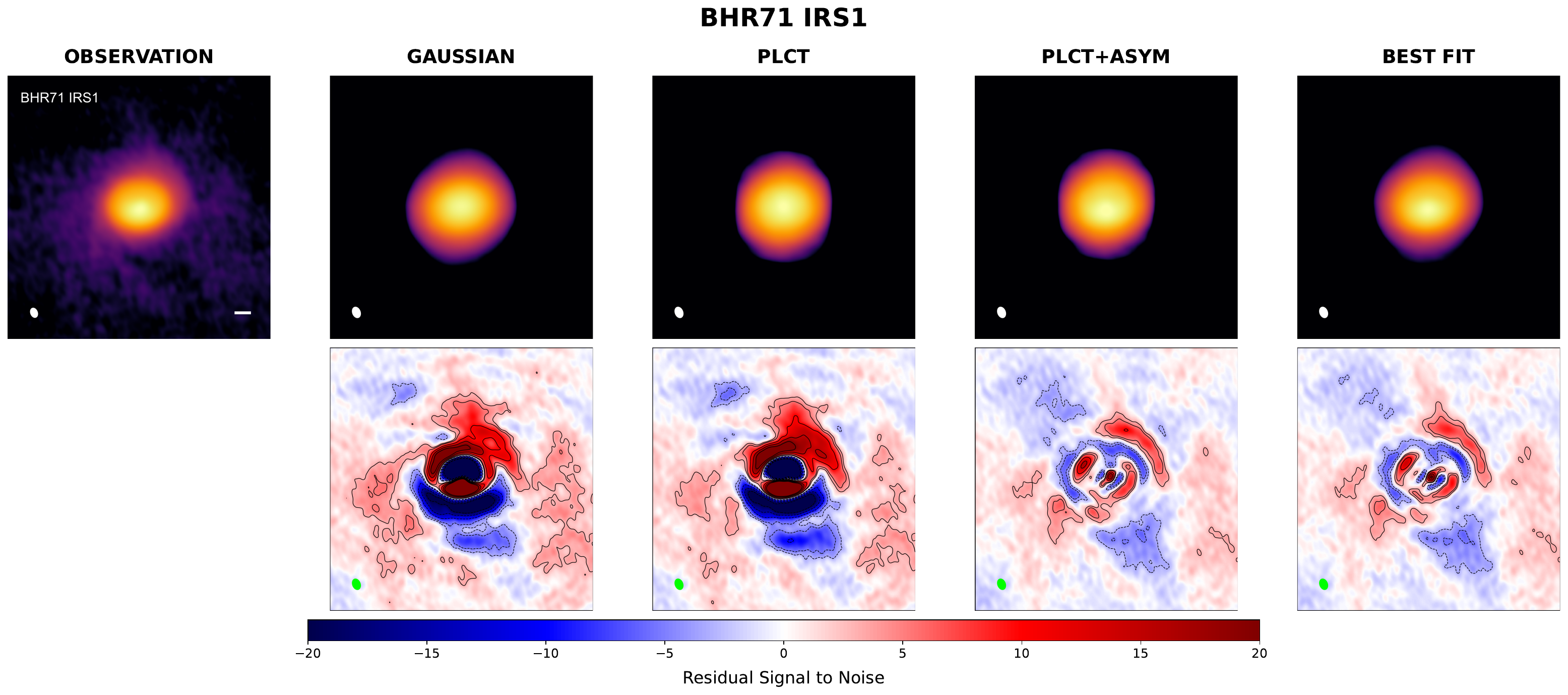}
\figsetgrpnote{The various fits for the eDisk targets are shown. For each source, the first column in the top row presents the ALMA eDisk observations, the second column shows the Gaussian fit, the third column displays the PLCT profile, the fourth column presents the asymmetric PLCT profile, and the fifth column shows the best-fit model. The color scale for the observations and models are same as Figure \ref{Fig1} and \cite{2023ApJ...951....8O}. The beam size and the scale of 20 au are shown in the lower left and lower right corners of each panel. The residuals are shown in the bottom row. The residual contours represent signal-to-noise ratios of [-20, -15, -10, -5, -3, 3, 10, 15, 20], based on the background rms values specific to each source (see Table~\ref{tab:obs}).}
\figsetgrpend

\figsetgrpstart
\figsetgrpnum{2.12}
\figsetgrptitle{Ced110 IRS4}
\figsetplot{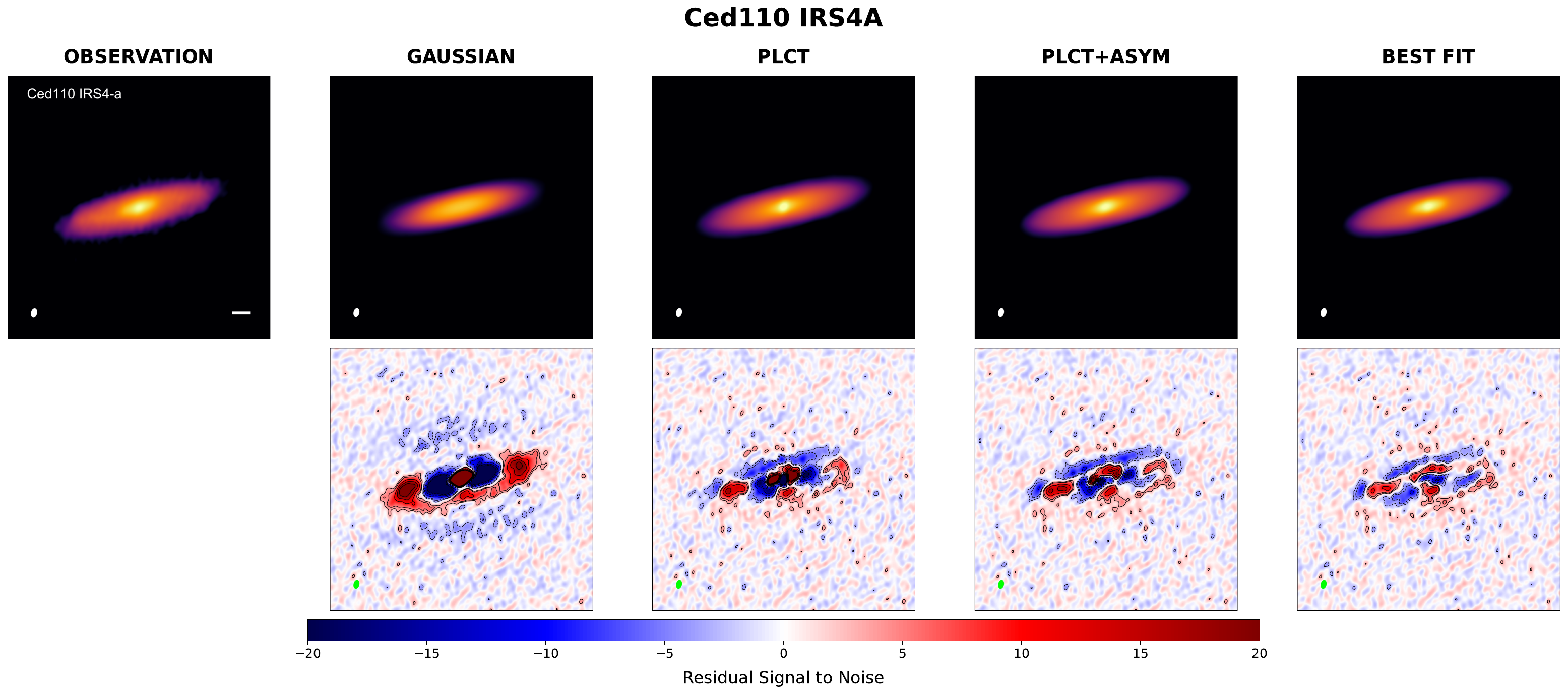}
\figsetgrpnote{The various fits for the eDisk targets are shown. For each source, the first column in the top row presents the ALMA eDisk observations, the second column shows the Gaussian fit, the third column displays the PLCT profile, the fourth column presents the asymmetric PLCT profile, and the fifth column shows the best-fit model. The color scale for the observations and models are same as Figure \ref{Fig1} and \cite{2023ApJ...951....8O}. The beam size and the scale of 20 au are shown in the lower left and lower right corners of each panel. The residuals are shown in the bottom row. The residual contours represent signal-to-noise ratios of [-20, -15, -10, -5, -3, 3, 10, 15, 20], based on the background rms values specific to each source (see Table~\ref{tab:obs}).}
\figsetgrpend

\figsetgrpstart
\figsetgrpnum{2.13}
\figsetgrptitle{Ced110 IRS4-b}
\figsetplot{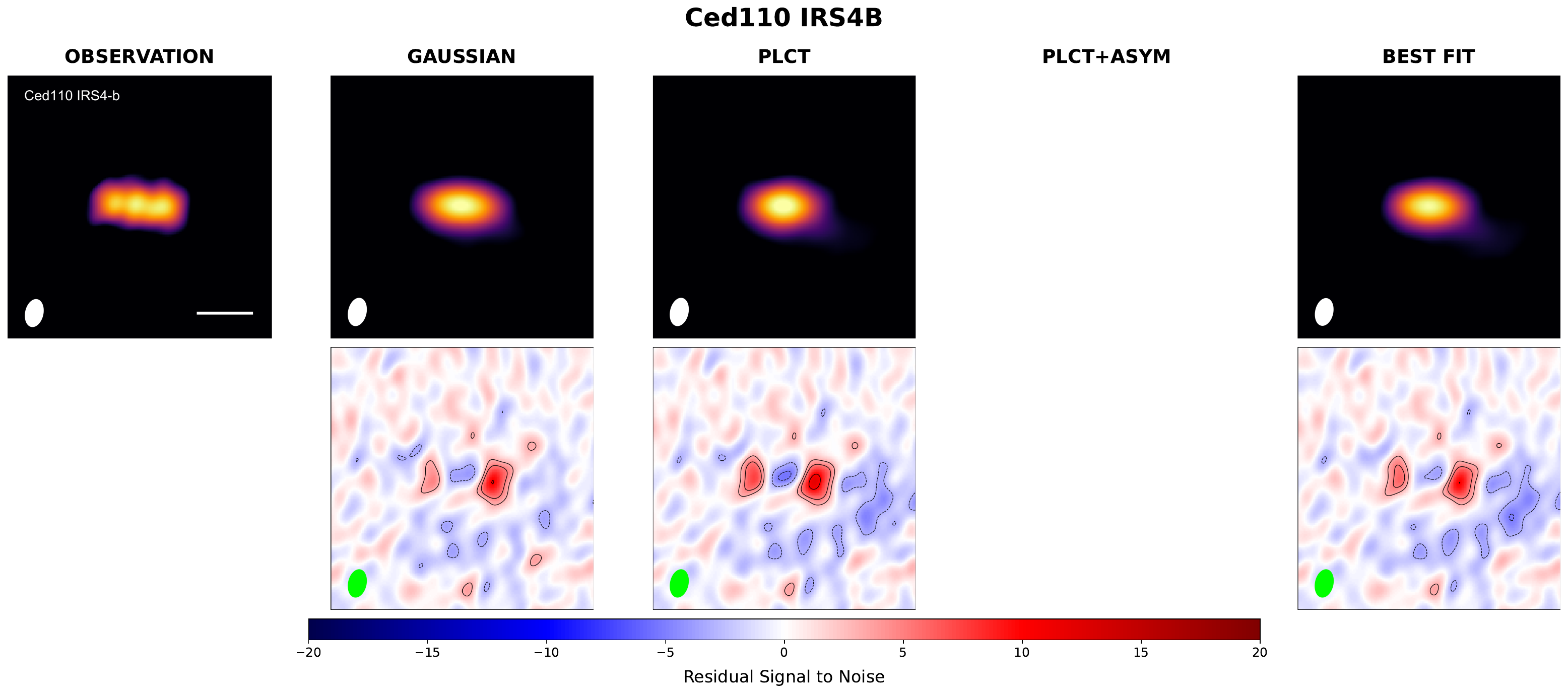}
\figsetgrpnote{The various fits for the eDisk targets are shown. For each source, the first column in the top row presents the ALMA eDisk observations, the second column shows the Gaussian fit, the third column displays the PLCT profile, the fourth column presents the asymmetric PLCT profile, and the fifth column shows the best-fit model. The color scale for the observations and models are same as Figure \ref{Fig1} and \cite{2023ApJ...951....8O}. The beam size and the scale of 20 au are shown in the lower left and lower right corners of each panel. The residuals are shown in the bottom row. The residual contours represent signal-to-noise ratios of [-20, -15, -10, -5, -3, 3, 10, 15, 20], based on the background rms values specific to each source (see Table~\ref{tab:obs}).}
\figsetgrpend

\figsetgrpstart
\figsetgrpnum{2.14}
\figsetgrptitle{IRS 7B}
\figsetplot{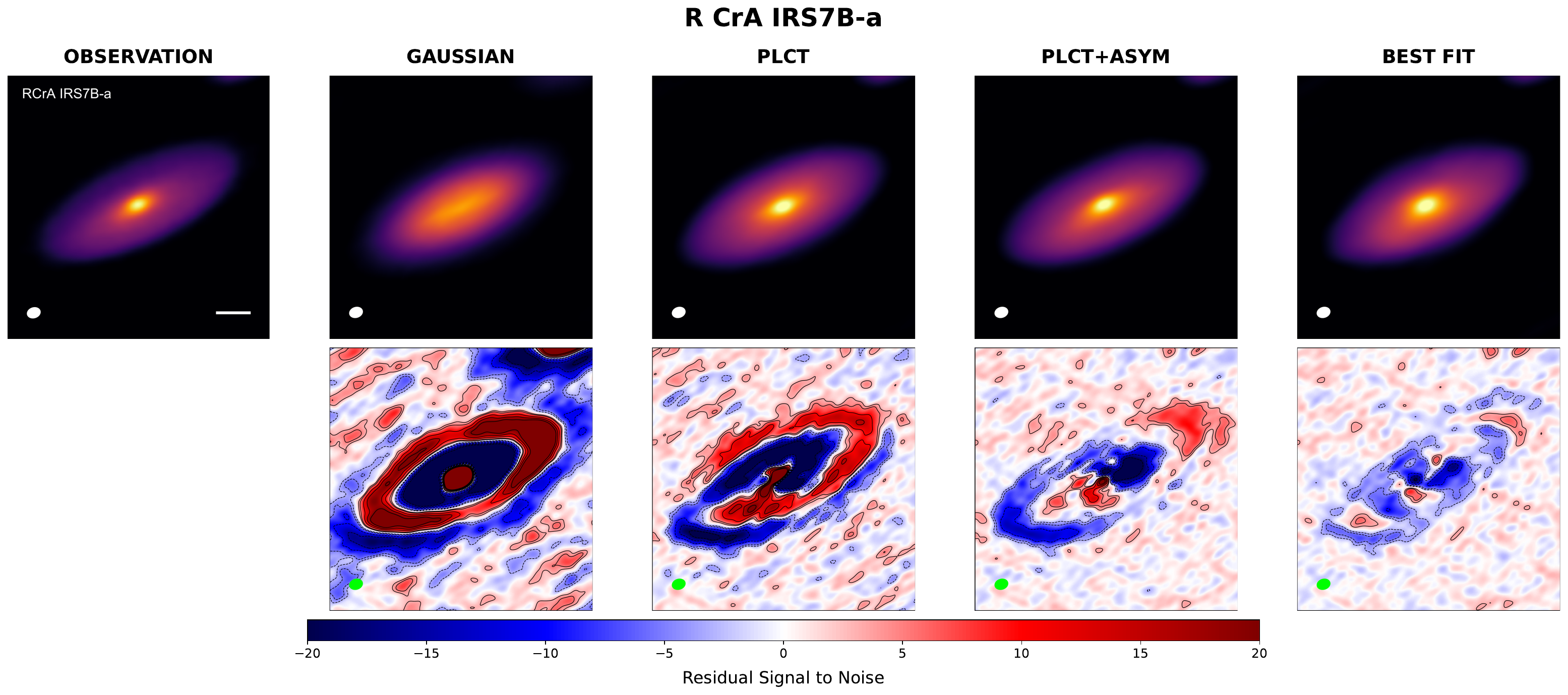}
\figsetgrpnote{The various fits for the eDisk targets are shown. For each source, the first column in the top row presents the ALMA eDisk observations, the second column shows the Gaussian fit, the third column displays the PLCT profile, the fourth column presents the asymmetric PLCT profile, and the fifth column shows the best-fit model. The color scale for the observations and models are same as Figure \ref{Fig1} and \cite{2023ApJ...951....8O}. The beam size and the scale of 20 au are shown in the lower left and lower right corners of each panel. The residuals are shown in the bottom row. The residual contours represent signal-to-noise ratios of [-20, -15, -10, -5, -3, 3, 10, 15, 20], based on the background rms values specific to each source (see Table~\ref{tab:obs}).}
\figsetgrpend

\figsetgrpstart
\figsetgrpnum{2.15}
\figsetgrptitle{IRS 7B-b}
\figsetplot{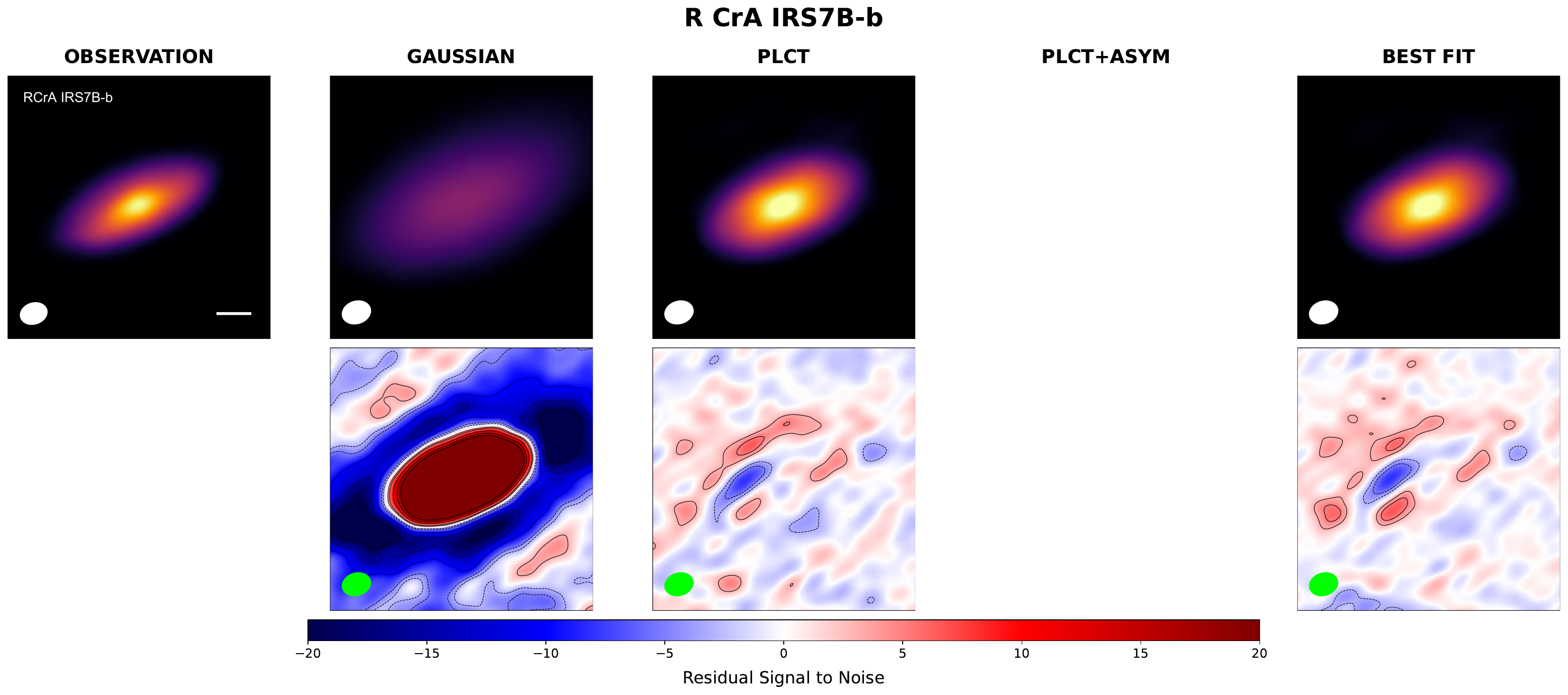}
\figsetgrpnote{The various fits for the eDisk targets are shown. For each source, the first column in the top row presents the ALMA eDisk observations, the second column shows the Gaussian fit, the third column displays the PLCT profile, the fourth column presents the asymmetric PLCT profile, and the fifth column shows the best-fit model. The color scale for the observations and models are same as Figure \ref{Fig1} and \cite{2023ApJ...951....8O}. The beam size and the scale of 20 au are shown in the lower left and lower right corners of each panel. The residuals are shown in the bottom row. The residual contours represent signal-to-noise ratios of [-20, -15, -10, -5, -3, 3, 10, 15, 20], based on the background rms values specific to each source (see Table~\ref{tab:obs}).}
\figsetgrpend

\figsetgrpstart
\figsetgrpnum{2.16}
\figsetgrptitle{IRAS 04169}
\figsetplot{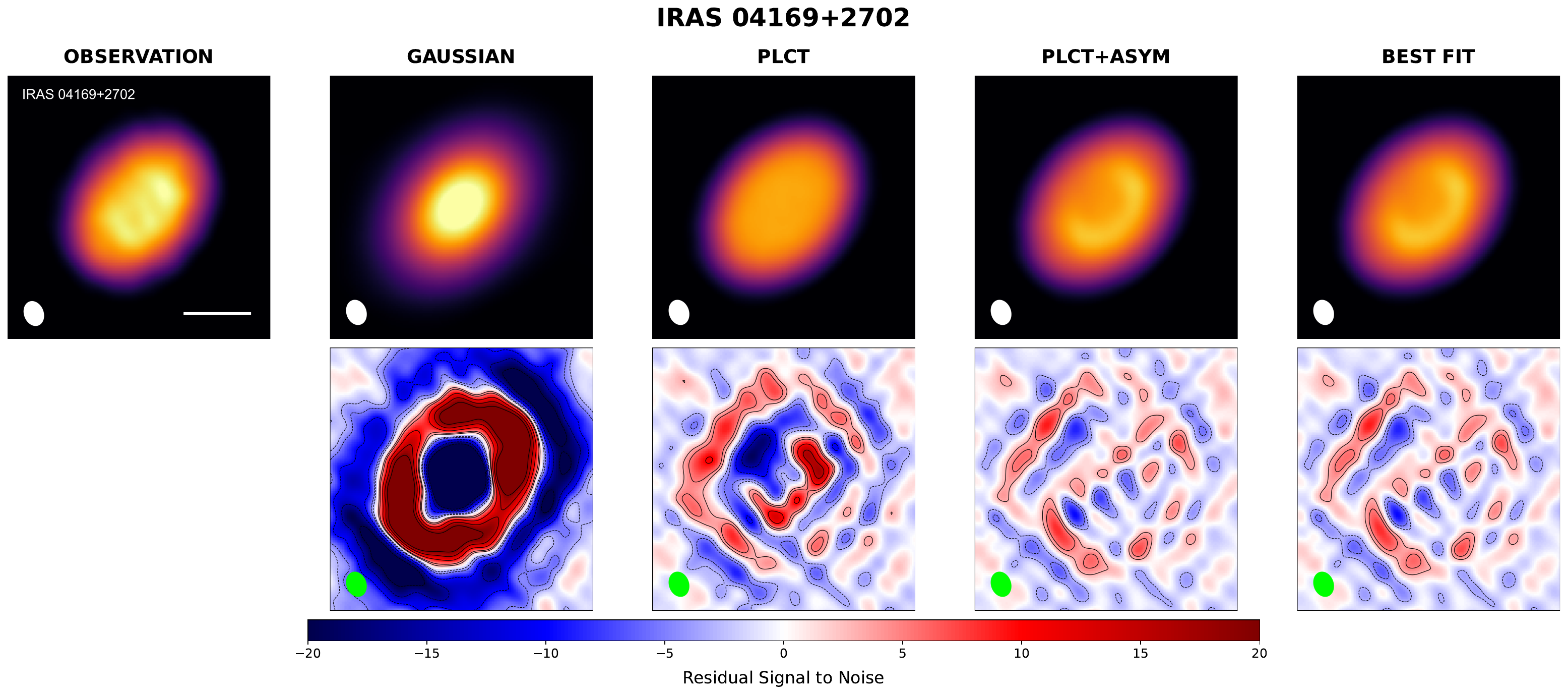}
\figsetgrpnote{The various fits for the eDisk targets are shown. For each source, the first column in the top row presents the ALMA eDisk observations, the second column shows the Gaussian fit, the third column displays the PLCT profile, the fourth column presents the asymmetric PLCT profile, and the fifth column shows the best-fit model. The color scale for the observations and models are same as Figure \ref{Fig1} and \cite{2023ApJ...951....8O}. The beam size and the scale of 20 au are shown in the lower left and lower right corners of each panel. The residuals are shown in the bottom row. The residual contours represent signal-to-noise ratios of [-20, -15, -10, -5, -3, 3, 10, 15, 20], based on the background rms values specific to each source (see Table~\ref{tab:obs}).}
\figsetgrpend

\figsetgrpstart
\figsetgrpnum{2.17}
\figsetgrptitle{TMC1A}
\figsetplot{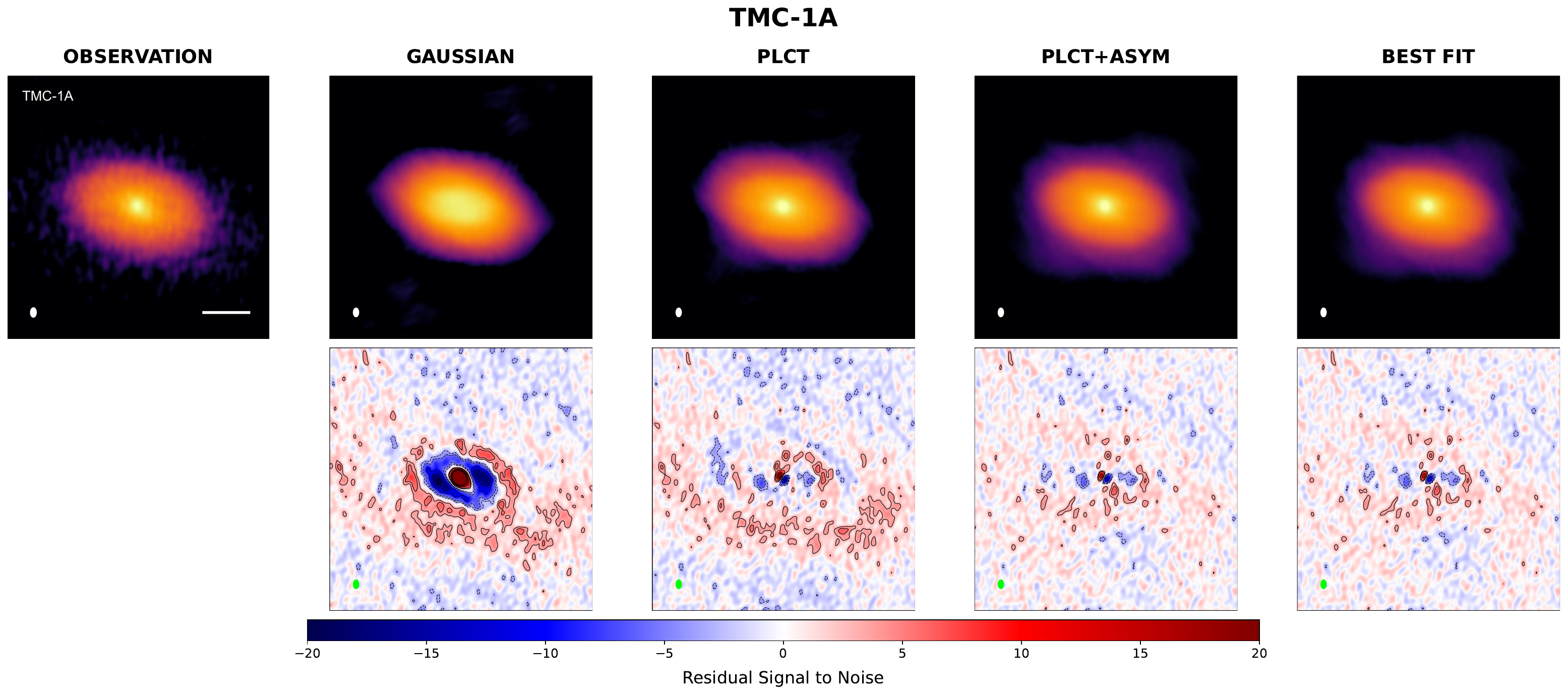}
\figsetgrpnote{The various fits for the eDisk targets are shown. For each source, the first column in the top row presents the ALMA eDisk observations, the second column shows the Gaussian fit, the third column displays the PLCT profile, the fourth column presents the asymmetric PLCT profile, and the fifth column shows the best-fit model. The color scale for the observations and models are same as Figure \ref{Fig1} and \cite{2023ApJ...951....8O}. The beam size and the scale of 20 au are shown in the lower left and lower right corners of each panel. The residuals are shown in the bottom row. The residual contours represent signal-to-noise ratios of [-20, -15, -10, -5, -3, 3, 10, 15, 20], based on the background rms values specific to each source (see Table~\ref{tab:obs}).}
\figsetgrpend

\figsetgrpstart
\figsetgrpnum{2.18}
\figsetgrptitle{Oph IRS43}
\figsetplot{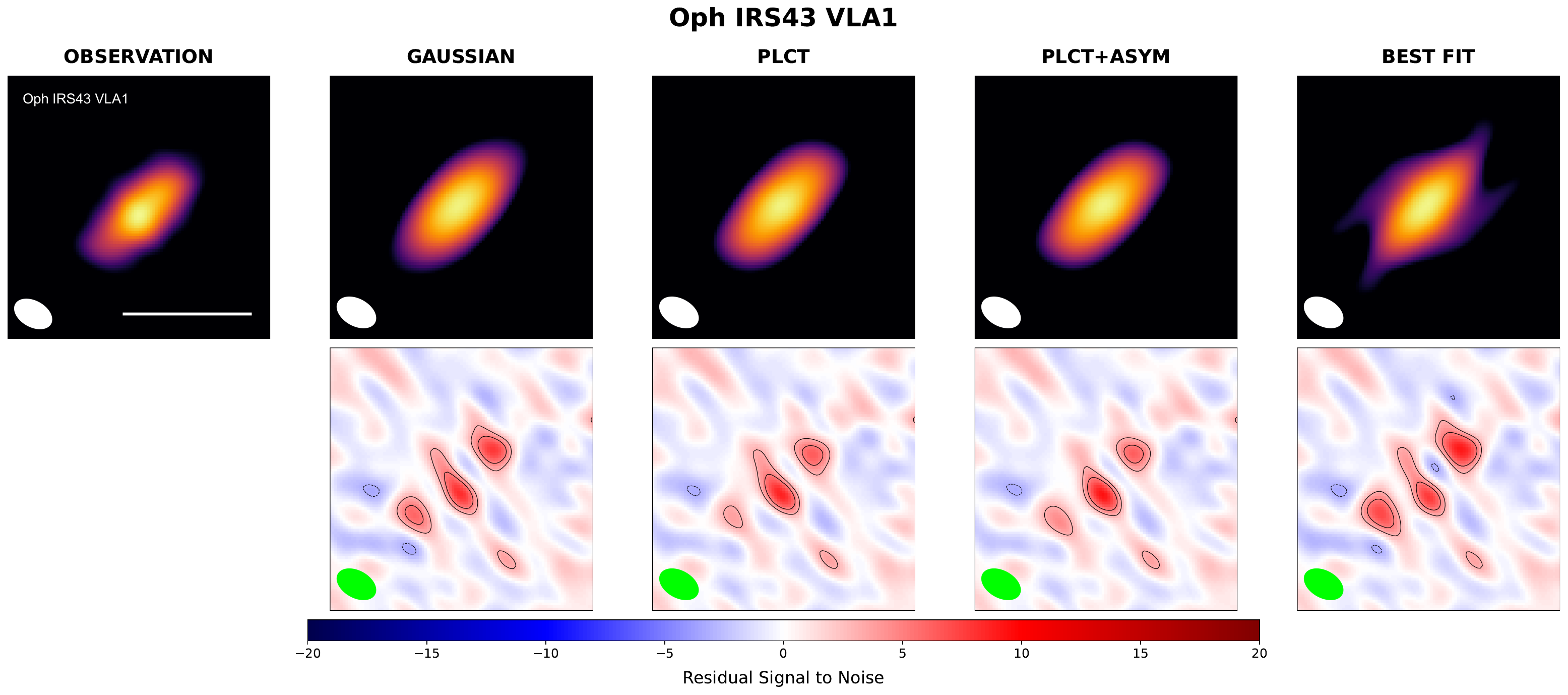}
\figsetgrpnote{The various fits for the eDisk targets are shown. For each source, the first column in the top row presents the ALMA eDisk observations, the second column shows the Gaussian fit, the third column displays the PLCT profile, the fourth column presents the asymmetric PLCT profile, and the fifth column shows the best-fit model. The color scale for the observations and models are same as Figure \ref{Fig1} and \cite{2023ApJ...951....8O}. The beam size and the scale of 20 au are shown in the lower left and lower right corners of each panel. The residuals are shown in the bottom row. The residual contours represent signal-to-noise ratios of [-20, -15, -10, -5, -3, 3, 10, 15, 20], based on the background rms values specific to each source (see Table~\ref{tab:obs}).}
\figsetgrpend

\figsetgrpstart
\figsetgrpnum{2.19}
\figsetgrptitle{Oph IRS43 VLA2}
\figsetplot{OphIRS43 VLA2_model_comparison.pdf}
\figsetgrpnote{The various fits for the eDisk targets are shown. For each source, the first column in the top row presents the ALMA eDisk observations, the second column shows the Gaussian fit, the third column displays the PLCT profile, the fourth column presents the asymmetric PLCT profile, and the fifth column shows the best-fit model. The color scale for the observations and models are same as Figure \ref{Fig1} and \cite{2023ApJ...951....8O}. The beam size and the scale of 20 au are shown in the lower left and lower right corners of each panel. The residuals are shown in the bottom row. The residual contours represent signal-to-noise ratios of [-20, -15, -10, -5, -3, 3, 10, 15, 20], based on the background rms values specific to each source (see Table~\ref{tab:obs}).}
\figsetgrpend

\figsetgrpstart
\figsetgrpnum{2.20}
\figsetgrptitle{L1489 IRS}
\figsetplot{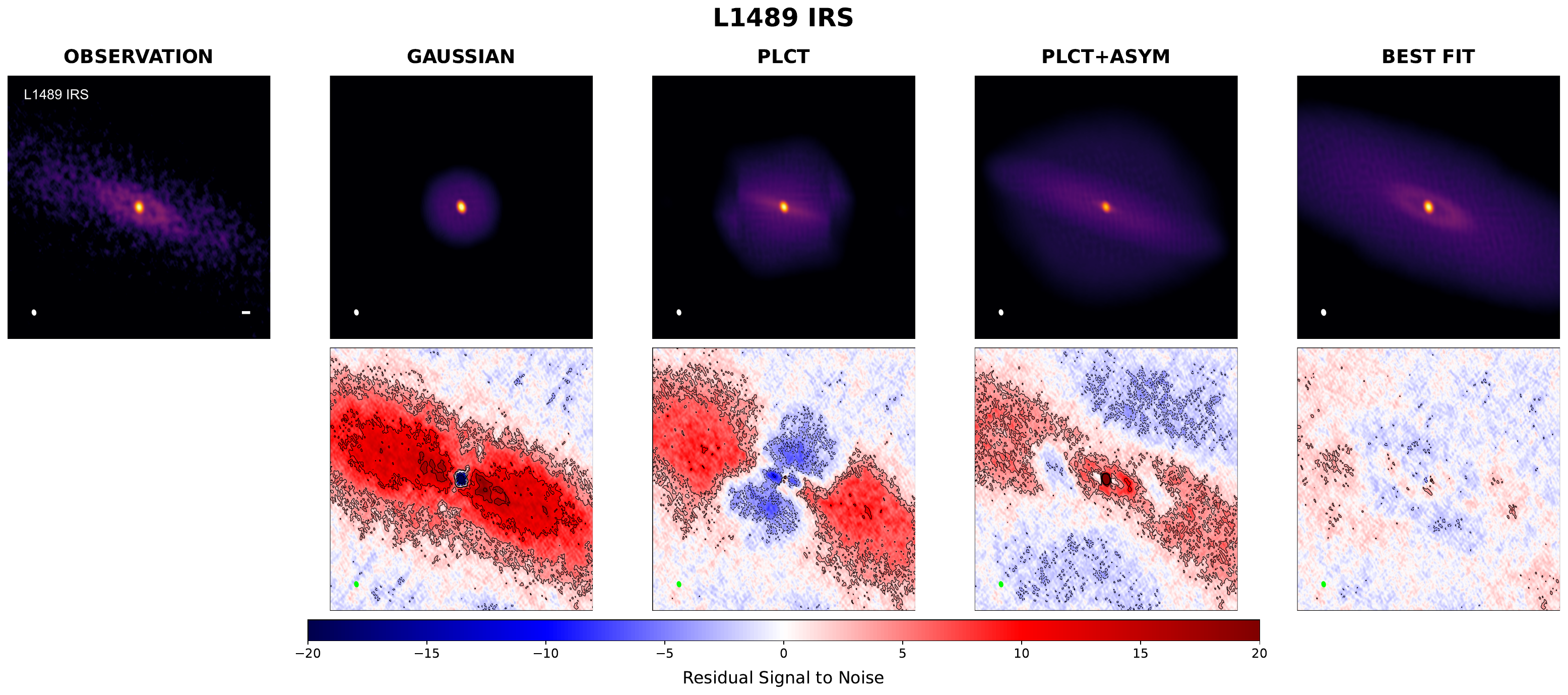}
\figsetgrpnote{The various fits for the eDisk targets are shown. For each source, the first column in the top row presents the ALMA eDisk observations, the second column shows the Gaussian fit, the third column displays the PLCT profile, the fourth column presents the asymmetric PLCT profile, and the fifth column shows the best-fit model. The color scale for the observations and models are same as Figure \ref{Fig1} and \cite{2023ApJ...951....8O}. The beam size and the scale of 20 au are shown in the lower left and lower right corners of each panel. The residuals are shown in the bottom row. The residual contours represent signal-to-noise ratios of [-20, -15, -10, -5, -3, 3, 10, 15, 20], based on the background rms values specific to each source (see Table~\ref{tab:obs}).}
\figsetgrpend

\figsetgrpstart
\figsetgrpnum{2.21}
\figsetgrptitle{Oph IRS63}
\figsetplot{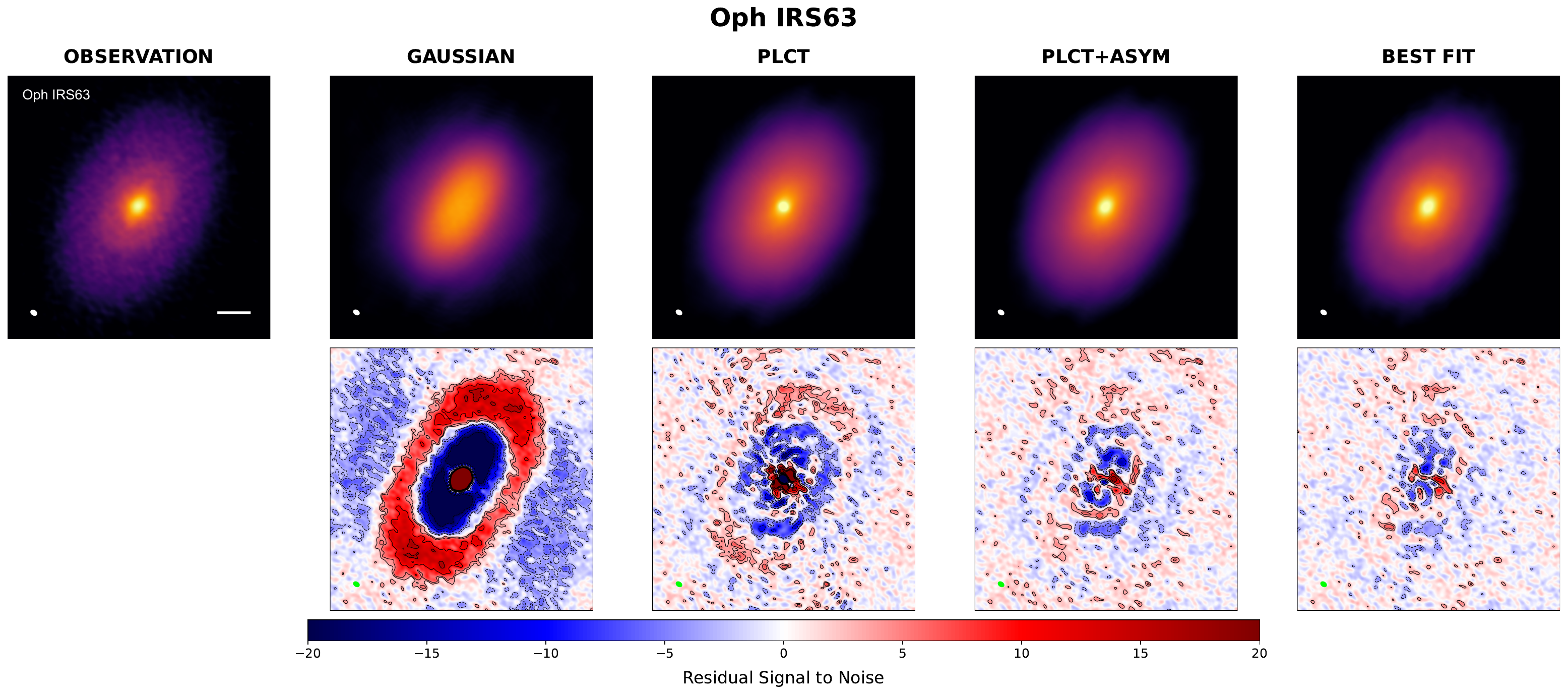}
\figsetgrpnote{The various fits for the eDisk targets are shown. For each source, the first column in the top row presents the ALMA eDisk observations, the second column shows the Gaussian fit, the third column displays the PLCT profile, the fourth column presents the asymmetric PLCT profile, and the fifth column shows the best-fit model. The color scale for the observations and models are same as Figure \ref{Fig1} and \cite{2023ApJ...951....8O}. The beam size and the scale of 20 au are shown in the lower left and lower right corners of each panel. The residuals are shown in the bottom row. The residual contours represent signal-to-noise ratios of [-20, -15, -10, -5, -3, 3, 10, 15, 20], based on the background rms values specific to each source (see Table~\ref{tab:obs}).}
\figsetgrpend

\figsetend

\addtocounter{figure}{-1}

\begin{figure*}
\centering
\includegraphics[width=0.85\linewidth]{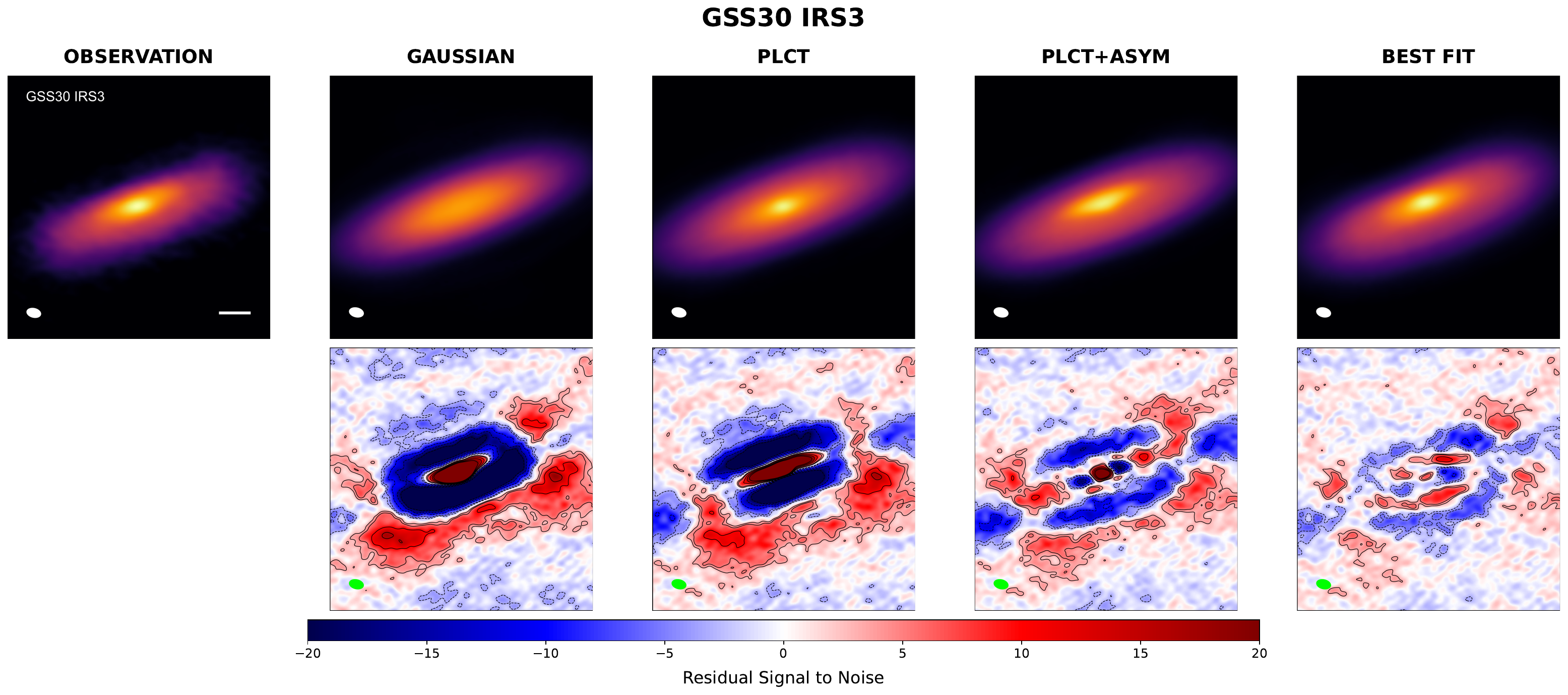}
\includegraphics[width=0.85\linewidth]{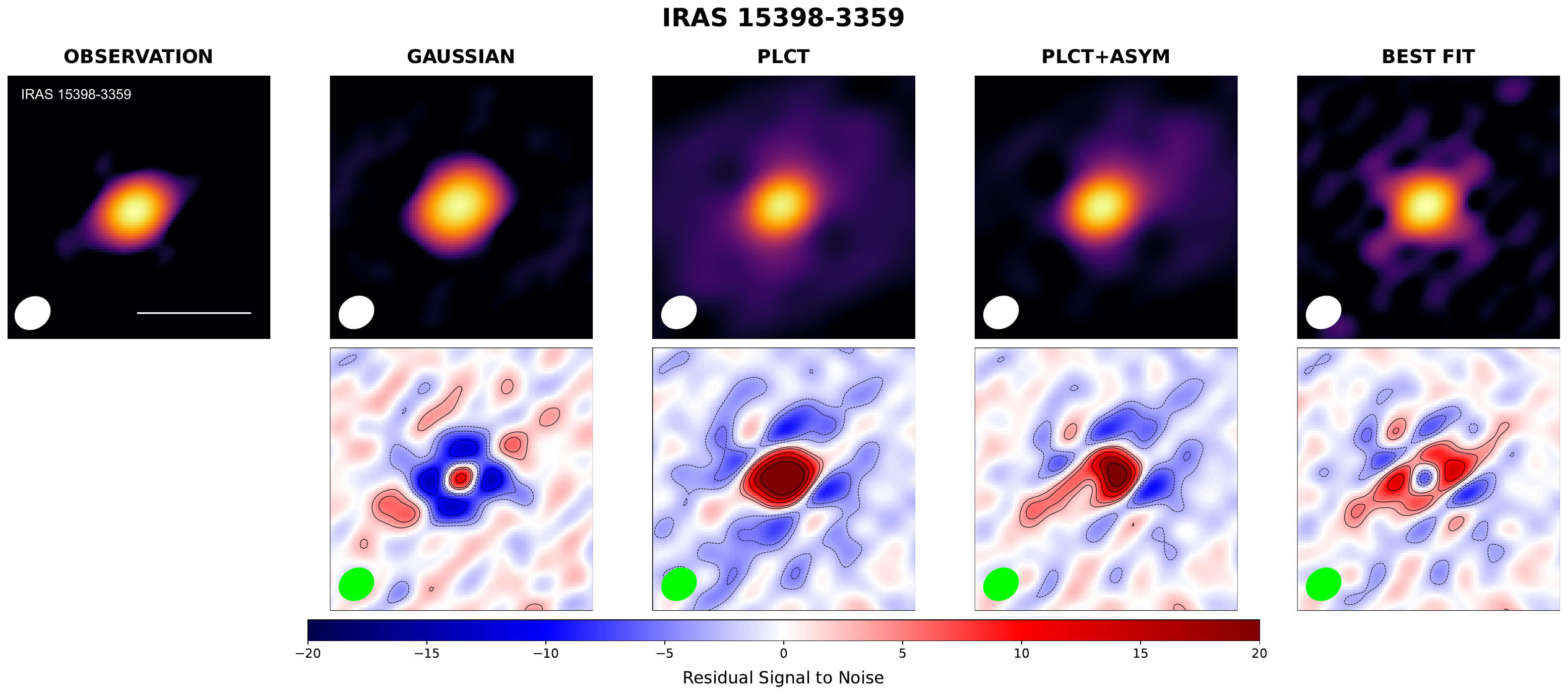}
\includegraphics[width=0.85\linewidth]{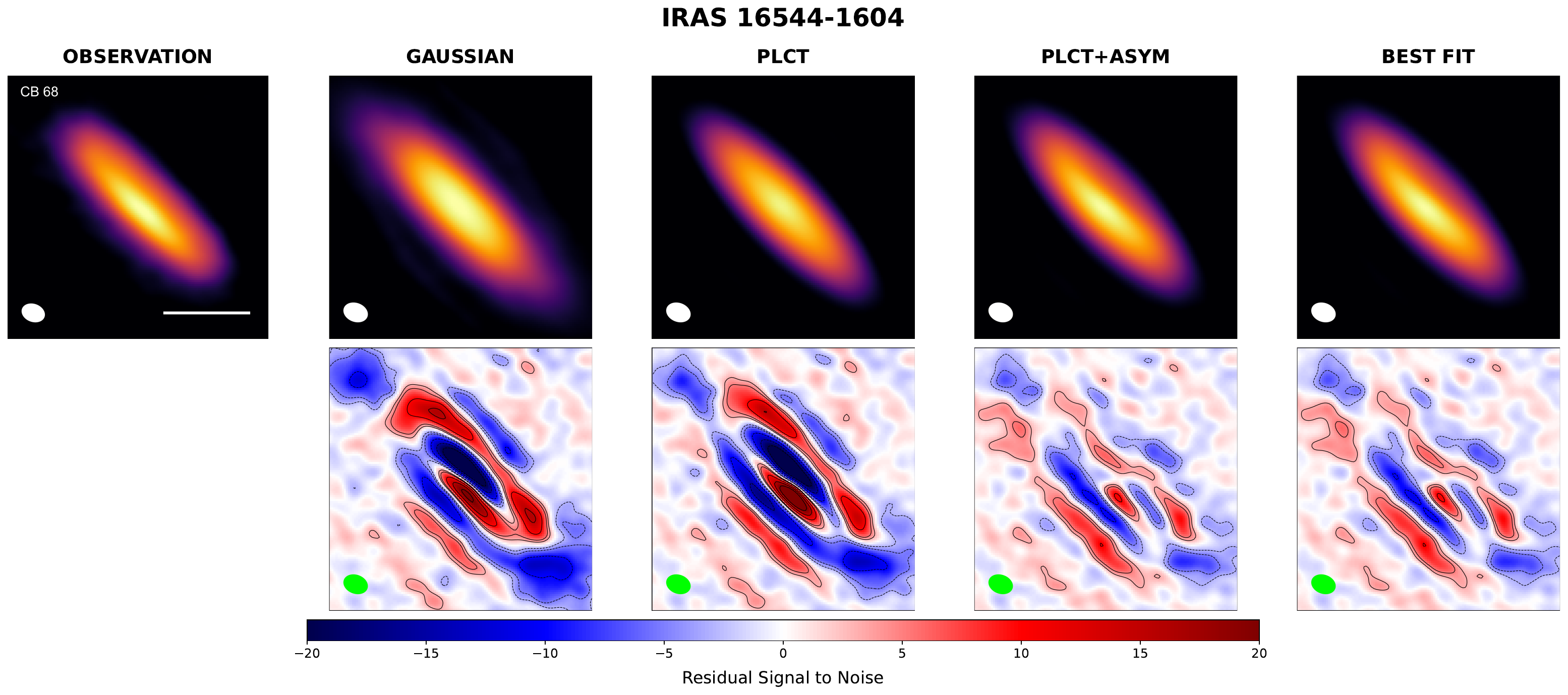}
\caption{Continuation of Figure \ref{Fig2}.}
\end{figure*}

\addtocounter{figure}{-1}

\begin{figure*}
\centering
\includegraphics[width=0.85\linewidth]{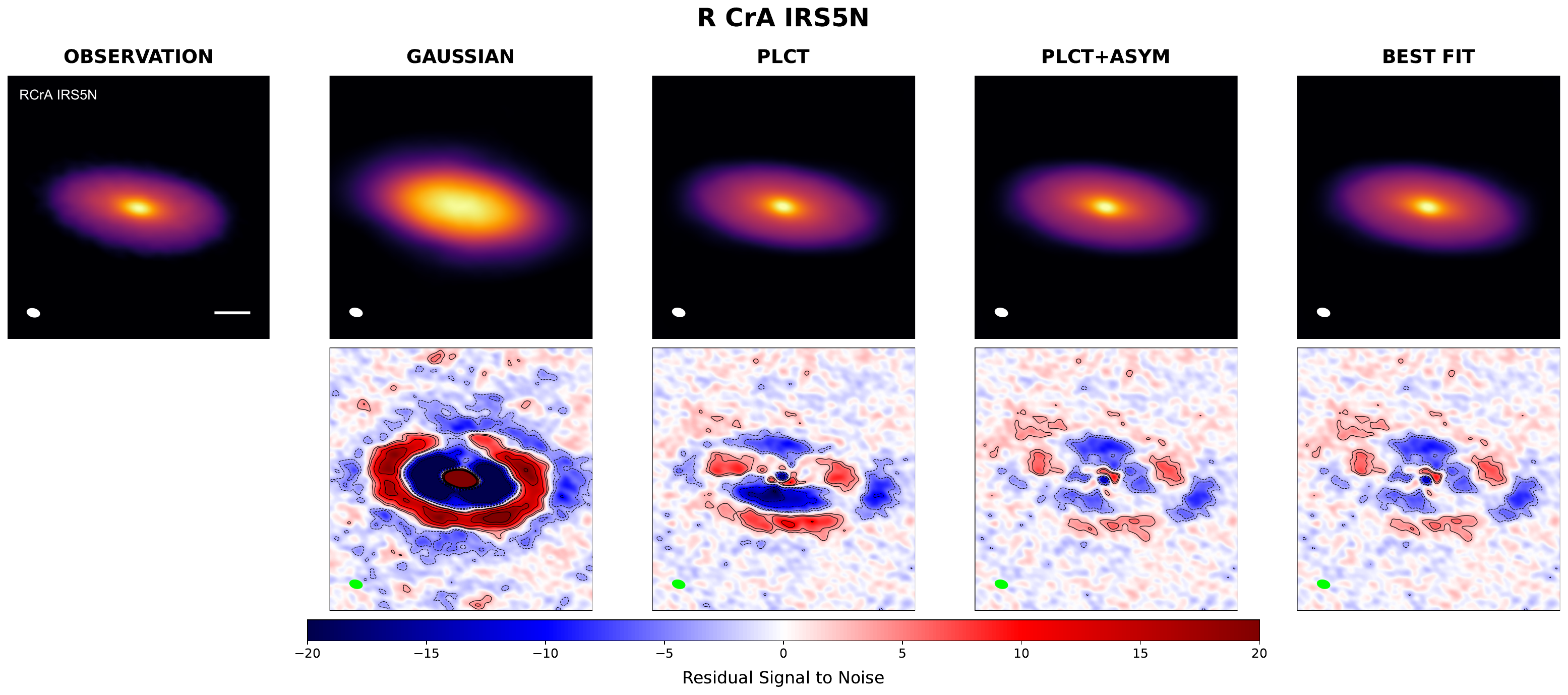}
\includegraphics[width=0.85\linewidth]{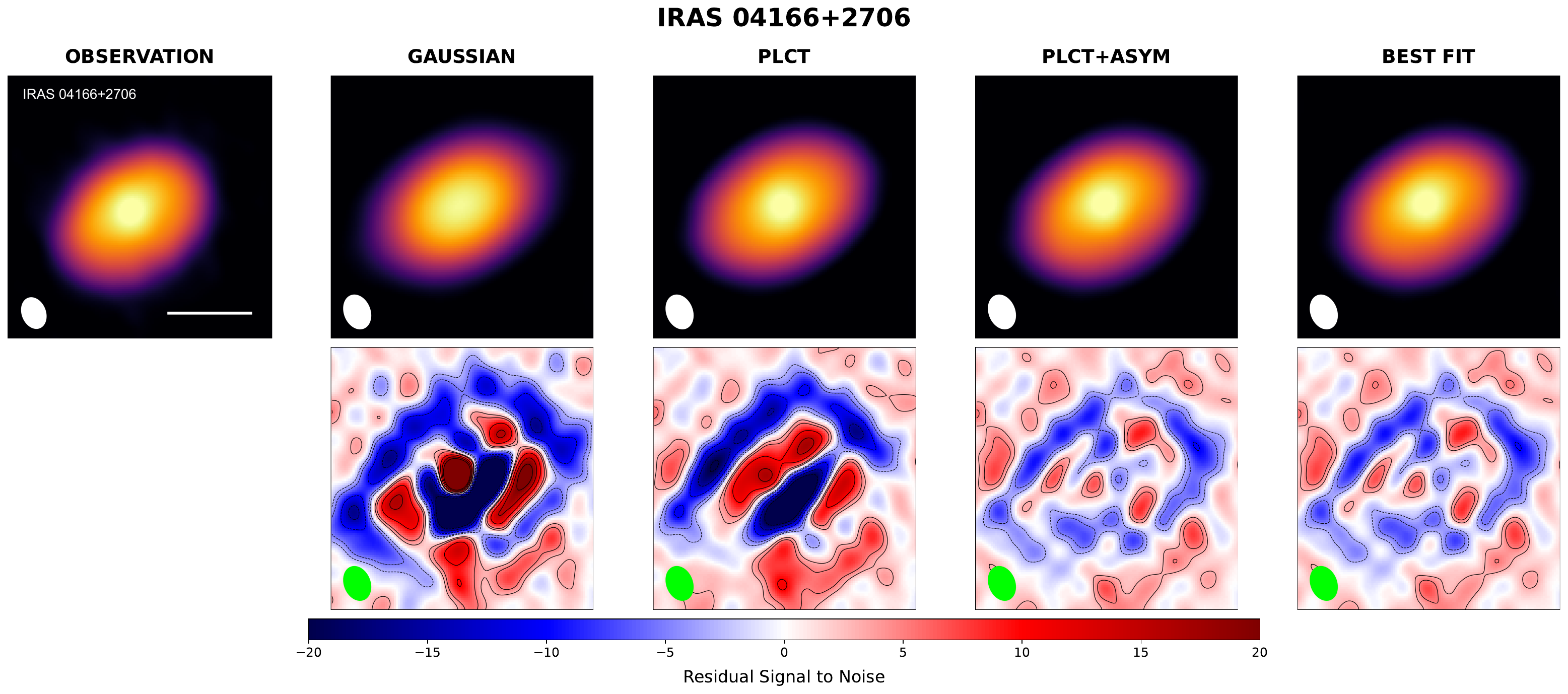}
\includegraphics[width=0.85\linewidth]{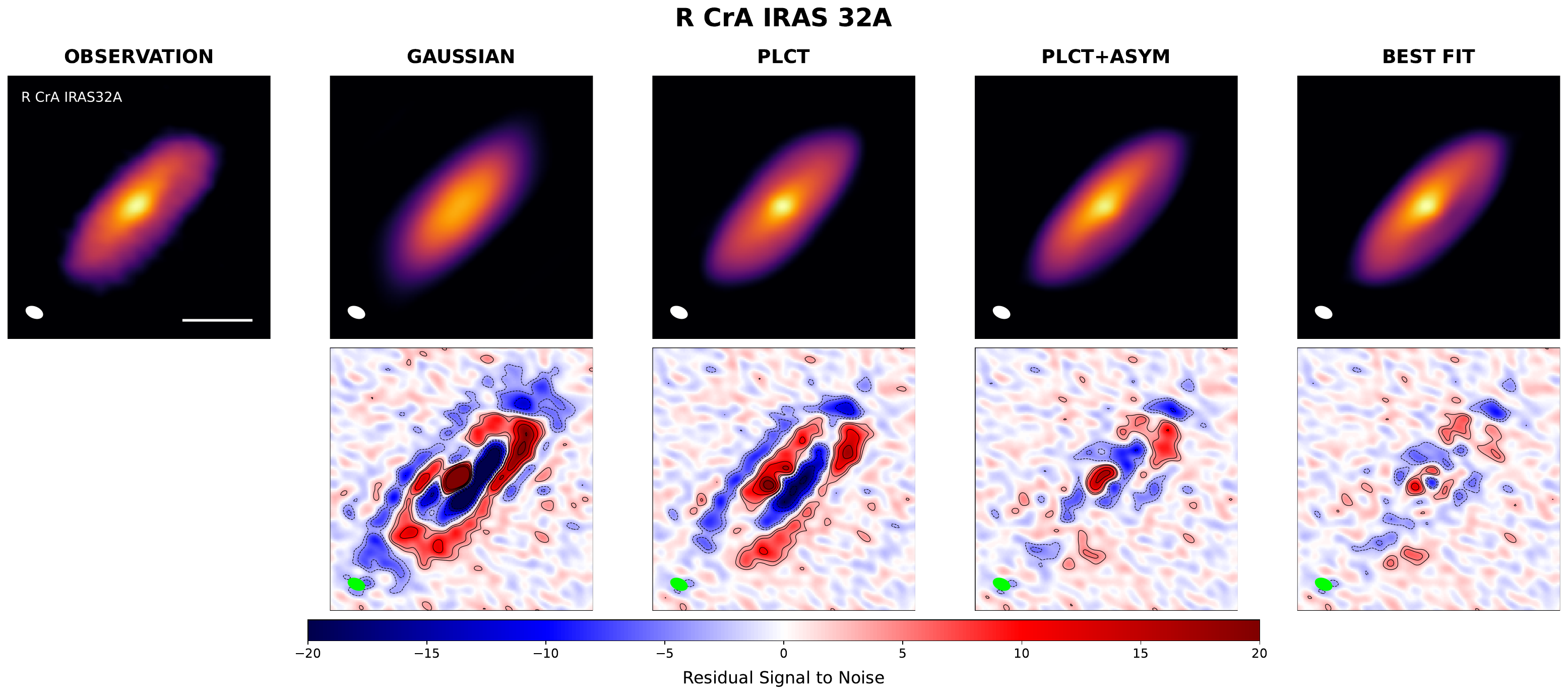}
\caption{Continuation of Figure \ref{Fig2}.}
\end{figure*}

\addtocounter{figure}{-1}

\begin{figure*}
\centering
\includegraphics[width=0.85\linewidth]{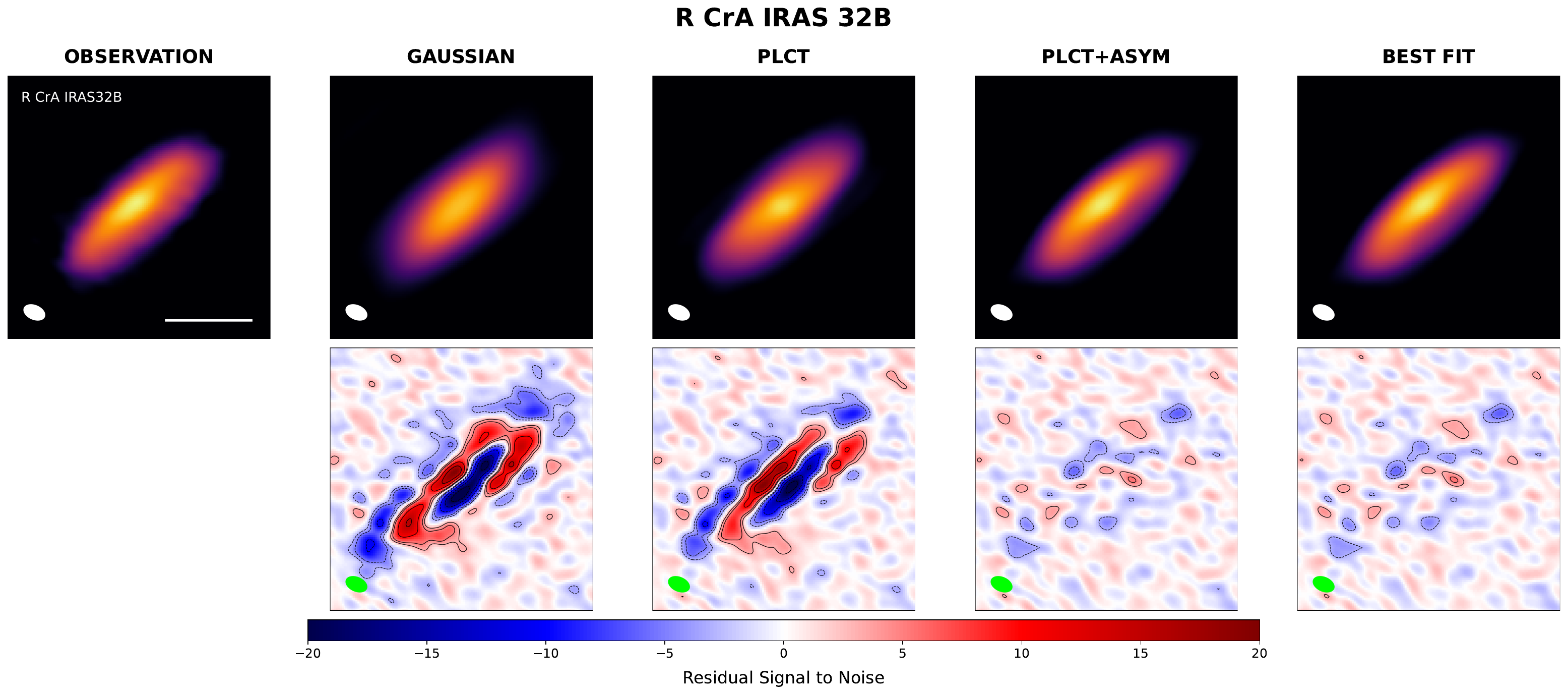}
\includegraphics[width=0.85\linewidth]{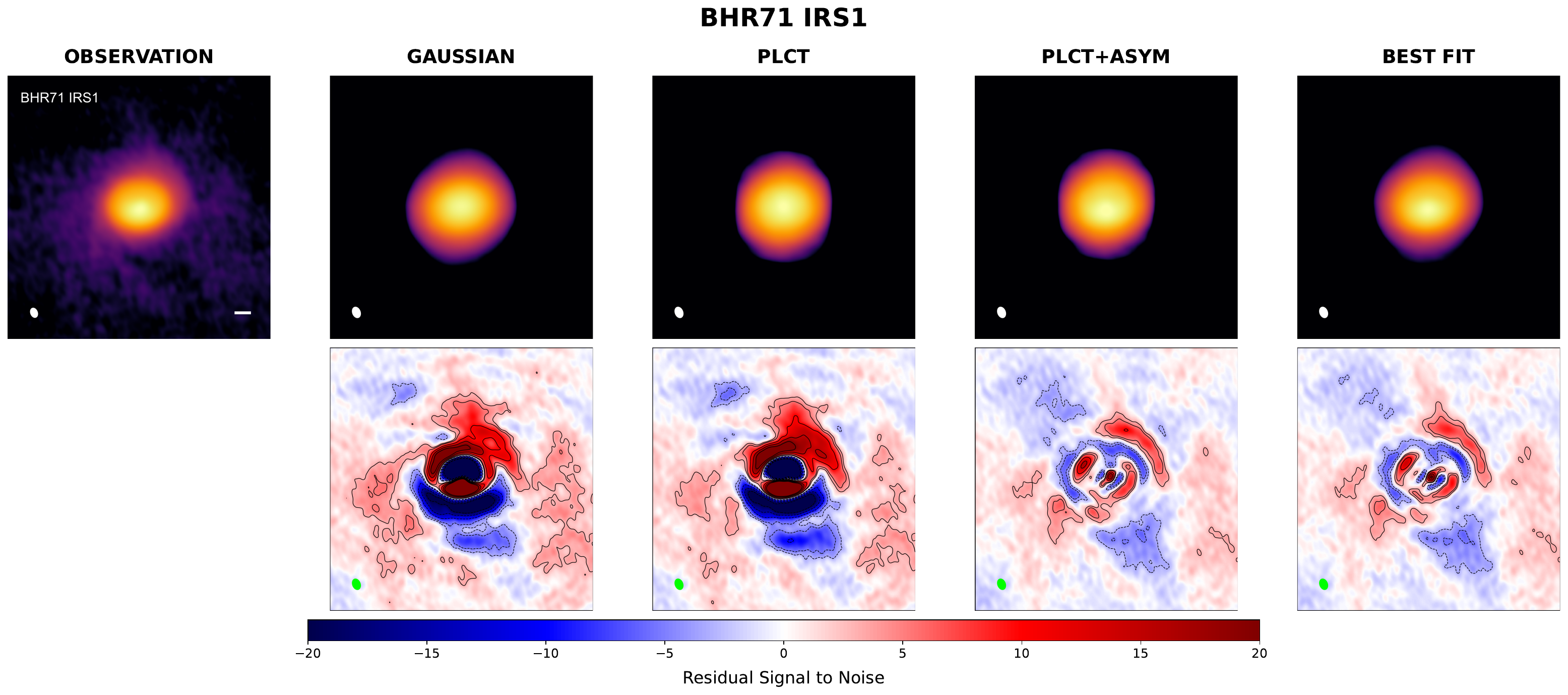}
\includegraphics[width=0.85\linewidth]{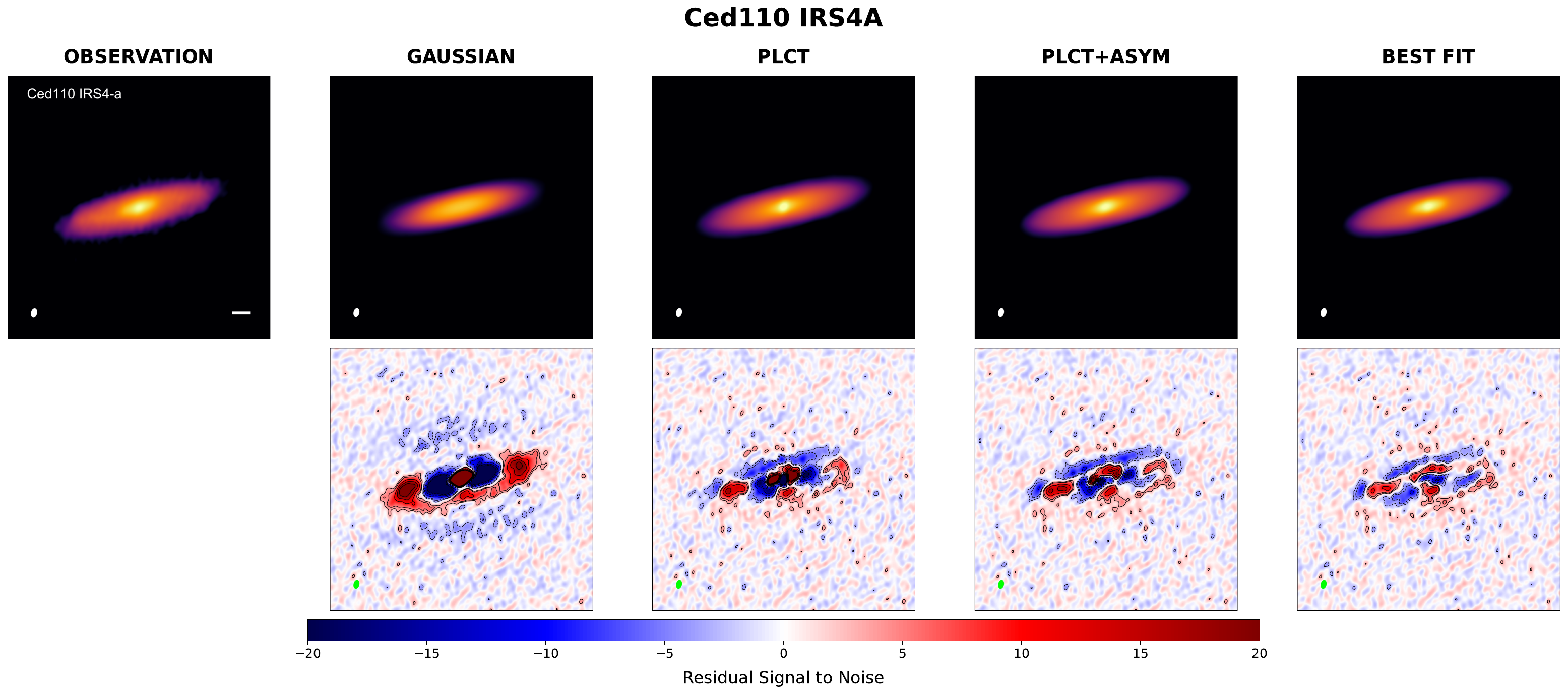}
\caption{Continuation of Figure \ref{Fig2}.}
\end{figure*}

\addtocounter{figure}{-1}

\begin{figure*}
\centering
\includegraphics[width=0.85\linewidth]{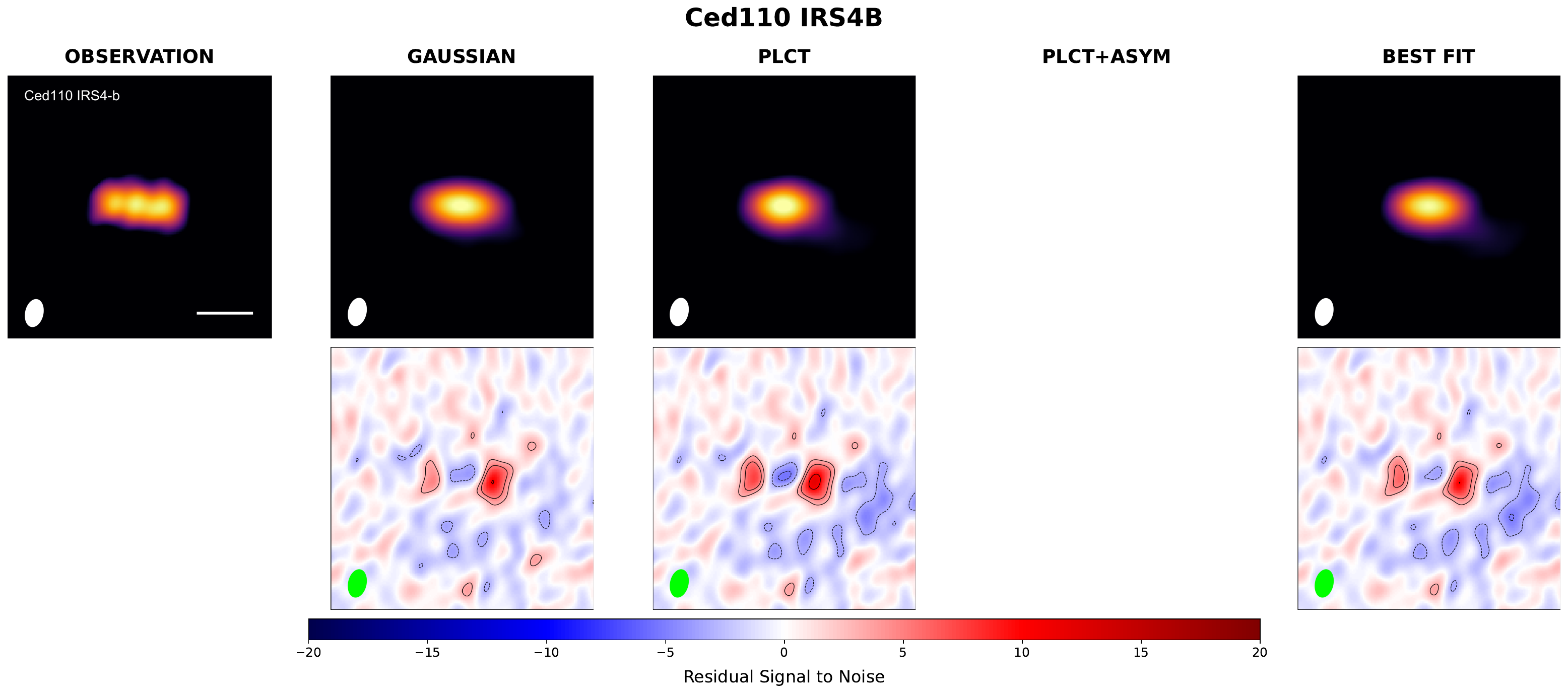}
\includegraphics[width=0.85\linewidth]{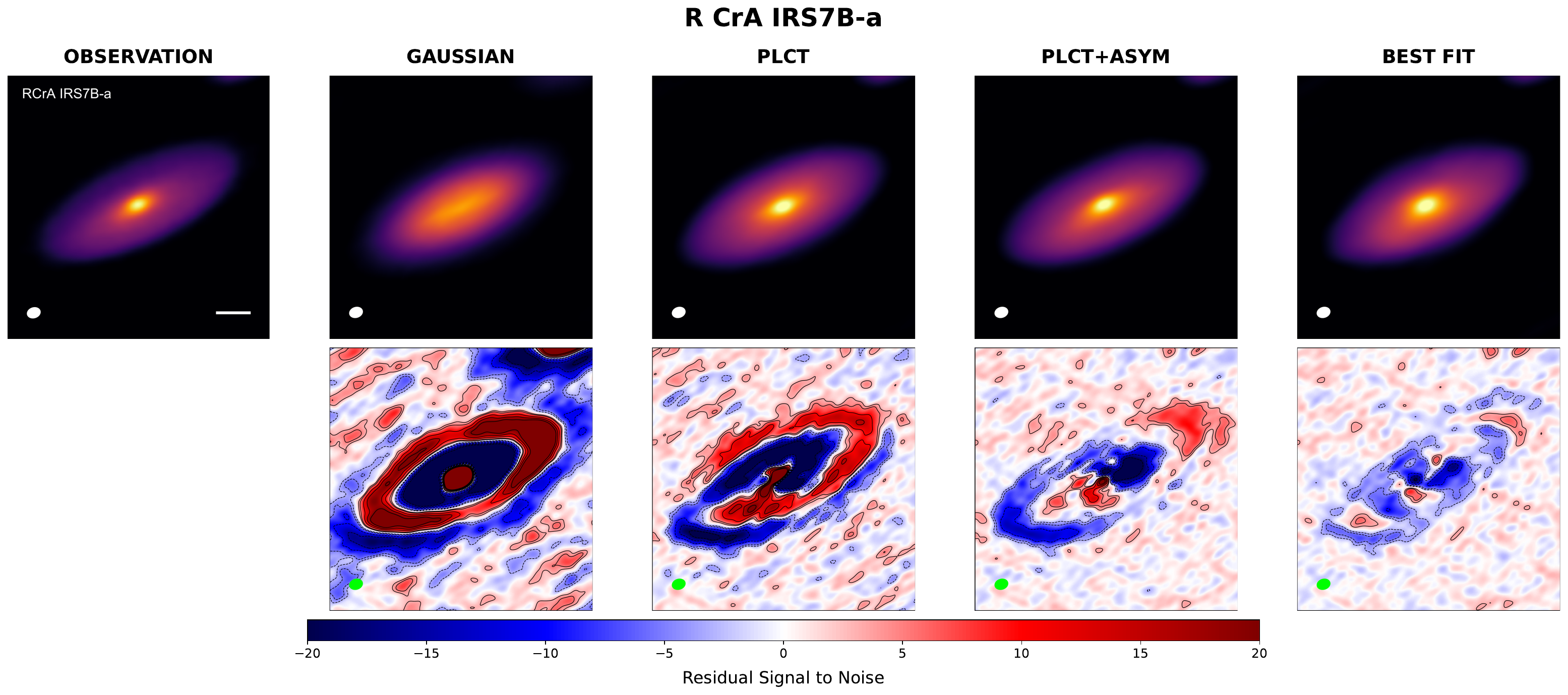}
\includegraphics[width=0.85\linewidth]{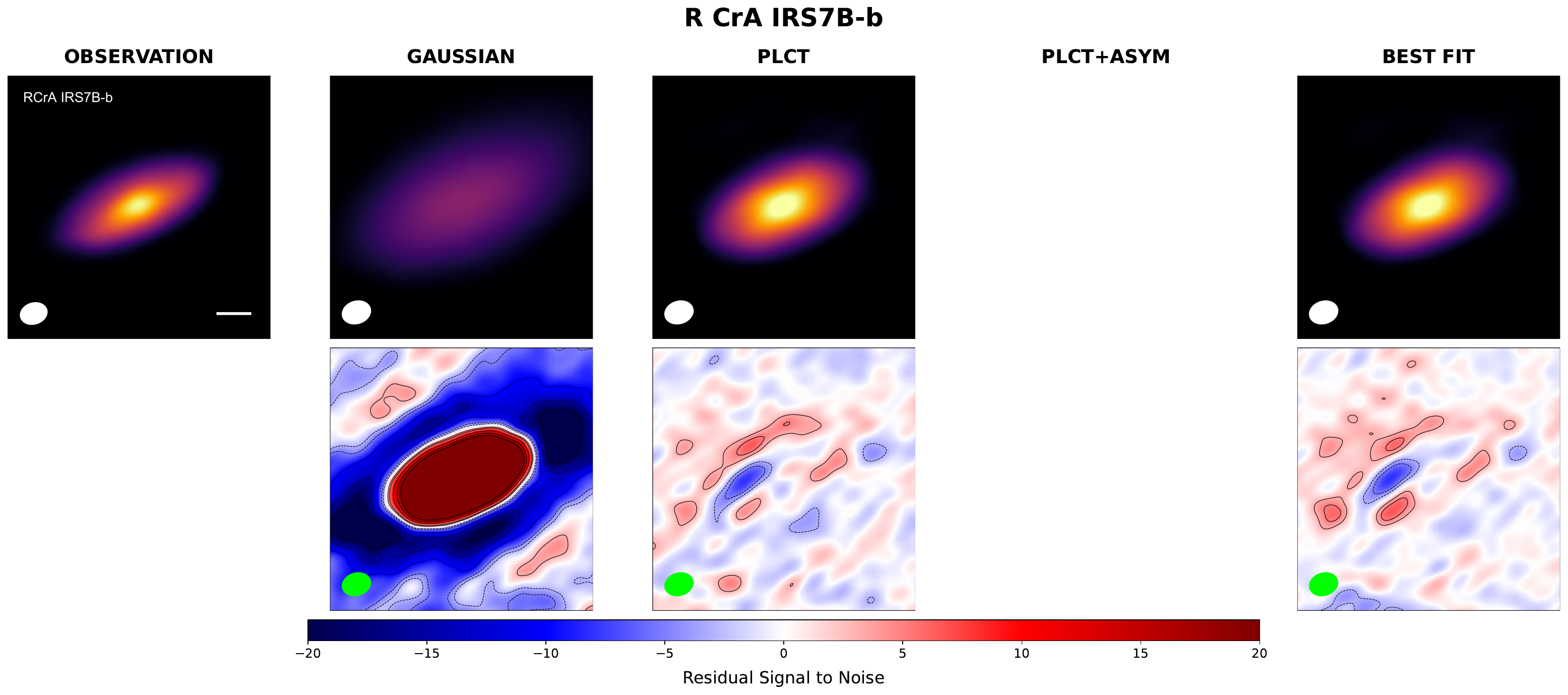}
\caption{Continuation of Figure \ref{Fig2}.}
\end{figure*}

\addtocounter{figure}{-1}

\begin{figure*}
\centering
\includegraphics[width=0.85\linewidth]{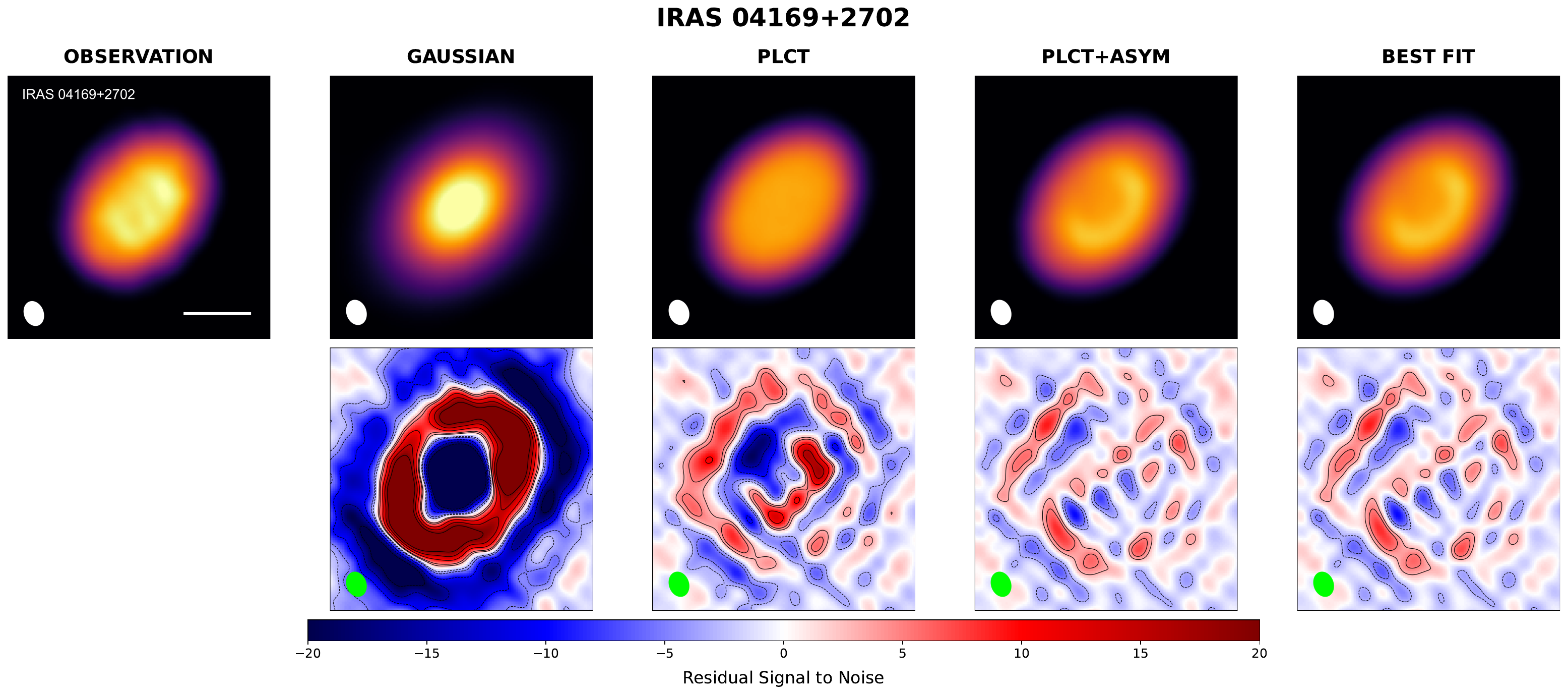}
\includegraphics[width=0.85\linewidth]{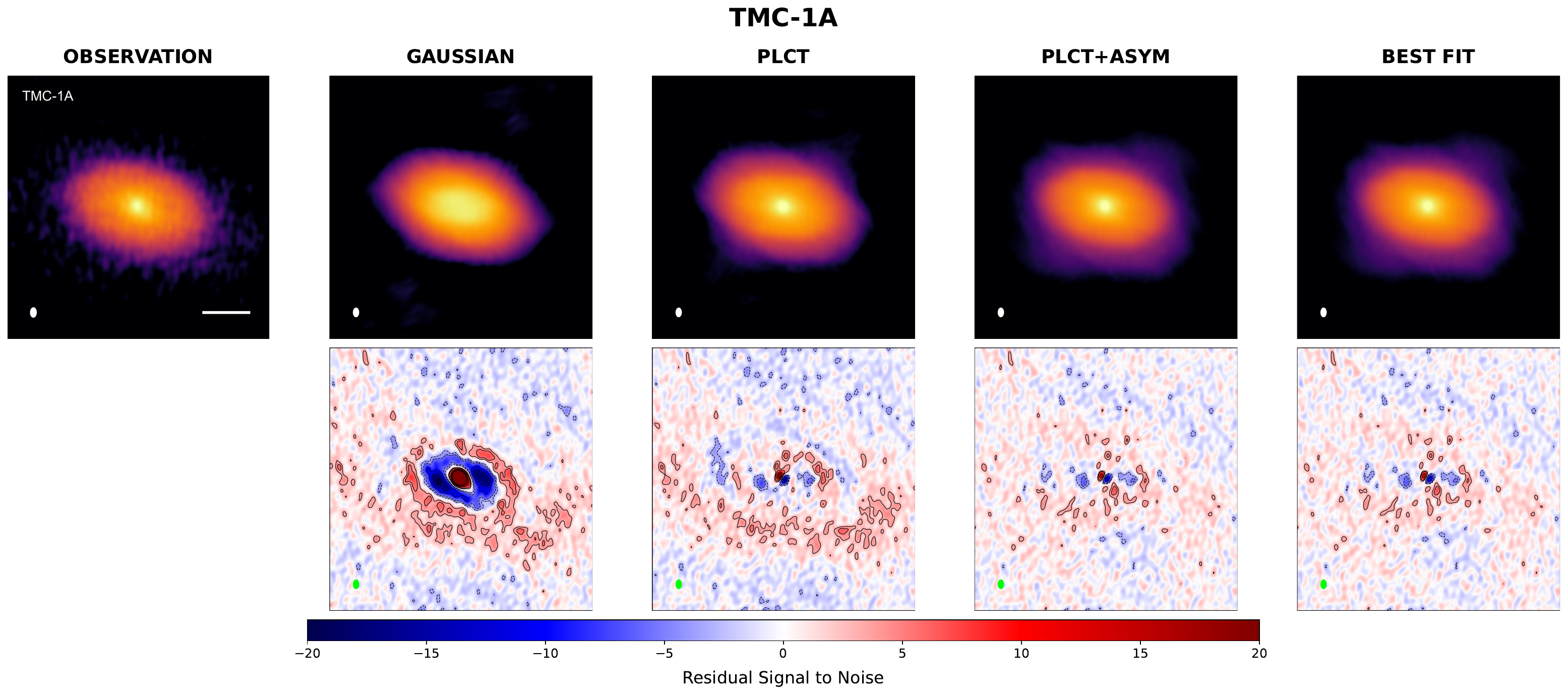}
\includegraphics[width=0.85\linewidth]{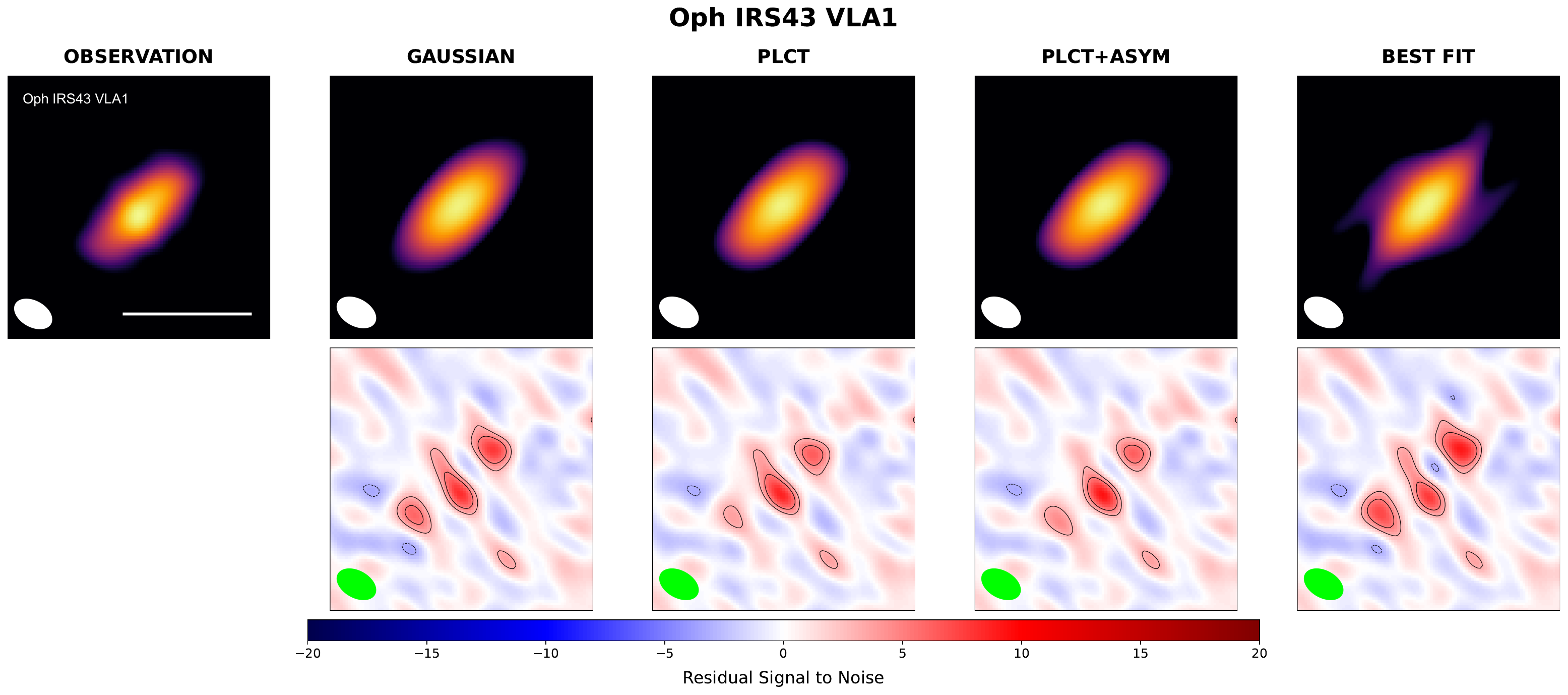}
\caption{Continuation of Figure \ref{Fig2}.}
\end{figure*}

\addtocounter{figure}{-1}

\begin{figure*}
\centering
\includegraphics[width=0.85\linewidth]{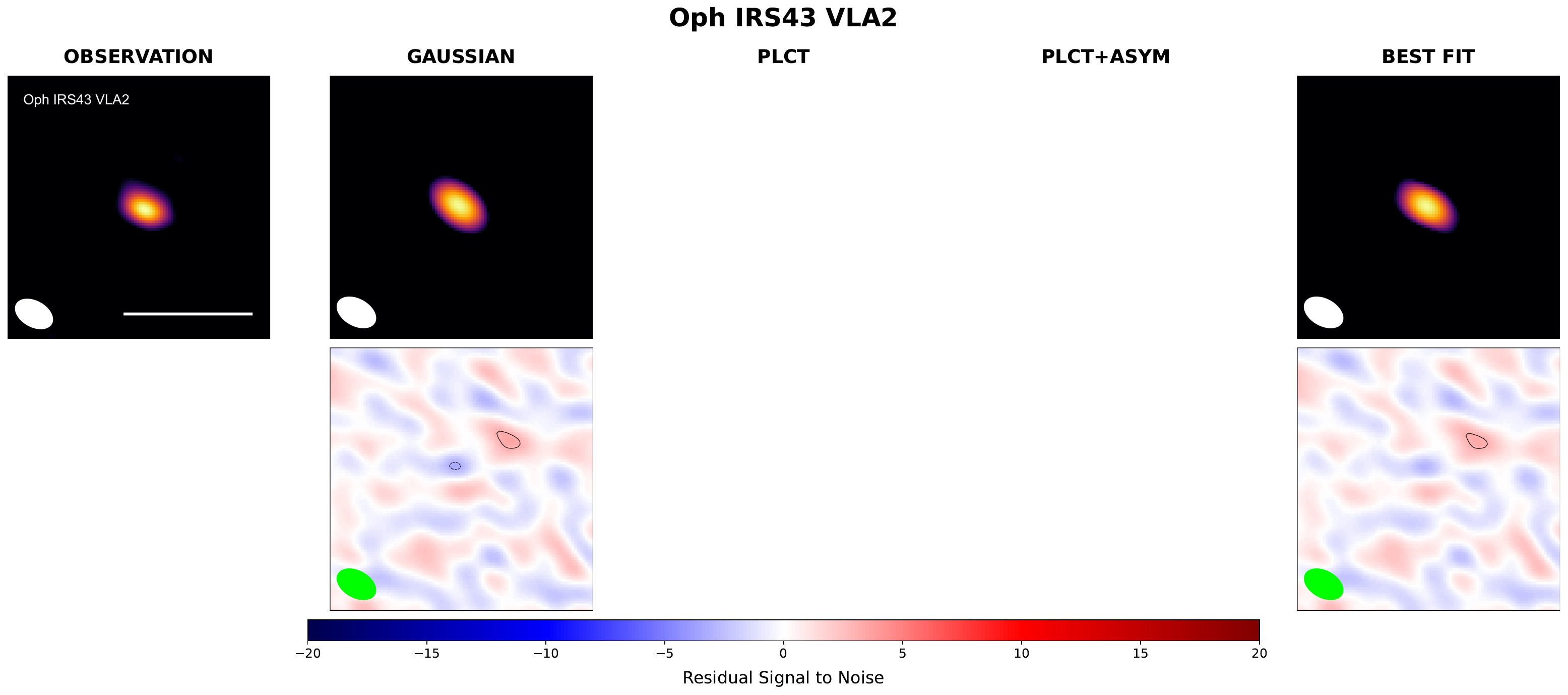}
\includegraphics[width=0.85\linewidth]{L1489IRS_model_comparison.pdf}
\includegraphics[width=0.85\linewidth]{OphIRS63_model_comparison.pdf}
\caption{Continuation of Figure \ref{Fig2}.}
\end{figure*}

\section{Results}

\subsection{Inclined Disks}

For all inclined protostellar {disks}, we tested a variety of models, including those described in Appendix B to analyze the data comprehensively. Ultimately, in this study, we have chosen to focus on and highlight three specific models for each source: (1) a 2D Gaussian, (2) a Power-Law Core with an Exponential Tail (PLCT), and (3) a Power-Law Core with an Exponential Tail that includes asymmetry (PLCT+asym). For the binary systems (all with inclined disks, {Figure \ref{fig:Binary}) in our sample, we fit the primary using the three models mentioned above i.e., Gaussian, PLCT and the PLCT profile with asymmetry while we only fit a simple Gaussian to the much weaker secondary in most cases.}

\subsubsection{Gaussian fit}
In Figure \ref{Fig2} column 2 {(as well as Figure \ref{fig:vis})}, we present the results of fitting a simple Gaussian function to the sample of inclined disks in the eDisk sample. The corresponding fit parameters for each source are listed in Table \ref{Table2}. We find that the Gaussian model provides a good description of the overall disk structure, yielding inclination angles, position angles, and disk radii that are generally consistent with previous observational results. As shown in Figure \ref{Fig2} column 2, the Gaussian model provides a reasonable first-order approximation of the disk structure, capturing the general radial decline in emission. However, despite its simplicity and utility, the Gaussian fits  do not well represent most of the eDisk sample with residuals more than 5~$\sigma$. The exceptions are Oph IRS43, as well as for the fainter and more compact secondary components in the binary systems of Ced 110 IRS4 and Oph IRS43.

The Gaussian model exhibits several significant limitations when applied to our sample. First, it systematically underestimates the intensity of the centrally peaked emission observed in many disks. This discrepancy is evident in the large residuals shown in Figure \ref{Fig2} column 2, where prominent central excesses remain after subtracting the model. These residuals suggest that the inner disk regions often host compact emission components that are not well described by a single Gaussian. Second, the Gaussian profile inherently assumes azimuthal symmetry, rendering it incapable of capturing brightness asymmetries associated with substructures such as rings, gaps, spirals, or localized enhancements possibly driven by planet-disk interactions or radiative transfer effects. Some of these asymmetries may not be due to real physical substructures but could be due to temperature gradients as discussed in \cite{2023ApJ...951....8O}. Furthermore, in the case of L1489 IRS, the Gaussian model fails dramatically. This source exhibits a complex radial morphology, with a bright central peak flanked by two distinct annular rings of emission at larger radii. The Gaussian fit cannot replicate this multi-component structure, resulting in significant deviations between the model and the observed profile. These shortcomings highlight the need for more sophisticated models to fully characterize the complex emission features observed in the eDisk sample.

\subsubsection{Power-Law Core with an Exponential Tail}

The Power-Law Core with an Exponential Tail (PLCT) profile, characterized by its two distinct components, provides a more flexible framework for modeling disk structures. The power-law core represents the inner protostellar disk, capturing the steep brightness gradient near the center, while the exponential tail describes the outer disk’s gradual decline, effectively accounting for the brighter inner regions that the Gaussian fit fails to model adequately.

In Figure \ref{Fig2} column 3, we present the best-fit results for the inclined disks using the PLCT profile. Table \ref{tab:PLCT} lists the best-fit parameters of the PLCT profile for our sample. Focusing on the parameters $\gamma_1$ and $\gamma_2$, which characterize the power-law core and the exponential tail, respectively, we observe a clear deviation from the standard PLCT profile with $\gamma_1 = \gamma_2$ as proposed by \cite{1974MNRAS.168..603L}. We note that for two sources B335 and IRAS 15398-3359, we do not obtain a good converged solution.  The PLCT profile represents a significant improvement over the simpler Gaussian model, as evidenced by the markedly reduced residuals shown in Figure \ref{Fig2}. Unlike the Gaussian profile, the PLCT fit more effectively captures the bright, compact central emission often observed in protostellar disks, resulting in a much closer match to the overall disk brightness distribution. This improved fidelity is particularly evident in sources with pronounced central peaks, where the Gaussian model tends to underestimate the core intensity.

The non-zero and distinct values of $\gamma_1$ and $\gamma_2$ further indicate a significant departure from Gaussianity in the disk profiles. This result implies that, to first order, no disk in our sample is well described by a pure Gaussian distribution as can be quantified with the increased Bayes factor (as discussed in Sec 5.4) for going from a Gaussian model to the PLCT profile. The PLCT profile provides a more physically motivated model for characterizing disk properties, capturing both the steep inner gradients and the extended outer emission more effectively.

An examination of the residual images for the PLCT profile (Figure \ref{Fig2} column 3) shows that, although the PLCT profile offers a significant improvement over the Gaussian fit, particularly in reproducing the inner disk's steep brightness distribution, it still falls short in capturing the full complexity of the disk structures. Specifically, the PLCT model cannot fully account for brightness asymmetries or localized overdensities, which are common features in protostellar/protoplanetary disks. Furthermore, the PLCT profile begins to reveal subtle substructures in some disks, such as those observed in Oph IRS63 and TMC-1A, where substructures have been previously detected\citep{2020Natur.586..228S,2021ApJ...920...71A,2023ApJ...954..190X}. In addition, the remaining non-zero residuals highlight asymmetric brightness structures within the disks that require further modeling to understand.

\subsubsection{PLCT with asymmetry}

In Figure \ref{Fig2} column 4, we show the best-fit models and corresponding residuals obtained using the asymmetric PLCT profile. Incorporating asymmetries into the PLCT model further reduces the residuals. The best-fit parameters are listed in Table \ref{tab:asy}. Several disks (e.g., IRAS 04169, R CrA IRAS 32A, and 32B, BHR71 IRS1 and IRAS 16544-1604) that previously exhibited prominent asymmetric structures in Figure \ref{Fig1} now show negligible residuals. However, some disks (e.g., L1489 IRS, IRS7B, Oph IRS63) still display significant residuals that cannot be accounted for by such simple models and thus require a more specialized treatment.

\subsection{Edge-on disks}
\begin{figure*}
\centering
\includegraphics[width=0.9\linewidth]{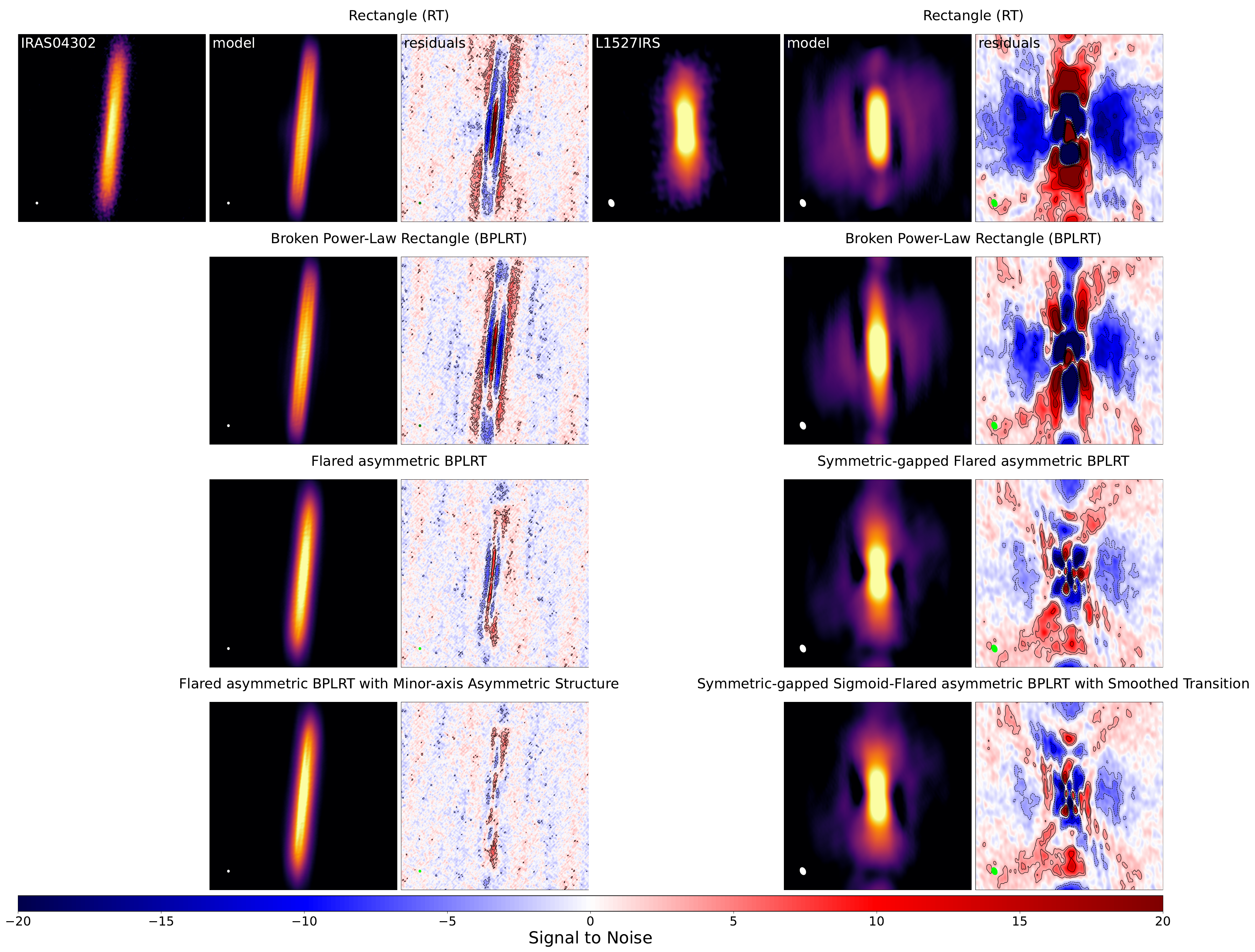}
\caption{Model fitting results for the two edge-on disks IRAS 04302 and L1527 IRS. From top to bottom, each row shows the model and residual maps, with corresponding model names indicated above. Residuals are plotted with same convention as Figure~\ref{Fig2}, using signal-to-noise contours based on source-specific background rms values.}
\label{Fig9}
\end{figure*}

For the two edge-on disks in our sample we test the models listed in \cite{2022ApJ...934...95S}. In Figure \ref{Fig9}, we present the fitting results of rectangle model, broken power-law rectangle model, and flared asymmetric broken power-law rectangle model, as well as its improved variants. The mathematical formulation of these models and improvements has been described in Section~\ref{sec:edge-on math}. The first two models resemble the Gaussian and standard power-law profiles with exponential tales applied to inclined disks, respectively. The rectangle model serves as a useful baseline, capturing the overall morphology of edge-on disks, providing a simple estimate of sizes and brightness, while its uniform intensity and fixed vertical extent fails to reproduce more complex brightness distributions as can been seen in Figure \ref{Fig9}. Next we implemented the broken power-law to introduce a smooth radial intensity gradient along the major axis, allowing for different inner and outer slopes. Still, its fixed vertical width limits its ability to reproduce the flaring structure as can been seen from the large residuals on either side. 

Building upon previous models, the flared broken power-law rectangle model introduces several key improvements. First, it allows for different outer gradients on either side of the major axis, providing the flexibility to capture asymmetric radial profiles. Second, the vertical width of the disk is allowed to vary with radial distance, in order to capture flaring structures, which is especially effective in reproducing L1527 IRS's morphology. For L1527 IRS, two symmetric vertical gaps are included to account for possible substructures. This model has much stronger ability to reproduce the observed disk structures and leads to a substantial reduction in residuals compared to the simpler models.

\subsection{Best-fit Models with Additional Structures} \label{sec:best-fit}
While the majority of the inclined sources in our sample can be well described by the PLCT profile with one asymmetry profile, some sources (BHR 71 IRS1, GSS 30 IRS3, IRAS 15398-3359, L1489 IRS, Oph IRS 43 VLA1, Oph IRS 63, and R CrA IRS 7B-a) require more complex structures to achieve an adequate fit (which we define as $\lesssim 5\sigma$ residuals). Based on residual and Bayes factor analysis , the developed structures include, but are not limited to, asymmetric radius profile, broken power-law profile, spiral arms, multiple rings, Gaussian envelopes with varying inclination. These models are further discussed in Appendix B and fitting results are described in Appendix C.
We summarize the best-fit models for all sources in Figure \ref{Fig2} column 5 and also in Table \ref{tab:bestfit_models}. For the binary sources we also show the bestfit of the two systems together in Figure \ref{fig:Binary} in the Appendix.

\begin{deluxetable*}{lll}[t]
\tablecaption{Description of the best-fit disk models for the eDisk sample \label{tab:bestfit_models}}
\tablehead{
\colhead{Name} & \colhead{Best-fit Model} & 
}
\startdata
BHR~71~IRS2 & PLCT & Eq \ref{eq:PLCT}\\
B335 & PLCT + asymmetric profile (sinusoidal) with center offset & Eq \ref{eq:PLCT}(1 + a $\cdot$ Eq \ref{eq:8}) \\
L1527~IRS & Symmetric gapped sigmoid-flared asymmetric BPLRT with smoothed transition & Eq \ref{eq:edge-on-bpl_sym_gapped} \\
IRAS~16253-2429 & PLCT + asymmetric profile (4th-order Gaussian) & Eq \ref{eq:PLCT}(1 + a $\cdot$ Eq \ref{eq7}) \\
IRAS 16544-1604 & PLCT + asymmetric profile (4th-order Gaussian) & Eq \ref{eq:PLCT}(1 + a $\cdot$ Eq \ref{eq7}) \\
GSS~30~IRS3 & Weighted broken power laws with asymmetric radius + asymmetric profile (4th-order Gaussian) & Eq \ref{eq:GSS30IRS3_bestfit} \\
IRAS~15398-3359 & PLCT + symmetric ring & Eq \ref{eq:PLCT_sym_ring} \\
R~CrA~IRS~5N & PLCT + asymmetric profile (4th-order Gaussian) & Eq \ref{eq:PLCT}(1 + a $\cdot$ Eq \ref{eq7}) \\
IRAS~04166+2706 & PLCT + asymmetric profile (4th-order Gaussian) & Eq \ref{eq:PLCT}(1 + a $\cdot$ Eq \ref{eq7}) \\
R~CrA~IRAS~32A & PLCT + asymmetric profile (sinusoidal) & Eq \ref{eq:PLCT}(1 + a $\cdot$ Eq \ref{eq:8}) \\
R~CrA~IRAS~32B & PLCT + asymmetric profile (sinusoidal) & Eq \ref{eq:PLCT}(1 + a $\cdot$ Eq \ref{eq:8}) \\
BHR~71~IRS1 & PLCT with asymmetric radius (linear) + asymmetric profile (4th-order Gaussian) & Eq \ref{eq:BHR71IRS1_bestfit} \\
Ced~110~IRS4A & PLCT + asymmetric profile (4th-order Gaussian) & Eq \ref{eq:PLCT}(1 + a $\cdot$ Eq \ref{eq7}) \\
Ced~110~IRS4B & Gaussian & Eq 1 \\
IRAS~04302+2247 & Flared asymmetric broken power-law rectangle model with minor-axis asymmetry & Eq \ref{eq:edge-on-bpl-ystructure} \\
R~CrA~IRS~7B-a & PLCT + single spiral + asymmetric profile (4th-order Gaussian) & Eq \ref{eq:IRS7B_bestfit} \\
R~CrA~IRS~7B-b & PLCT & Eq \ref{eq:PLCT} \\
IRAS~04169+2702 & PLCT + asymmetric profile (4th-order Gaussian) & Eq \ref{eq:PLCT}(1 + a $\cdot$ Eq \ref{eq7}) \\
TMC--1A & PLCT + asymmetric profile (4th-order Gaussian) & Eq \ref{eq:PLCT}(1 + a $\cdot$ Eq \ref{eq7}) \\
Oph~IRS~43 VLA1 & PLCT + Gaussian disk & Eq \ref{eq:OphIRS43_bestfit} \\
Oph~IRS~43 VLA2 & PLCT & Eq \ref{eq:PLCT}\\
L1489~IRS & PLCT + intermediate Gaussian with asymmetric ring + outer Gaussian & Eq \ref{eq:L1489IRS_bestfit} \\
Oph~IRS~63 & Weighted broken power laws with two symmetric Gaussian gaps & Eq \ref{eq:OphIRS63_bestfit} \\
\enddata
\end{deluxetable*}

\section{Discussion}
Our analysis of the uv-plane visibilities, supported by inspection of the image-plane residuals, shows that most protostellar disks exhibit significant brightness asymmetries and clear departures from the standard PLCT profile. However, unlike Class II disks, we do not find evidence for prominent gaps or rings in most of protostellar disks. While our models successfully capture the overall morphology of protostellar disks and account for large-scale asymmetries, we face a significant limitation: it remains challenging to determine whether these asymmetries represent genuine physical structures or if they are artifacts arising from radiative transfer effects in optically thick regions of the disk. This ambiguity represents a major drawback in our current modeling approach, as it hinders our ability to definitively interpret the observed features. Despite these limitations, the disk profiles generated by our models are essential inputs for more sophisticated radiative transfer simulations. They provide a robust analytical framework for dust distribution, serving as critical reference profiles that can be further refined to improve our understanding of disk structure and evolution in the protostellar phase.

\subsection{Comparing disk radii using different models}

To enable a consistent comparison across different models, we systematically analyze the results derived from the Gaussian, PLCT, and best-fit profiles. A key aspect of this comparison is adopting a uniform definition of disk radius that can be applied to all model profiles. Even in previous eDisk studies, the definition of disk radius varies: \citet{2023ApJ...951....8O} define the disk radius as the 1$\times$FWHM width along the major axis of a deconvolved 2D Gaussian fit in the image plane using CASA, whereas \citet{2024ApJ...969..125Y} define it as 2$\times$FWHM of the 2D Gaussian or as the radius enclosing 90\% of the total flux based on the 2D Gaussian fit. 

In this work, we define the disk radius as the radius enclosing 95\% of the total disk flux density, motivated by theoretical studies of Class II disks \citep{2019MNRAS.486.4829R} and several observational studies \citep[e.g.,][]{2019ApJ...882...49L,2024PASJ...76..437Y}. This definition can be evaluated directly from the analytical expressions of each model profile, as well as measured from the corresponding model image, allowing for a consistent comparison regardless of the underlying functional form. For a simple Gaussian surface-brightness profile, the radius enclosing 95\% of the total flux corresponds to $r_{95} \simeq 2.448\,\sigma$. For more complex models, such as the PLCT, edge-on disk models and the best fit models, the 95\% flux radius is determined by measuring it directly from the corresponding model image. We have listed the disk radii computed from various models in Table \ref{tab:disk_radii}.

\begin{figure*}
\centering
\includegraphics[width=0.5\linewidth]{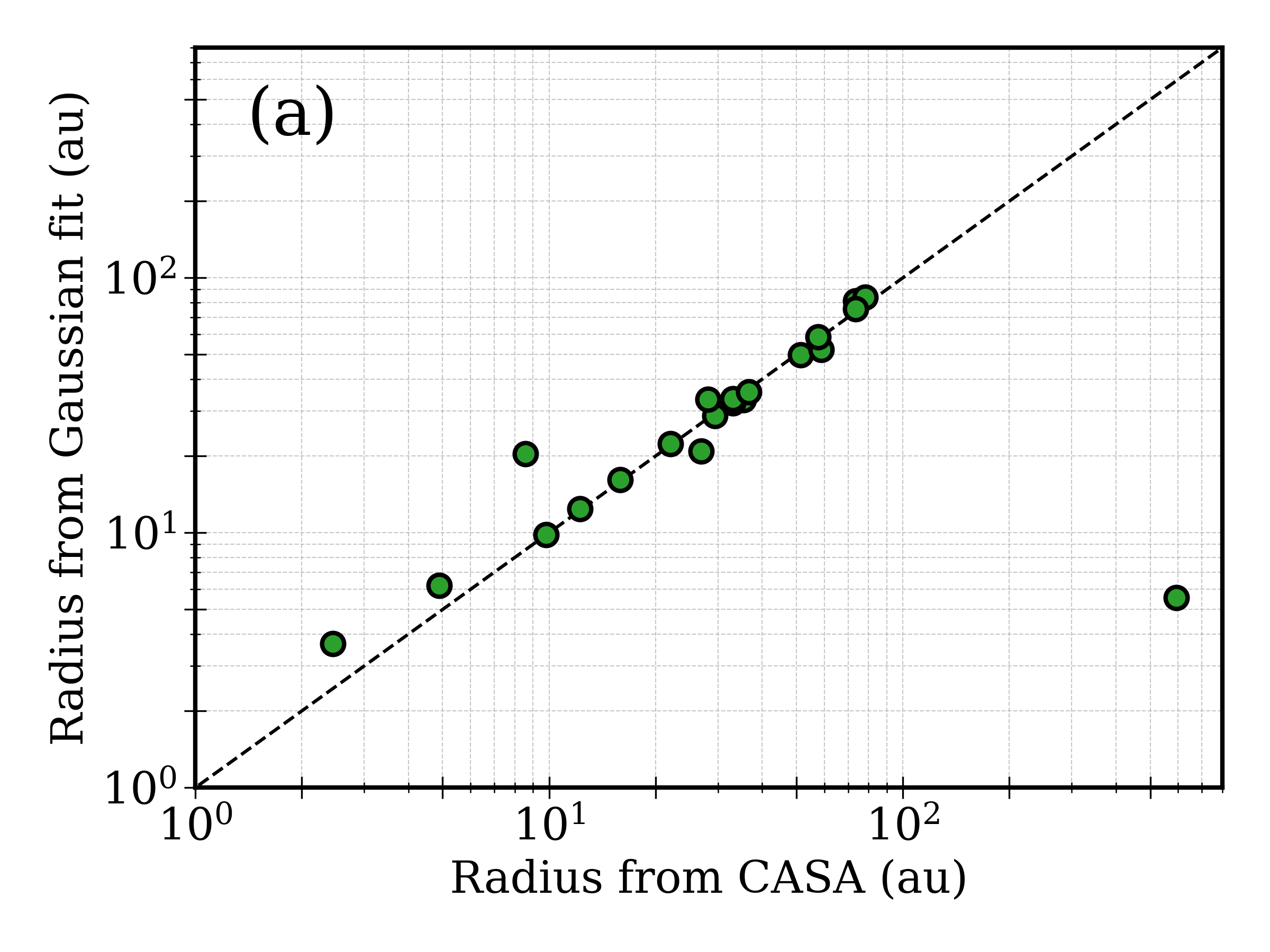}\includegraphics[width=0.5\linewidth]{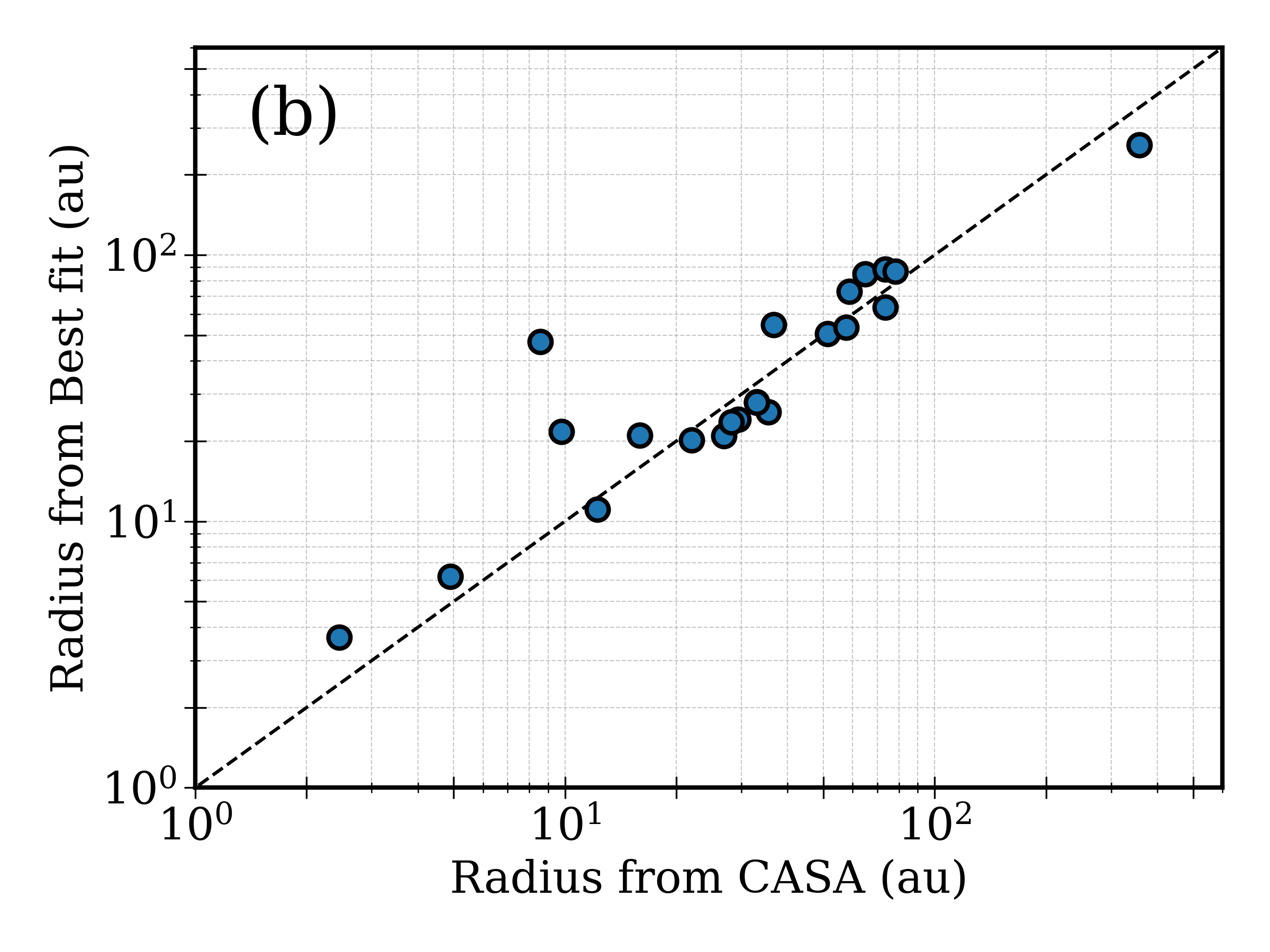}
\caption{We compare the radii from 2-dimensional Gaussian fitting in the image plane using the CASA task \textit{imfit} with those obtained in this work. Panel (a) shows the comparison using radii derived from Gaussian fits, while panel (b) shows the comparison using the best-fit model.}
\label{Fig11}
\end{figure*}

In Figure~\ref{Fig11}(a), we compare disk radii derived from 2D Gaussian fits performed in the image plane using CASA with those obtained from our parametric modeling in the visibility plane. For the majority of sources, the radii inferred from the two independent methods are in good agreement, indicating that the simpler image-plane approach provides a reasonable estimate of disk size in most cases. However, for L1489 IRS the 2D Gaussian fit in the uv-plane fails to recover a reliable disk radius. In Figure~\ref{Fig11}(b), we further compare the disk radii derived from the best-fit visibility-plane models with those inferred from the deconvolved 2D Gaussian fits in the image plane using CASA. Once again, we find that the two radius estimates are largely consistent across the sample, with only minor scatter.


\subsection{Disk radii vs Stellar mass}


More evolved Class II disks have shown a strong correlation with disk radii and stellar mass \citep{2020ARA&A..58..483A,2018ApJ...865..157A}. 
Both hydrodynamical (HD) and magneto-hydrodynamical (MHD) simulations have shown a correlation between disk radii and stellar masses \citep[e.g.,][]{1984ApJ...286..529T, 1998ApJ...509..229B, 2016ApJ...830L...8H, 2021A&A...648A.101L}. 
\citet{2024ApJ...969..125Y} analyzed the eDisk sample and found a statistically significant correlation between the current protostellar mass and the disk radius. However, the protostellar masses used in that study were derived from different papers. To address this limitation, we re-examined the relationship between disk radius and protostellar mass using a set of stellar masses that have been derived in a homogeneous manner (from Aso et al. 2026).

\begin{figure}
\centering
\includegraphics[width=1\linewidth]{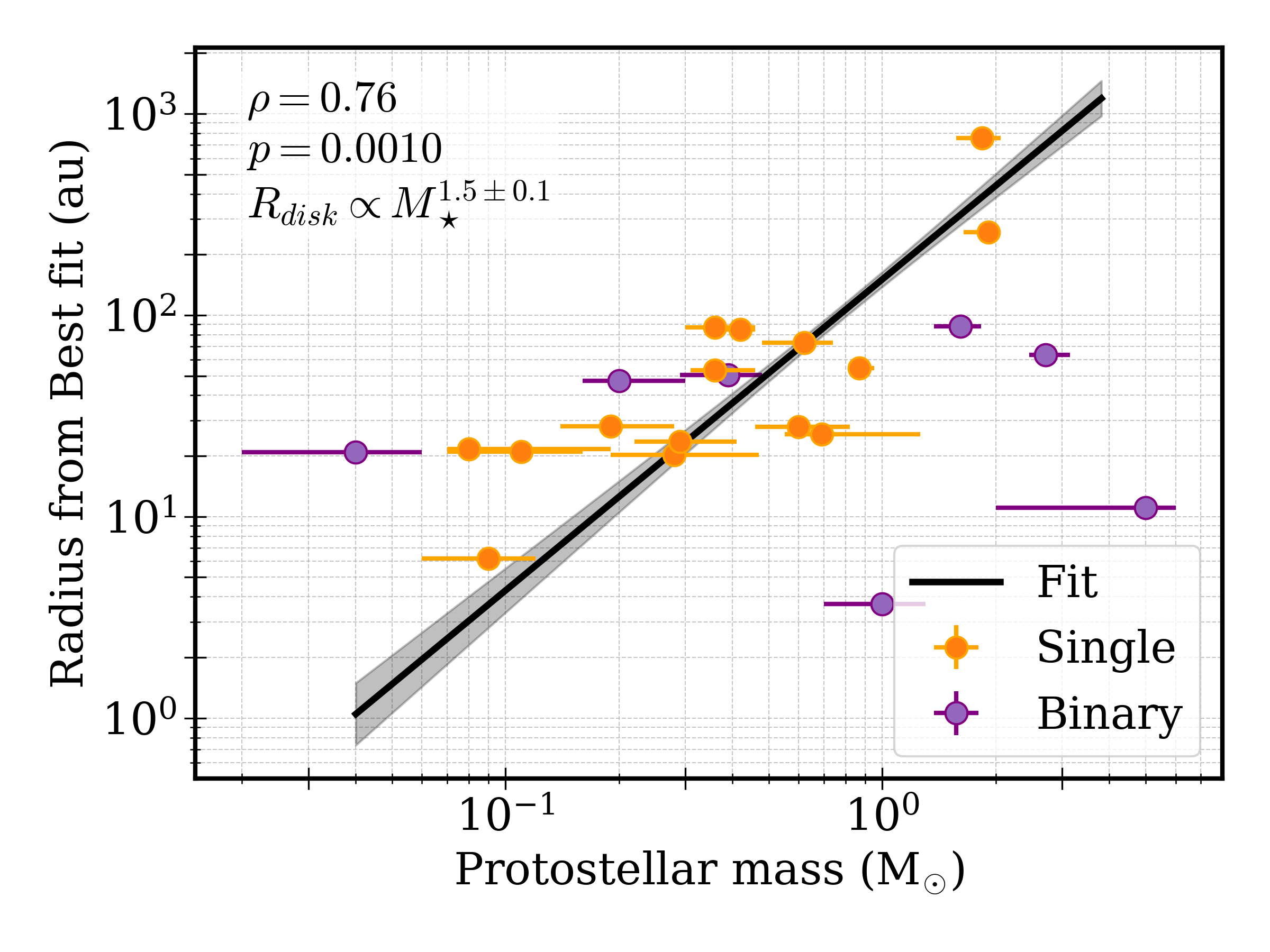}
\caption{The protostellar mass-radius relationship based on the best fit radii for our sample.}
\label{Fig12}
\end{figure}

In Figure~\ref{Fig12}, we present the {best-fit disk radius} (listed in Table \ref{tab:disk_radii}) as a function of stellar mass for our sample. We assume a 10\% error in the best-fit values. In this analysis, we restrict ourselves to the population of isolated protostars. Consistent with the findings of \citet{2024ApJ...969..125Y}, we identify a clear positive correlation between stellar mass and disk radius. This trend is quantified by a Spearman rank correlation coefficient of $\rho = 0.76$, with a corresponding $p$-value of 0.001, indicating that the correlation is statistically significant. We note that this correlation may not arise solely from disk evolution i.e., the idea that more evolved sources host larger disks as suggested by HD and MHD simulations. The sources in our sample are not necessarily destined to become solar-type stars, and the observed differences in stellar mass may partly reflect intrinsic variations in the final stellar masses of the sources themselves.

We fit a power-law relation to the isolated protostar population and find that the disk radius scales strongly with stellar mass, following $R_{\rm disk} \propto M_{\star}^{1.5 \pm 0.1}$. This dependence is notably steeper than that observed for Class~II disks, where previous studies have found a relation of approximately $R_{\rm disk} \propto M_{\star}^{0.9}$ (e.g., \citealt{2020ARA&A..58..483A,2018ApJ...865..157A}). However there are only a very small number of sources in intermediate mass range ($M_{\star}>1 M_{\odot}$) which could bias our results. In addition, we observe a systematic offset between different source populations: at a given stellar mass, disks associated with binary systems tend to be smaller than those around isolated protostars. This behavior is consistent with expectations from dynamical truncation in multiple systems and suggests that binarity plays an important role in regulating disk sizes during the protostellar phase.

\subsection{Comparison to Previous Results on Disk Substructure}

Substructures in several of our targets have been previously identified in the literature, including initial results from the eDisk survey as well as independent studies. Among these, Oph IRS 63, L1489 IRS and TMC-1A stand out as particularly well-studied examples where some hints of substructures have previously been detected. In the case of Oph IRS 63, our residual plots clearly reveal the presence of concentric ring-like features, which are in strong agreement with earlier findings (see Figure \ref{Fig2}). \citet{2020Natur.586..228S} analyzed 1.3 mm ALMA continuum observations at a spatial resolution of 5 au and reported the detection of two bright rings separated by two dark gaps. Building upon this, \citet{2023AJ....166..184M} modeled the continuum emission using a PLCT profile combined with two Gaussian rings and an additional inner Gaussian component. Their model successfully reproduced the ring locations observed by \citet{2020Natur.586..228S}. In our study, we conduct an independent, detailed modeling of Oph IRS 63 (presented in Appendix~\ref{IRS63}). Our fits recover the positions of substructures consistent with both \citet{2020Natur.586..228S} and \citet{2023AJ....166..184M}, further validating the robustness of the observed ring-gap morphology.

The disk around L1489 IRS has been extensively studied at millimeter wavelengths prior to the eDisk survey \citep{2014ApJ...793....1Y, 2020ApJ...893...51S, 2022ApJ...933...23O}. These studies revealed an extended disk structure consisting of a bright, compact inner component surrounded by a more tenuous outer disk, which is misaligned with the inner disk. In addition, \citet{2022ApJ...933...23O} reported tentative evidence for the presence of an inner gap or ring. The high-angular-resolution eDisk observations confirm and extend these earlier results, clearly resolving a complex, multi-component disk morphology. The eDisk data reveals a bright compact inner disk, an intermediate disk, and an outer more tenuous disk, with misaligned inclinations and position angles \citep{2023ApJ...951...11Y}. When using the simple PLCT model fitting, we find that only the compact central source in L1489 IRS is well-fitted, while extended positive residuals remain at larger radii. Therefore the best-fit model included a compact PLCT core representing the inner disk and the central source, and two extended Gaussian components.

\begin{deluxetable*}{lccccccc}
\tablecaption{Comparison of Observed and Model-derived Radii. The $R_{\rm model}$ values correspond to radii enclosing 95\% of the total flux (2.448 $\sigma$). Similarly the $R_{\rm Gaussian}$, $R_{\rm PLCT}$, $R_{\rm PLCT,asy}$ also correspond to radii enclosing 95\% of the total model flux. $R_{\rm Bestfit, 95}$, $\mathbf{R_{\rm Bestfit, 90}}$ and $\mathbf{R_{\rm Bestfit, 68}}$  correspond to the radii enclosing 95\%, 90\% and 68\% of the total model flux in the best fit model. $^\dagger$ Edge-on disks}
\tablehead{
\colhead{Name} &
\colhead{$R_{\rm observe}$} &
\colhead{$R_{\rm Gaussian}$} &
\colhead{$R_{\rm PLCT}$} &
\colhead{$R_{\rm PLCT,asy}$} &
\colhead{$R_{\rm Bestfit, 95}$} &
\colhead{$R_{\rm Bestfit, 90}$} &
\colhead{$R_{\rm Bestfit, 68}$} \\ \colhead{---} &
\colhead{(au)} &
\colhead{(au)} &
\colhead{(au)} &
\colhead{(au)} &
\colhead{(au)} &
\colhead{(au)} &
\colhead{(au)}
} 
\label{tab:disk_radii}
\startdata
BHR71 IRS2 & 8.6 & 20.4 & 47.2 & & 47.2 & 31.7 & 12.2 \\
B335 & 9.8 & 9.8 & 23.3 & 21.7 & 21.7 & 16.5 & 7.7 \\
L1527 IRS$^\dagger$ & 64.9 & & & & 85.0 & 68.6 & 35.0 \\
IRAS 16253-2429 & 15.9 & 16.1 & 19.4 & 21.0 & 21.0 & 17.7 & 11.1 \\
IRAS 16544-1604 & 33.0 & 32.2 & 29.0 & 28.1 & 28.1 & 25.6 & 19.3 \\
GSS30 IRS3 & 78.3 & 84.0 & 89.1 & 86.4 & 86.7 & 75.4 & 51.7 \\
IRAS 15398-3359 & 4.9 & 6.2 & 202.2 & 260.2 & 6.2 & 5.3 & 3.9 \\
R CrA IRS5N & 57.5 & 58.6 & 54.0 & 53.4 & 53.4 & 49.2 & 37.3 \\
IRAS 04166+2706 & 22.0 & 22.4 & 20.2 & 20.2 & 20.2 & 18.2 & 13.9 \\
R CrA IRAS 32A & 33.0 & 33.5 & 28.2 & 28.2 & 27.9 & 26.0 & 20.8 \\
R CrA IRAS 32B & 28.2 & 33.3 & 24.1 & 23.5 & 23.5 & 21.5 & 16.8 \\
BHR71 IRS1 & 51.4 & 49.9 & 46.5 & 47.8 & 50.6 & 44.3 & 31.1 \\
Ced110 IRS4A & 73.4 & 81.2 & 90.8 & 88.2 & 88.2 & 79.1 & 59.4 \\
Ced110 IRS4B & 26.9 & 20.9 & & & 20.9 & 19.3 & 13.5 \\
IRAS 04302+2247$^\dagger$ & 357.4 & & & & 258.8 & 224.8 & 148.1 \\
R CrA IRS7B-a & 73.4 & 75.7 & 64.4 & 62.9 & 63.5 & 59.4 & 47.2 \\
R CrA IRS7B-b & 29.4 & 28.8 & 24.1 & & 24.1 & 22.4 & 17.9 \\
IRAS 04169+2702 & 35.5 & 33.3 & 26.0 & 25.7 & 25.7 & 24.2 & 19.6 \\
TMC-1A & 36.7 & 35.6 & 33.5 & 54.6 & 54.6 & 40.8 & 24.7 \\
Oph IRS43 VLA1 & 12.2 & 12.4 & 11.1 & 11.1 & 11.1 & 10.2 & 8.2 \\
Oph IRS43 VLA2 & 2.4 & 3.7 & & & 3.7 & 3.1 & 2.3 \\
L1489 IRS & 593.6 & 5.6 & 291.7 & 291.7 & 756.3 & 142.6 & 113.8 \\
Oph IRS63 & 58.8 & 52.3 & 78.6 & 72.7 & 72.9 & 65.4 & 47.5 \\
\enddata
\end{deluxetable*}

For TMC-1A, \cite{2021ApJ...920...71A} and \citet{2023ApJ...954..190X} identified non-axisymmetric residuals suggestive of spiral substructures in high-resolution ALMA continuum data. In our analysis, we also observe deviations from axisymmetry in the residuals with hints of spiral features when fitting a symmetric PLCT profile (see Figure \ref{Fig2} column 3). However, these spirals are far less prominent than those observed in \citet{2023ApJ...954..190X}. To better match the observed morphology, we introduce an asymmetric component into our model, which significantly improves the fit. These results indicate that while the PLCT framework captures the bulk of the disk emission, there are still asymmetries in the disk. 

\cite{2026PASJ..tmp...39T} reported the possible presence of spiral-like structures in the disk of IRAS 16544-1604, which we also tentatively recover in the residuals of the PLCT profile (Figure \ref{Fig2} column 3). Tentative spiral-like residuals are also seen in IRS7B-a and IRAS 04169 after subtraction of the PLCT profile (Figure \ref{Fig2} column 3). However, higher-resolution and deeper multi-wavelength observations are required to robustly confirm these features and test their physical origin.

\cite{2022ApJ...934...95S} investigated high-resolution Karl G. Jansky Very Large Array (VLA) observations of L1527 IRS at wavelengths of 7 mm, 1.3 cm, and 2 cm. They detected the edge-on disk at all three wavelengths and reported a pronounced brightness asymmetry, with the southern side brighter than the northern side. Their analytic modeling favored an asymmetric, flared disk at 7 mm, consistent with our modeling of the eDisk observations. In addition, they investigated the presence of gaps within the disk; however, similar to our analysis, they were unable to confirm the existence of substructures with statistical significance.

\subsection{Comparison of Models Based on $\log Z$}
The quantity $\log Z$ provided by \texttt{dynesty} corresponds to the log Bayesian evidence, which measures the overall ability of a model to explain/fit the data. In the following, we compare models using $\Delta \log Z$, defined relative to the Gaussian model, such that all $\Delta \log Z$ values are presented as $\log Z_{\mathrm{model}} - \log Z_{\mathrm{Gaussian}}$. As a baseline, the Gaussian model has $\Delta \log Z = 0$. 

\begin{figure*}
 \centering
 \includegraphics[width=0.7\linewidth]{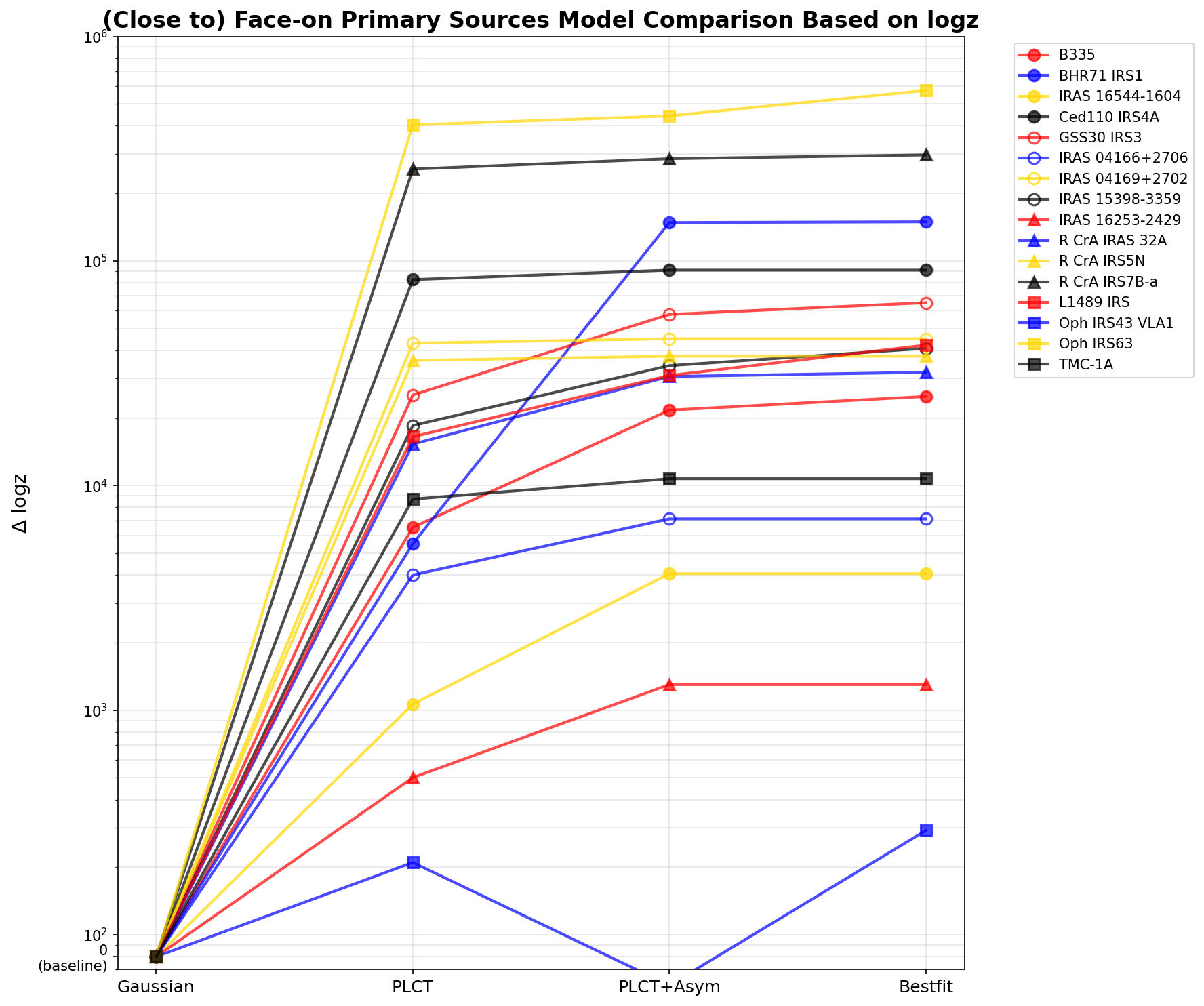}
 \caption{Comparison of $\Delta \log Z$ values relative to the Gaussian model for all inclined disk primary sources on log scale. PLCT models, PLCT models with a 4th-order Gaussian asymmetric profile, and the best-fit models are shown.}
 \label{fig:dlogz_all_sources}
\end{figure*}

\begin{figure*}
 \centering
 \includegraphics[width=1\linewidth]{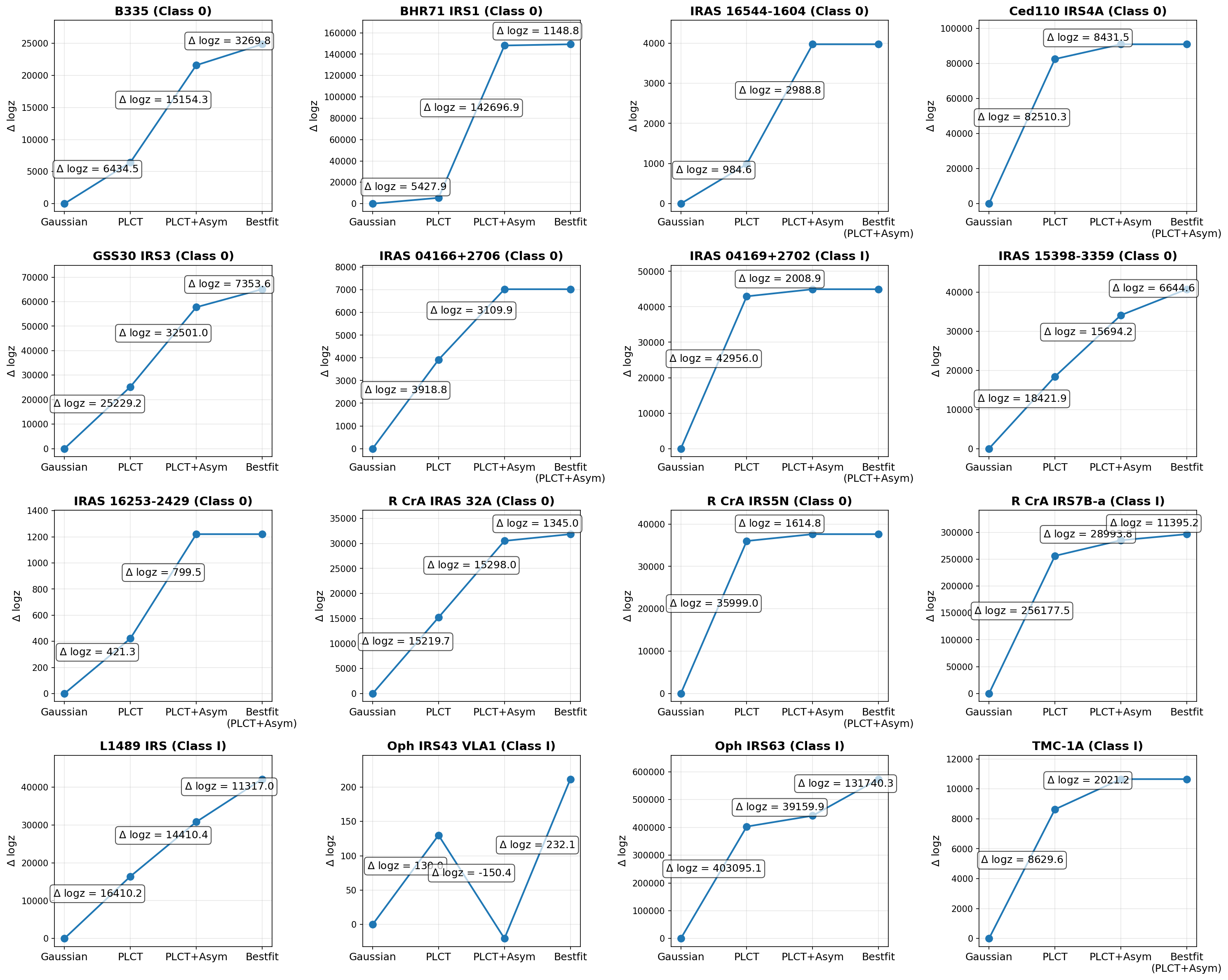}
 \caption{Same as Figure~\ref{fig:dlogz_all_sources}, but shown separately for each individual source on linear scale.}
 \label{fig:dlogz_single_source}
\end{figure*}

We compare the Gaussian, PLCT, PLCT models with a 4th-order Gaussian asymmetric profile, and the final best-fit models using $\Delta \log Z$. We present the results with all inclined disk shown together in Fig.~\ref{fig:dlogz_all_sources} to illustrate the overall trends across the sample, and shown separately for each individual source in Fig.~\ref{fig:dlogz_single_source}. 

For all sources, the transition from a Gaussian model to a PLCT model provides a significant improvement in $\Delta \log Z$. For most sources, the introduction of an asymmetric profile provides an additional improvement in $\Delta \log Z$, and for a substantial fraction of the sample, the improvement from the asymmetric profile is comparable to, or even exceeds that from the Gaussian to PLCT transition.

As shown in the single-source comparison (Fig.~\ref{fig:dlogz_single_source}), this trend is particularly pronounced among the Class~0 sources. With the exception of IRS5N and Ced110 IRS4A, all other Class~0 sources exhibit a strong preference for models that include an asymmetric profile. In contrast, for the Class~I sources, although introducing an asymmetric profile generally improves the fit, the corresponding increase in $\Delta \log Z$ is typically smaller than that achieved when transitioning from a Gaussian to a PLCT model. This behavior may suggest that Class~0 disks exhibit stronger brightness asymmetries than Class~I disks. \cite{2023ApJ...951....8O} suggested that embedded disks in the eDisk sample show more pronounced brightness asymmetries than Class~II disks due to the larger scale height and optical depth, potentially reflecting the evolutionary progression from embedded disks to more evolved Class~II systems. If such an evolutionary trend exists, one would naturally expect Class~I disks, representing an intermediate stage between Class~0 and Class~II, to exhibit fewer asymmetries than Class~0 disks, consistent with the pattern suggested by our results.

Notable exceptions among Class~I sources are L1489~IRS and Oph~IRS~43~VLA1, which both have a bright large-scale envelope. In L1489~IRS, the envelope and the warped disk structure affect the performance of the asymmetric profile. From its best-fit model, we find that the asymmetric structure is attributed to the intermediate warped disk, while the central disk does not exhibit apparent asymmetry. In Oph~IRS~43, after separating the Gaussian envelope component from the compact disk emission, the best-fit model for Oph~IRS~43 VLA1 is a PLCT profile without significant asymmetric structure. However the system is barely resolved with ALMA. Overall, these results suggest that Class~0 sources are more likely to favor an asymmetric profile than Class~I sources do. We interpret the asymmetric intensity profiles along the disk minor axes are as dust flaring and optically-thick 1.3-mm emission \citep{2023ApJ...951....8O,2024ApJ...964...24T}.

\subsection{When do gaps and rings form in the protostellar disk?}

In comparison with the protostellar sample from eDisk, Class II sources observed in programs such as DSHARP \citep{2018ApJ...869L..41A} and ODISEA \citep{2019MNRAS.482..698C} exhibit readily identifiable ring and spiral substructures in continuum emission, with only a few exceptions. The relative absence of such visually discernible features in the eDisk disks is noteworthy. This systematic difference between embedded protostars and more evolved Class II systems suggests that prominent substructures are either not yet well developed during the embedded (Class 0/I) phase or remain observationally {inconspicuous at that  same} 1.3 mm observing wavelength. It further implies that the formation and development of disk substructures may occur rapidly during the evolutionary transition from the Class 0/I to the Class II phase \citep[also see][]{2023ApJ...951....8O}. 

However, the apparent absence of prominent, sharp substructures in the continuum emission of Class 0/I protostars may not necessarily reflect their intrinsic absence. An alternative explanation is that the dust continuum emission at 225 GHz is at least partially optically thick and observations at longer wavelength are required. Under such conditions, underlying radial or azimuthal variations in surface density could be obscured. Indeed, several sources in the eDisk sample exhibit relatively high brightness temperatures \citep{2023ApJ...951....8O}, suggesting that the continuum emission in these systems may be optically thick and thus capable of masking substructures that would otherwise be detectable. {Disk inclination may also play a pivotal role in the detection of substructures. As shown by \cite{2026PASJ..tmp...39T}, spiral structures generated by gravitational instability become increasingly difficult to identify as the disk is viewed at higher inclinations.}

In our analysis of the eDisk sample, only two sources (L1489 IRS and Oph IRS 63) exhibit clear and well-defined gap and ring substructures, being consistent with the initial assessment based on visual inspections \citep{2023ApJ...951....8O}. For the remaining objects, the disk emission appears largely smooth at the spatial resolution and sensitivity of our observations, with no unambiguous evidence for resolved substructure. This low incidence is consistent with recent studies of young disks in Orion, Ophiuchus, Corona Australis, and Chamaeleon (e.g., \citealt{2020ApJ...902..141S, 2025PASJ...77..572S, 2025A&A...700A.235H}).

The two sources in our sample that show clear evidence for rings and gaps are also among the most evolved systems. Our sample does not include flat-spectrum sources, which may further restrict the range of evolutionary stages probed. The relatively small number of detected substructures may therefore reflect intrinsic diversity in disk properties and evolutionary state. Gaps and rings may not be ubiquitous during the earliest phases of disk evolution and may instead develop on later timescales. Moreover, the use of SED-based class as a proxy for evolutionary stage can complicate this interpretation. Several of the protostellar disk sources reported to host rings and gaps in \citealt{2020ApJ...902..141S, 2025PASJ...77..572S, 2025A&A...700A.235H} are detected at optical wavelengths \citep[e.g.,][]{2023JApA...44...92N}, suggesting that these systems are more evolved, have largely dispersed their envelopes, and are more closely analogous to Class~II disks. 

\subsection{Disk truncation}

The sharpness of a disk’s outer boundary is a key diagnostic of its evolutionary stage and the dominant physical processes shaping its structure. In an idealized picture, the disk ``edge" is set by the centrifugal barrier, where the angular momentum of infalling material balances gravity. In practice, however, this boundary is continually blurred by turbulent accretion streamers, infall shocks, and magnetic braking effects \citep[e.g.,][]{2014Natur.507...78S,2017MNRAS.467L..76S,2022MNRAS.517..213J}. As a result, in the early embedded phases (Class 0 and I), the transition between the disk and the surrounding envelope is  not well defined although there is recent development identifying a transition zone from an envelope to a disk \citep{2026ApJ..1001..166D}.

In contrast, by the Class II stage, the outer disk boundary becomes much more distinct, particularly in the dust continuum. Observations at (sub-)mm wavelengths show that mm-sized dust grains in Class II disks often exhibit a very sharp outer truncation \citep{2012ApJ...744..162A,2013A&A...557A.133D,2019ApJ...882...49L}. This pronounced edge is commonly attributed to radial drift \citep{2014ApJ...780..153B,2015ApJ...808..102K,2024ARA&A..62..157B}: as grains grow to millimeter sizes, they decouple partially from the gas. Because the gas is pressure-supported and orbits slightly sub-Keplerian, dust grains experience a headwind that removes angular momentum, causing them to spiral inward \citep{1977MNRAS.180...57W}. This process efficiently depletes the outer disk of large grains, producing a sharp cutoff in the (sub-)mm emission. 

{However, recent deep observations of Class II sources \citep{2025ApJ...984L...9C,2026A&A...710A.238D} have revealed extended, low-surface-brightness continuum emission, often termed  ``halos", which stands in tension with predictions from classical radial drift models. These halos harbor a substantial fraction of the total dust mass, highlighting their importance in mitigating the disk mass-budget problem. While ongoing, late-stage envelope infall can sustain such structures in some systems, the detection of millimeter- to centimeter-sized grains in others suggests that unresolved dust traps also play a critical role in retaining outer dust reservoirs \citep{2026A&A...710A.238D}.  }

\begin{figure*}
 \centering
 \includegraphics[width=0.5\linewidth]{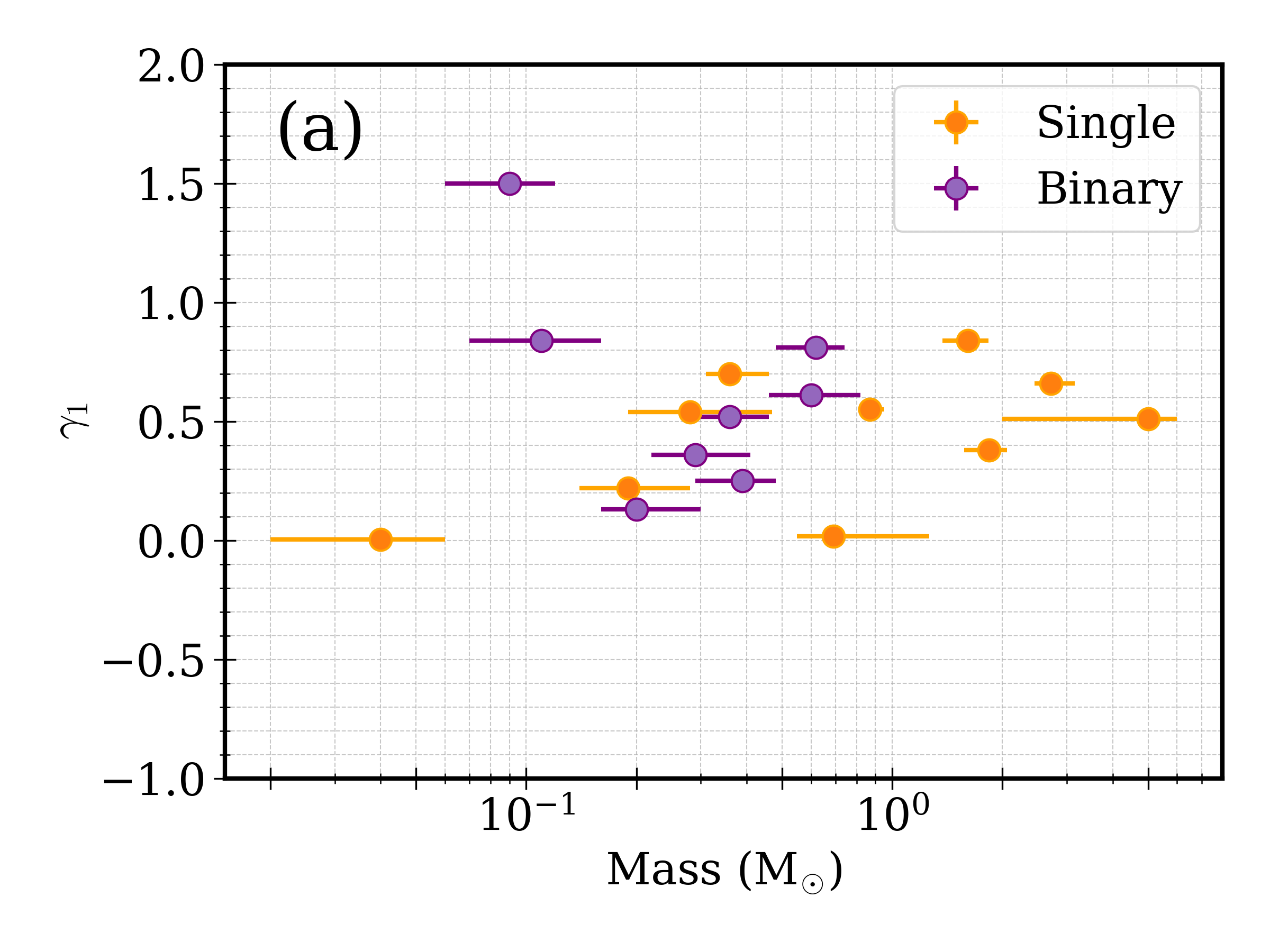}\includegraphics[width=0.5\linewidth]{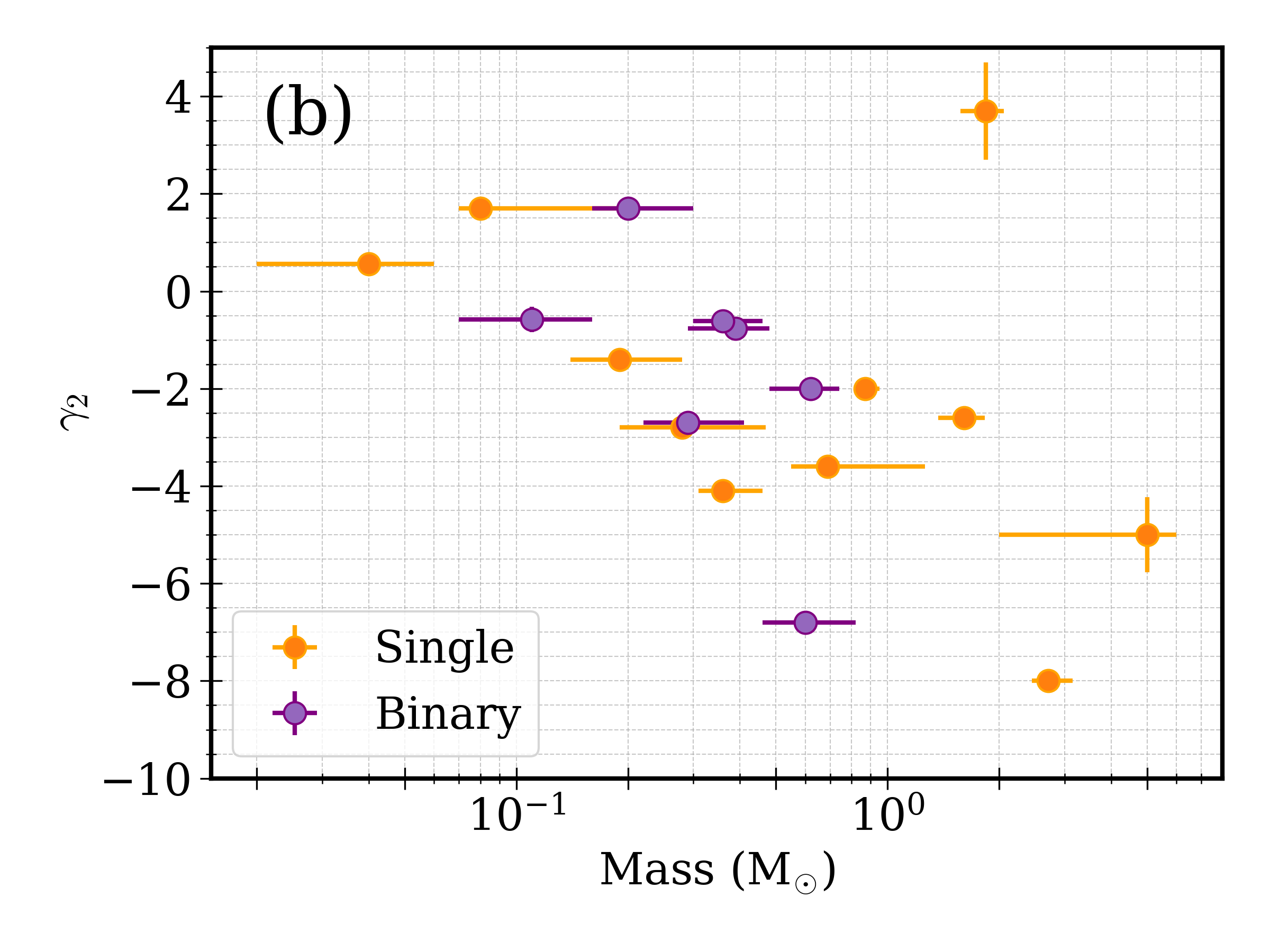}
 \caption{Parameters $\gamma_1$ and $\gamma_2$ as a function of protostellar mass. We also separate the single and binary systems in our sample as orange and purple solid circles. }
 \label{fig:gamma}
\end{figure*}

In our models, the disk sharpness is primarily controlled by the parameter $\gamma_2$, which governs the exponential taper in smooth disk surface density profiles, while $\gamma_1$ describes the inner disk structure. In Figure~\ref{fig:gamma}, we show $\gamma_1$ and $\gamma_2$ as a function of protostellar mass, distinguishing between single and binary systems. We find that the $\gamma_2$ values derived in this work are systematically lower than those reported in \cite{2019A&A...628A..95M} and \cite{2019ApJ...882...49L}, where values span $\sim 0.3$ to 50, while $\gamma_1$ remains broadly consistent across studies. This suggests that protostellar (Class 0/I) disks are, on average, less sharply truncated than Class II disks.

Previous work on Class II disks has also shown that $\gamma_2$ tends to be smaller in single-star systems compared to binaries \citep{2019A&A...628A..95M}, whereas $\gamma_1$ shows little systematic variation. Consistent with this trend, Figure~\ref{fig:gamma} indicates that $\gamma_1$ is similar for both single and binary systems, while $\gamma_2$ shows a marginal tendency to be lower for single stars at a given mass. However, the limited size of our current sample prevents any statistically robust conclusions regarding these trends.

\section{Summary and Conclusions}

In this work, we present a systematic analysis of protostellar disks in the eDisk sample carried out directly in the visibility plane. We employ a hierarchical modeling approach, progressing from simple Gaussian profiles to more flexible power-law core with exponential tail (PLCT) models, including asymmetric and higher-complexity extensions. We show that simple Gaussian models generally fail to reproduce both the centrally peaked emission and the extended outer structure observed in most protostellar disks. In contrast, the PLCT framework provides a substantially improved description of the radial brightness distribution. The inclusion of azimuthal asymmetries further reduces residuals in several sources, indicating that departures from axisymmetry are common even at early evolutionary stages.

Despite these modeling advances, only two disks (L1489 IRS and 
Oph IRS63) in our sample exhibit clear and well-defined gap and ring substructures. The majority of sources appear smooth at the spatial resolution and sensitivity of our observations, in contrast to studies of young disks in nearby star-forming regions that report a higher incidence of substructures. The disks that do show rings and gaps are among the most evolved systems in our sample, and the absence of flat-spectrum sources limits the range of evolutionary stages probed. This suggests that prominent gaps and rings are not ubiquitous during the earliest phases of disk evolution, but instead emerge over a range of timescales and become observationally apparent only  towards the end of the protostellar phase.

Using a uniform definition of disk radii based on the radius enclosing 95\% of the total continuum flux, we identify a positive correlation between disk radii and stellar masses, consistent with previous observational studies. The disk radii derived from our best-fit models follow a power-law relation of the form
$R_{\rm disk} \propto M_{\star}^{1.5 \pm 0.1}$. In addition, we find that for a given stellar mass, disks associated with binary systems are systematically smaller than those surrounding isolated protostars. This trend suggests that multiplicity plays a significant role in regulating disk size, likely through dynamical truncation or angular momentum redistribution during the protostellar phase.

Finally, the physical origin of brightness asymmetries remains ambiguous, potentially reflecting a combination of genuine substructures and radiative transfer effects in optically thick disks. The analytical disk profiles derived here provide robust and flexible inputs for future radiative transfer modeling. Together, these results place new constraints on the emergence of disk substructures in the protostellar phase and offer a unified framework for comparing disk sizes, truncation behavior, and morphology across early stages of disk evolution. 

In the future, we will focus on detailed radiative transfer modeling to determine whether the observed brightness asymmetries arise from genuine disk substructures or from radiative transfer and opacity effects in optically thick emission. Several disks in our sample may be optically thick at 1.33~mm, motivating comparisons with multi-wavelength continuum observations, particularly at 3~mm where the emission is expected to be more optically thin. Differences in morphology and contrast between wavelengths will provide direct constraints on dust opacity and radiative transfer effects. In addition, we will explore super-resolution imaging techniques, including the Sparse Modeling Imaging, to probe substructures below the nominal angular resolution and assess the robustness of inferred asymmetries. Together, these efforts will enable a more definitive interpretation of disk morphology during the protostellar phase.

\section{Data availability}

The calibrated measurement sets and FITS files for both the continuum and spectral line are available at the eDisk archive  and can also be available on request. 

\section{Acknowledgment}

N.O. and M.N. acknowledge support from the National Science and Technology Council (NSTC) in Taiwan through the grant NSTC 113-2112-M-001-037 and 114-2112- M-001-019, and the Academia Sinica Investigator Project Grant (AS-IV-114-M02). J.X., A.K., and L.W.L. acknowledge support from NSF AST-2108794. P.D.S acknowledges support from NSF AST-2305482. J.P.W. acknowledges support from the NSF AST-2107841. J.J.T. acknowledges support from NASA XRP 80NSSC22K1159. N.T.P acknowledges financial support from the Vietnam National Foundation for Science and Technology Development (NAFOSTED) under grant number 103.99-2024.36. S.T. is supported by JSPS KAKENHI grant Nos. JP21H04495 and JP24K00674 and by NAOJ
ALMA Scientific Research grant No. 2022-20A. This paper makes use of the following ALMA data: ADS/JAO.ALMA\#2019.1.00261.L. ALMA is a partnership of ESO (representing its member states), NSF (USA) and NINS (Japan), together with NRC (Canada), MOST and ASIAA (Taiwan), and KASI (Republic of Korea), in cooperation with the Republic of Chile. The Joint ALMA Observatory is operated by ESO, AUI/NRAO and NAOJ. The National Radio Astronomy Observatory and Green Bank Observatory are facilities of the U.S. National Science Foundation operated under cooperative agreement by Associated Universities, Inc.

\appendix

\section{Visibilities}

\begin{figure*}
    \centering
    \includegraphics[width=0.5\linewidth]{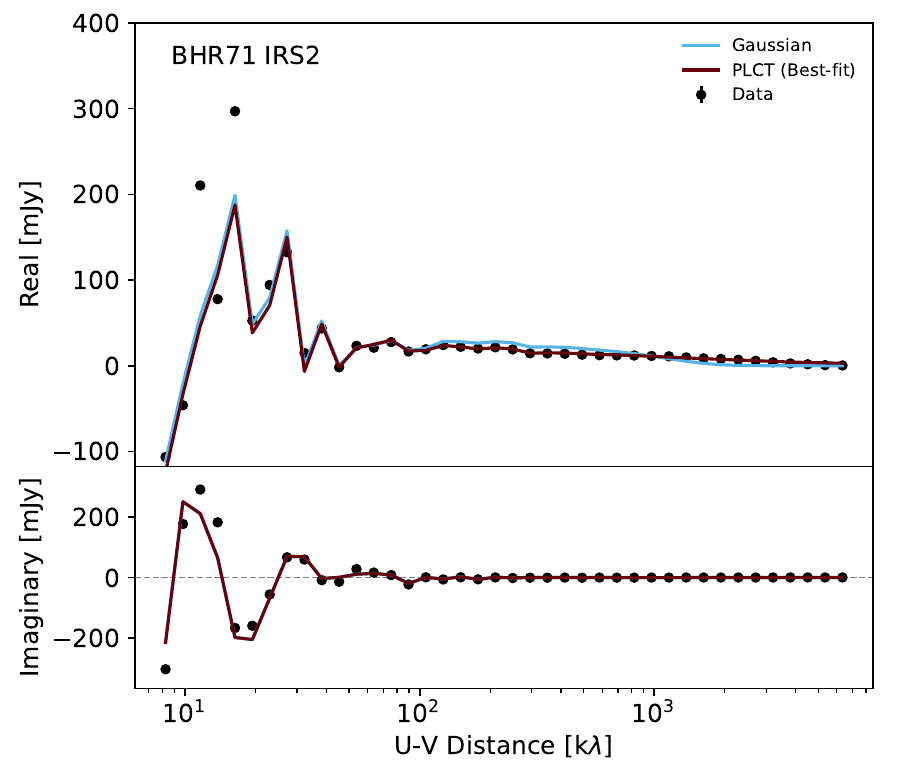}\includegraphics[width=0.5\linewidth]{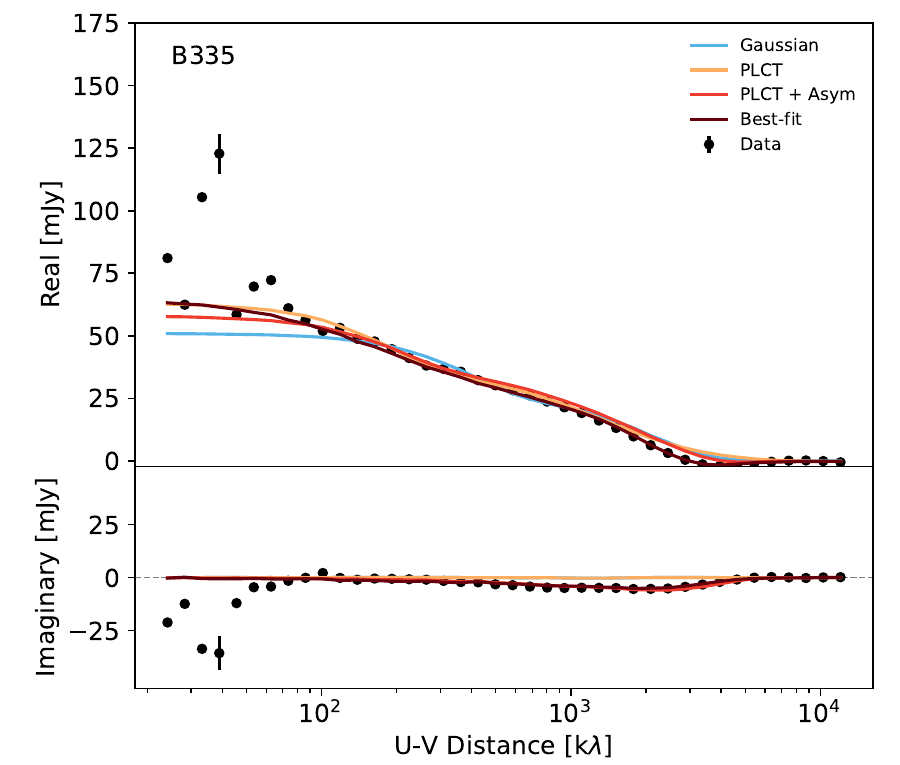} 

    \includegraphics[width=0.5\linewidth]{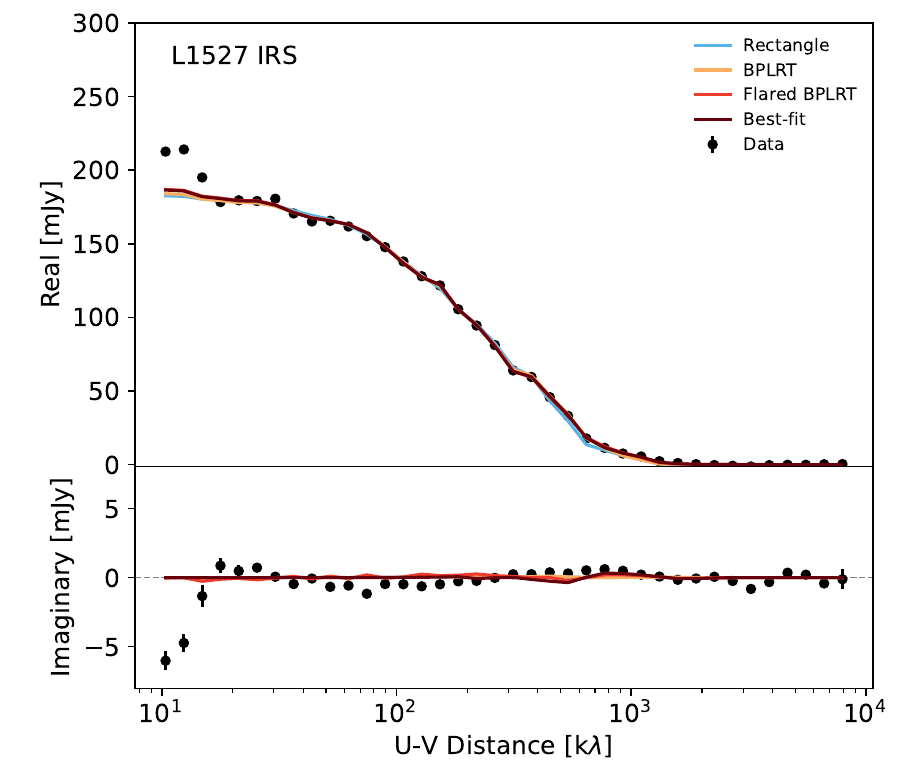}\includegraphics[width=0.5\linewidth]{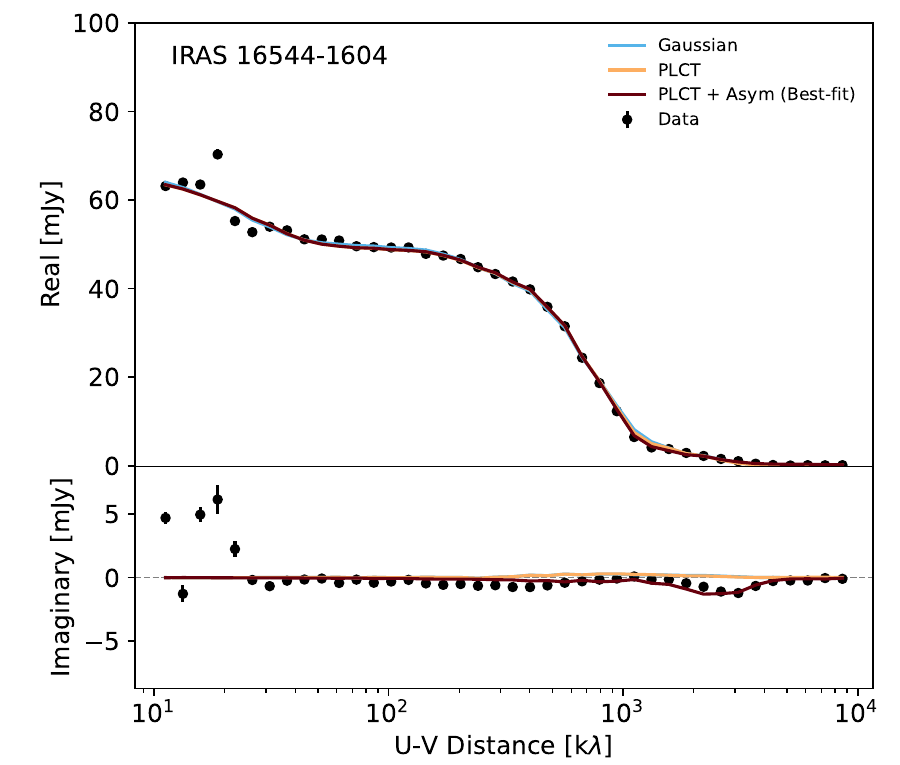} 

    \includegraphics[width=0.5\linewidth]{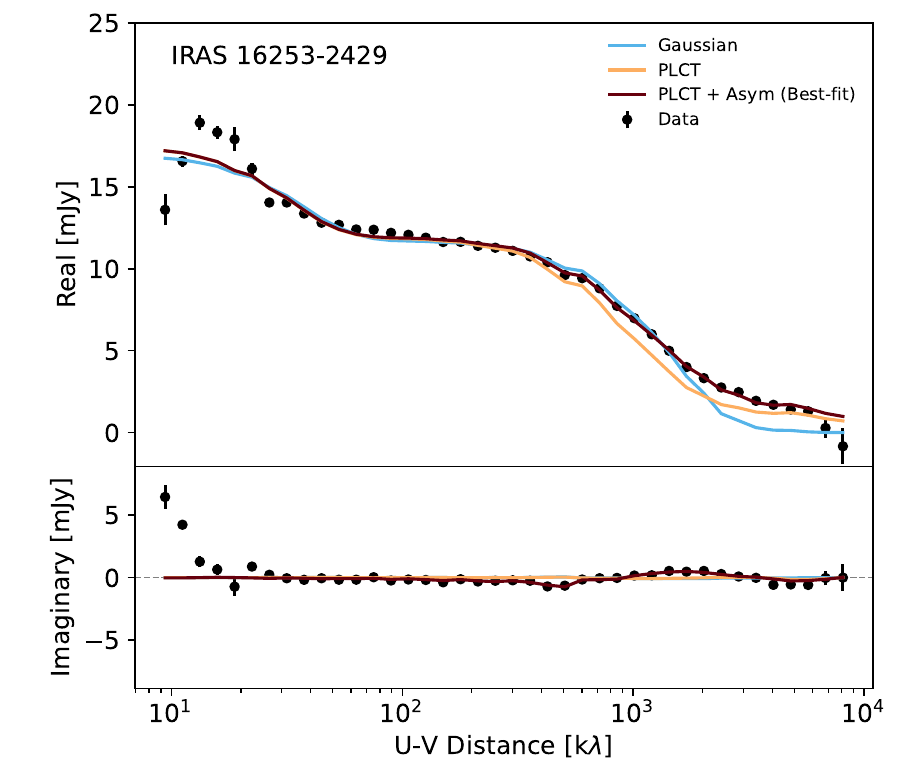}\includegraphics[width=0.5\linewidth]{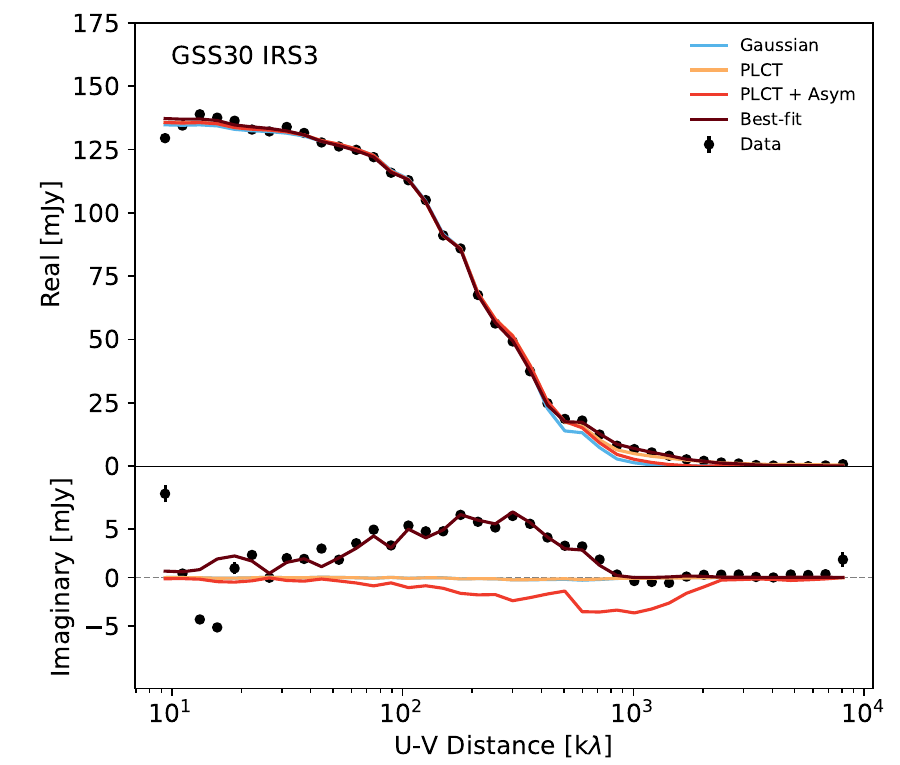} 
    
    \caption{Real and imaginary components of the visibilities as a function of baseline length for the observed data (black points) and the best-fitting profiles from the various models (colored curves). }
    \figsetstart
    \figsetnum{9}
    \figsettitle{225 GHz Visibility Profiles}
    \figsetgrpstart
    \figsetgrpnum{9}
    \figsetgrptitle{Class 0 and Class I Sources}
    \label{fig:vis}
\end{figure*}

Figure \ref{fig:vis} presents the real and imaginary components of the visibilities as a function of baseline length, together with the best-fitting profiles from the various models discussed in the main text.

\begin{figure*}
    \centering
    \includegraphics[width=0.5\linewidth]{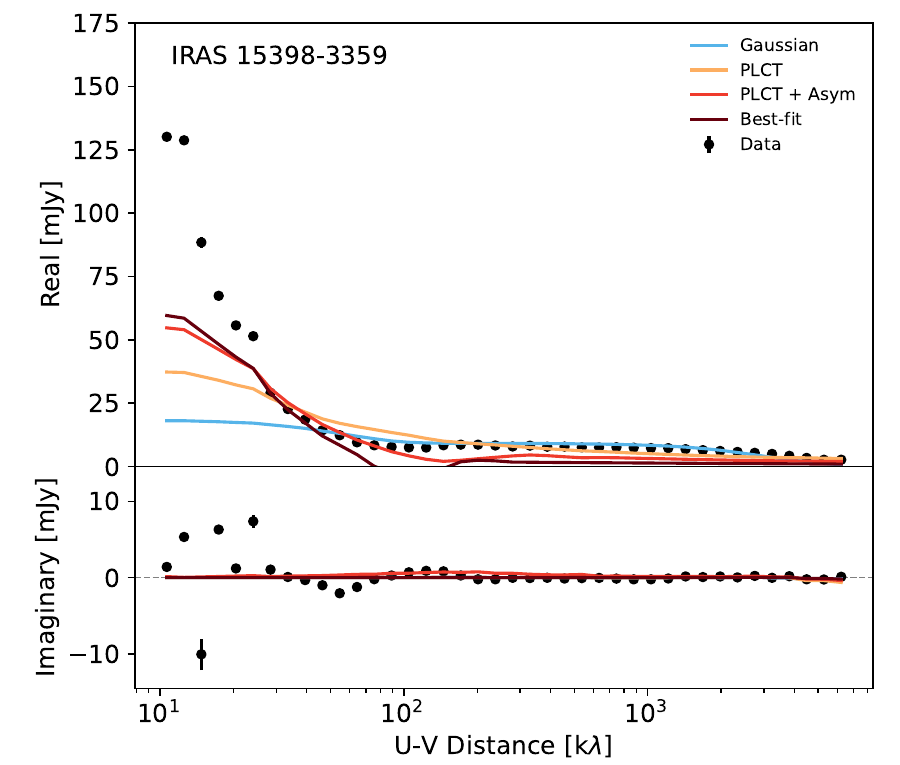}\includegraphics[width=0.5\linewidth]{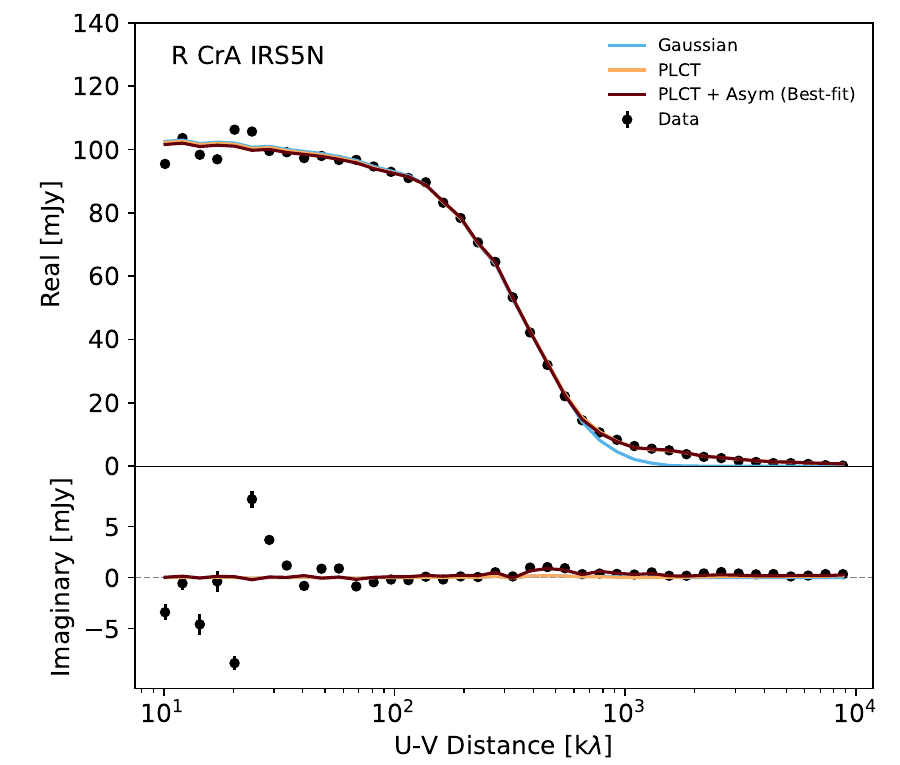} 

    \includegraphics[width=0.5\linewidth]{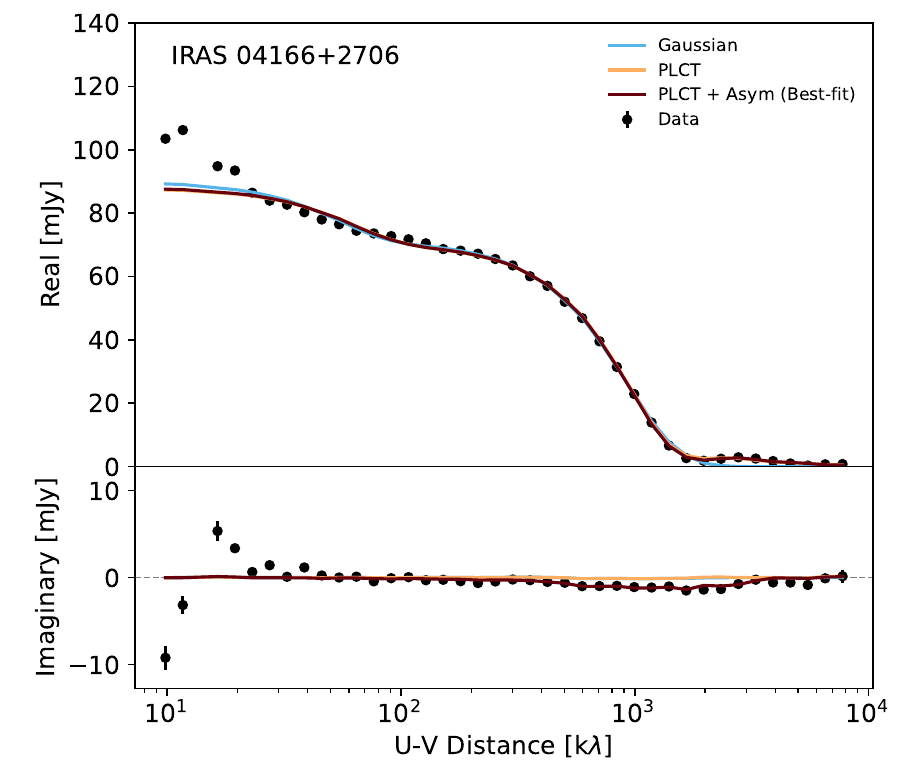}\includegraphics[width=0.5\linewidth]{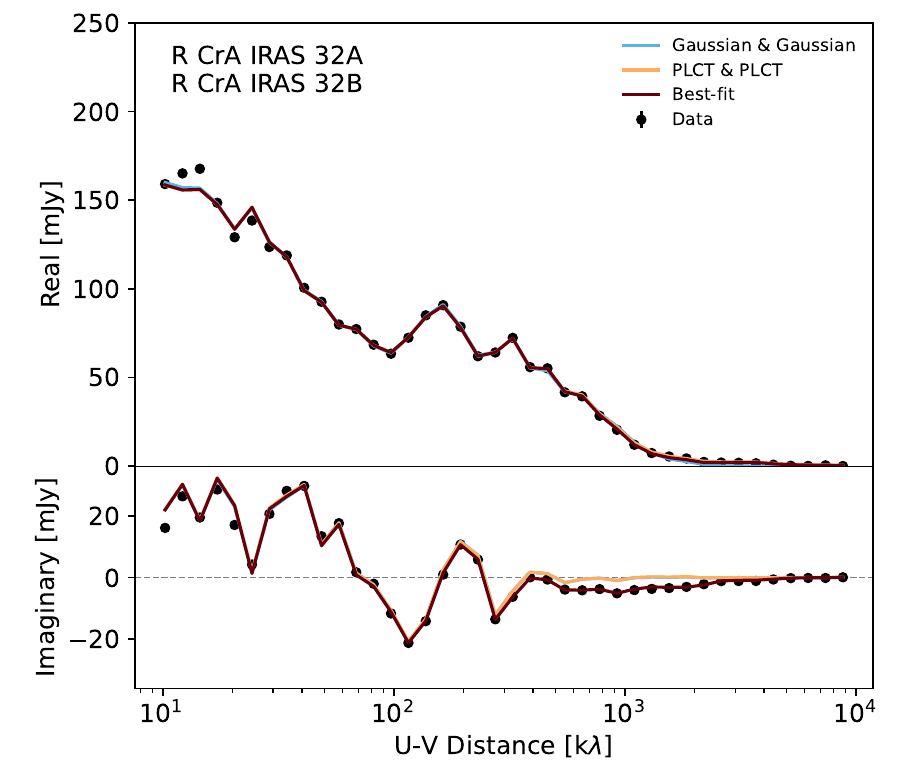} 

    \includegraphics[width=0.5\linewidth]{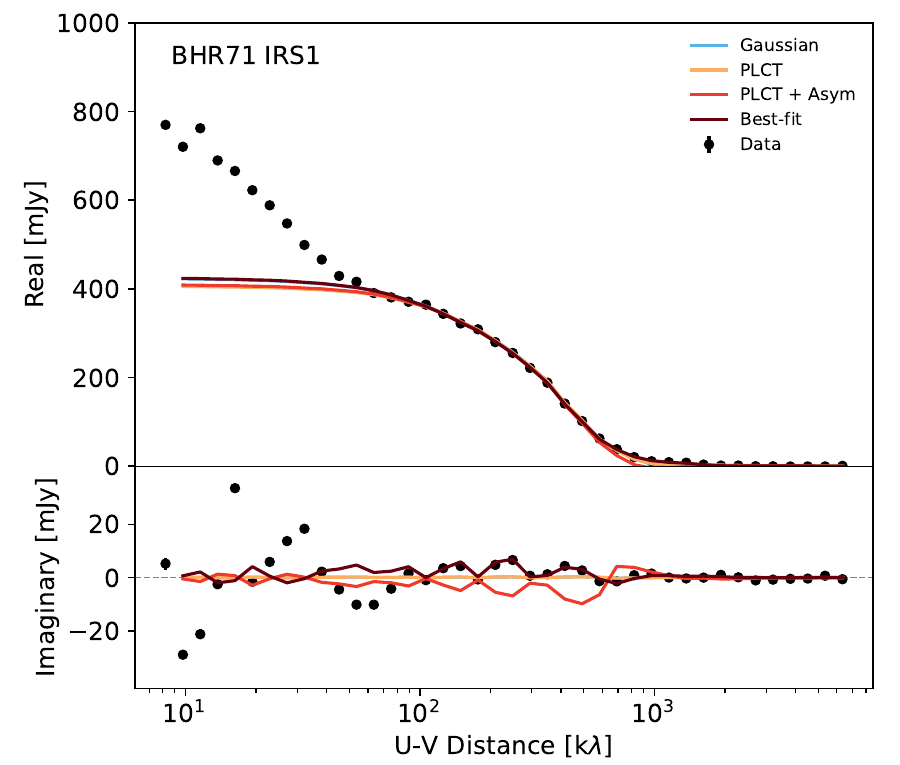}\includegraphics[width=0.5\linewidth]{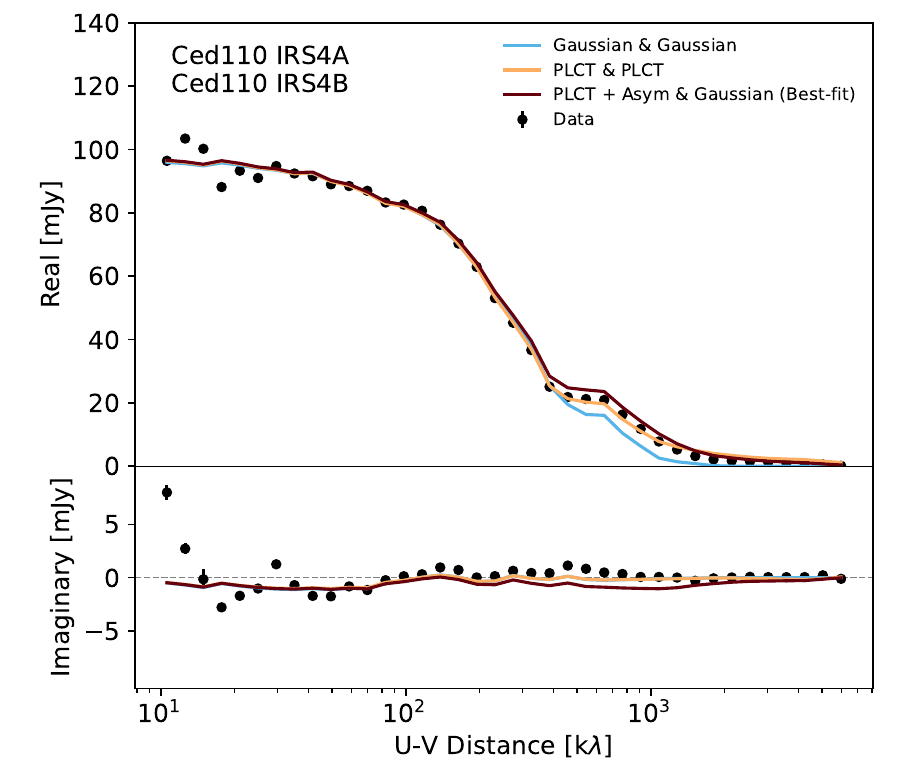} 

\end{figure*}

\begin{figure*}
    \centering
    \includegraphics[width=0.5\linewidth]{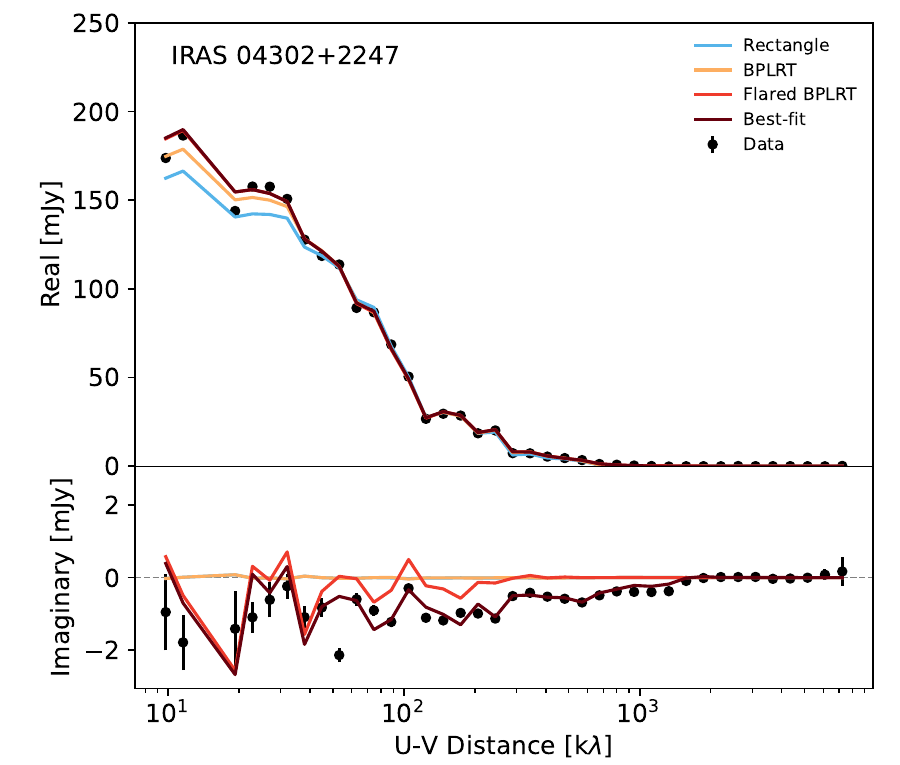}\includegraphics[width=0.5\linewidth]{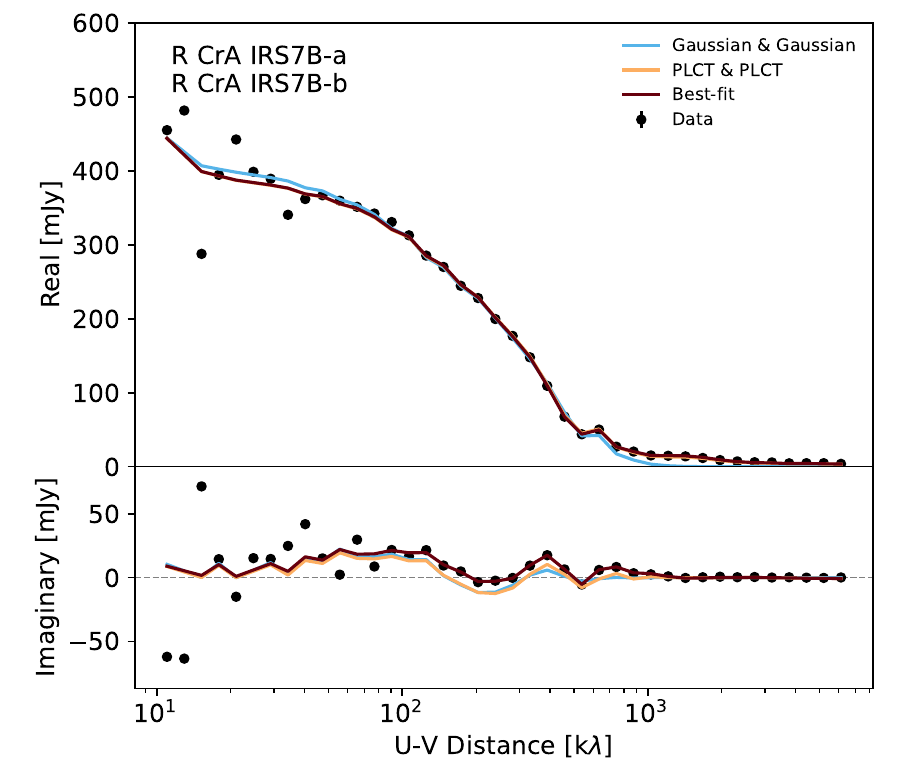} 

    \includegraphics[width=0.5\linewidth]{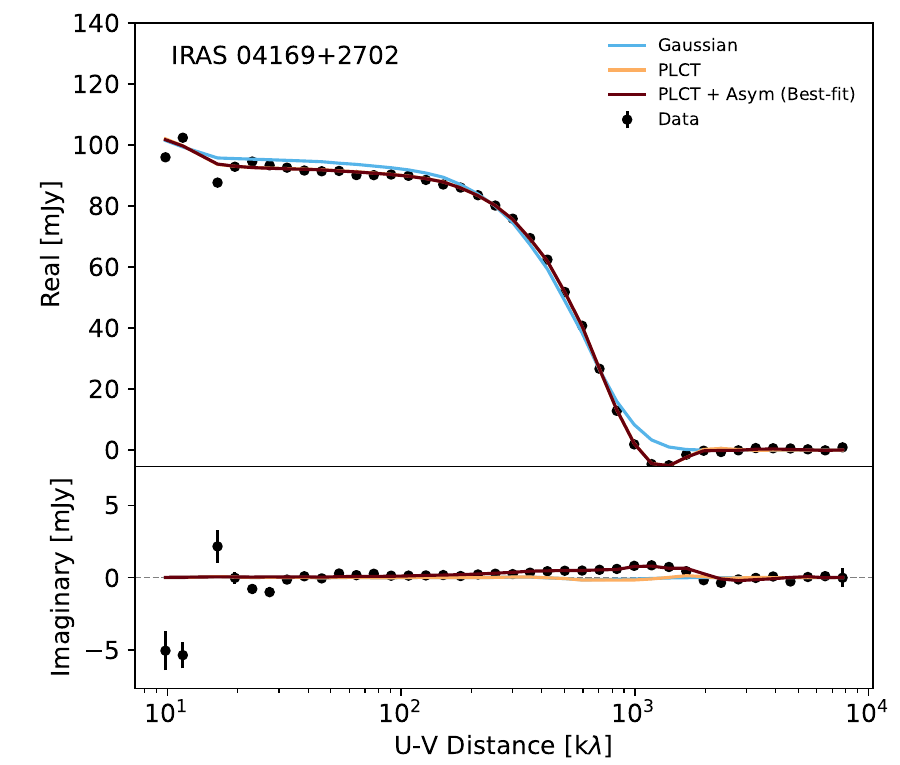}\includegraphics[width=0.5\linewidth]{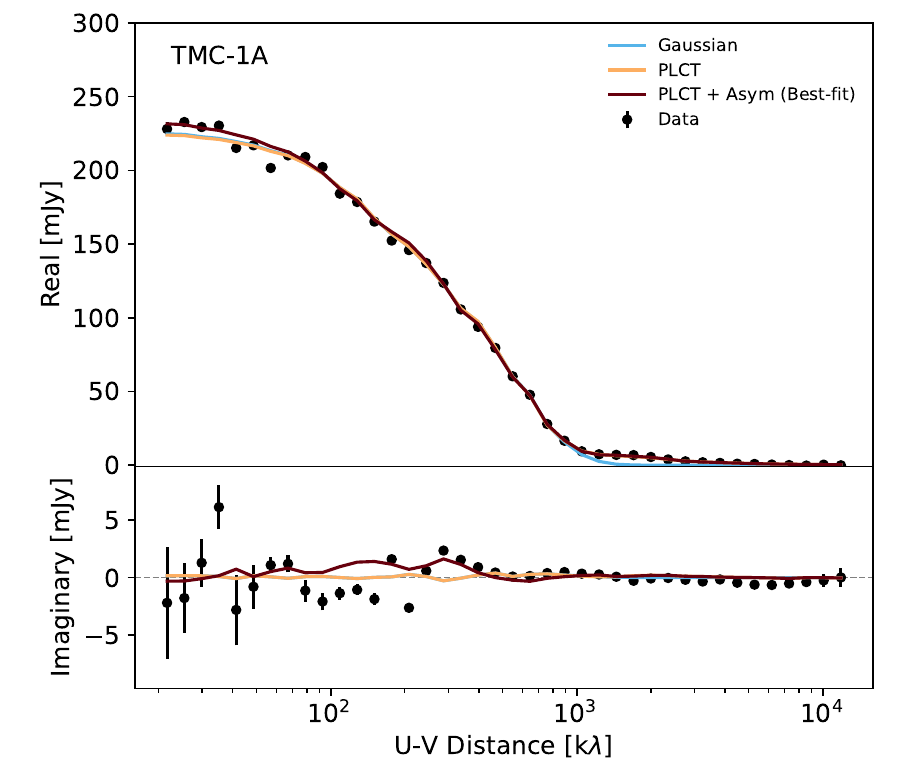} 

    \includegraphics[width=0.5\linewidth]{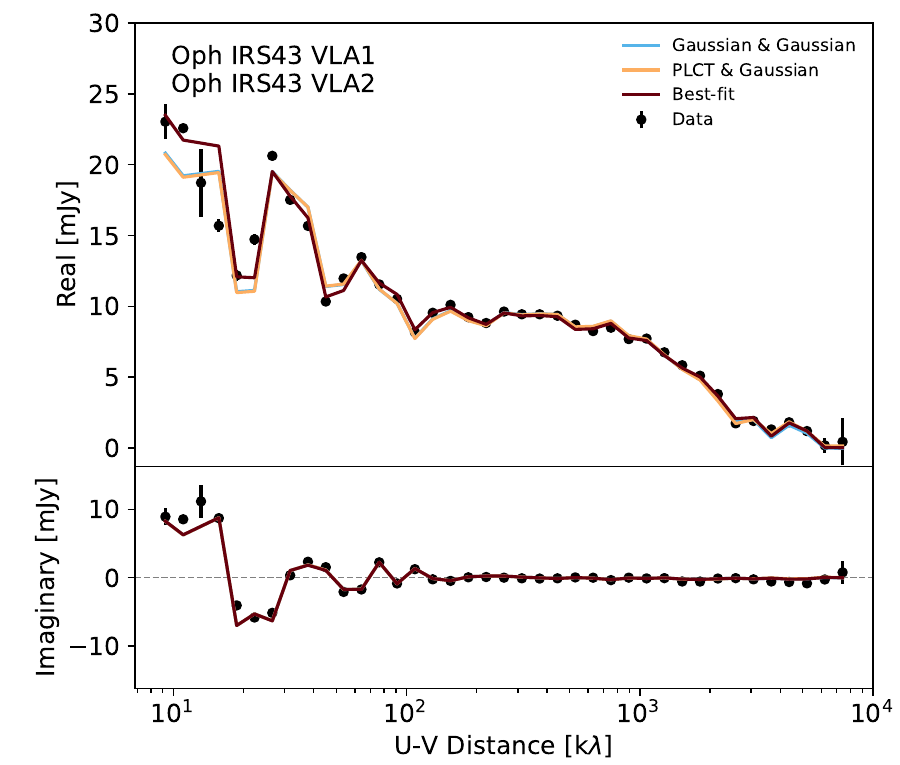}\includegraphics[width=0.5\linewidth]{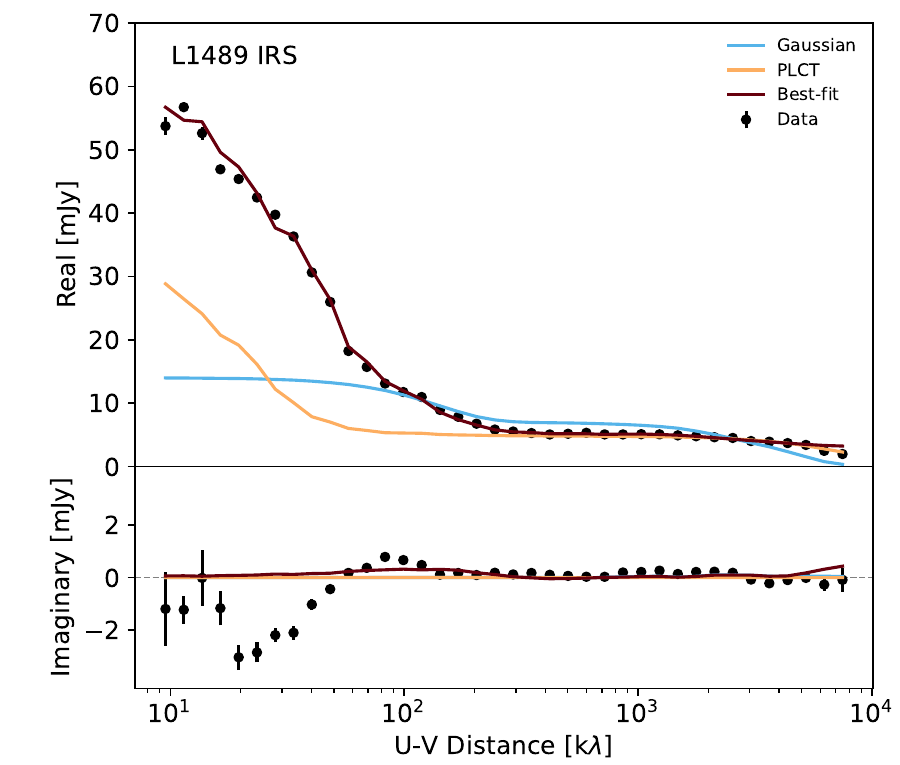}

\end{figure*}

\begin{figure*}
    \centering
    \includegraphics[width=0.5\linewidth]{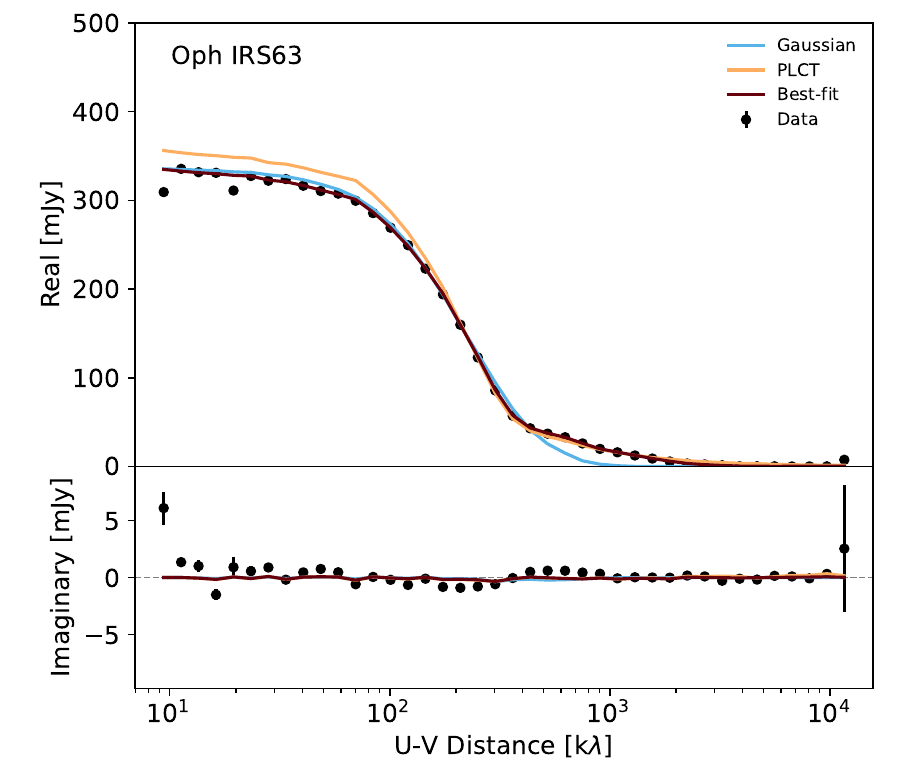}
\end{figure*}

\begin{figure*}

\figsetplot{BHR71_IRS2_225GHz_visibilities.pdf}

\figsetplot{B335_225GHz_visibilities.pdf}

\figsetplot{L1527IRS_225GHz_visibilities.pdf}

\figsetplot{CB68_225GHz_visibilities.pdf}

\figsetplot{IRAS16253_225GHz_visibilities.pdf}

\figsetplot{GSS30IRS3_225GHz_visibilities.pdf}

\figsetplot{IRAS15398_225GHz_visibilities.pdf}

\figsetplot{IRS5N_225GHz_visibilities.pdf}

\figsetplot{IRAS04166_225GHz_visibilities.pdf}

\figsetplot{IRAS32_225GHz_visibilities.pdf}

\figsetplot{BHR71_IRS1_225GHz_visibilities.pdf}

\figsetplot{Ced110IRS4_225GHz_visibilities.pdf}

\figsetplot{IRAS04302_225GHz_visibilities.pdf}

\figsetplot{IRS7B_225GHz_visibilities.pdf}

\figsetplot{IRAS04169_225GHz_visibilities.pdf}

\figsetplot{TMC1A_225GHz_visibilities.pdf}

\figsetplot{OphIRS43_225GHz_visibilities.pdf}

\figsetplot{L1489IRS_225GHz_visibilities.pdf}

\figsetplot{OphIRS63_225GHz_visibilities.pdf}

\figsetend
\end{figure*}

\begin{figure*}
    \centering
    \includegraphics[width=0.5\linewidth]{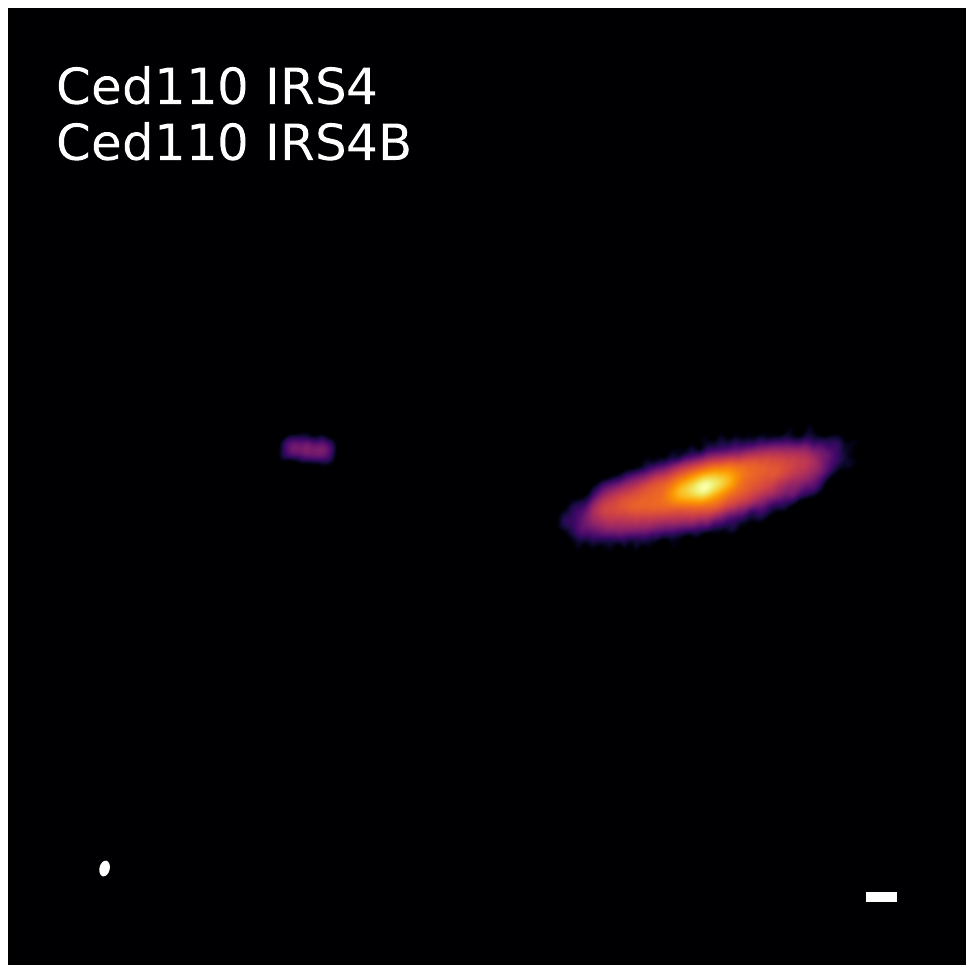}\includegraphics[width=0.5\linewidth]{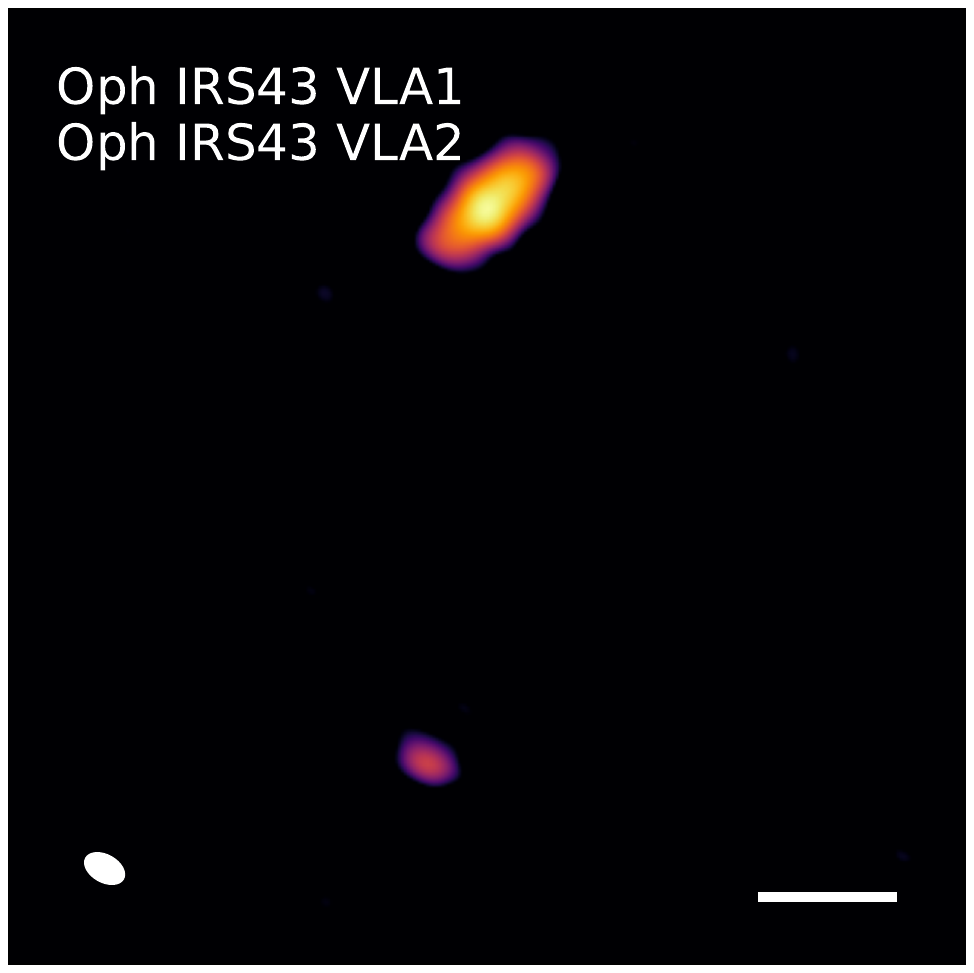} 
    \includegraphics[width=0.5\linewidth]{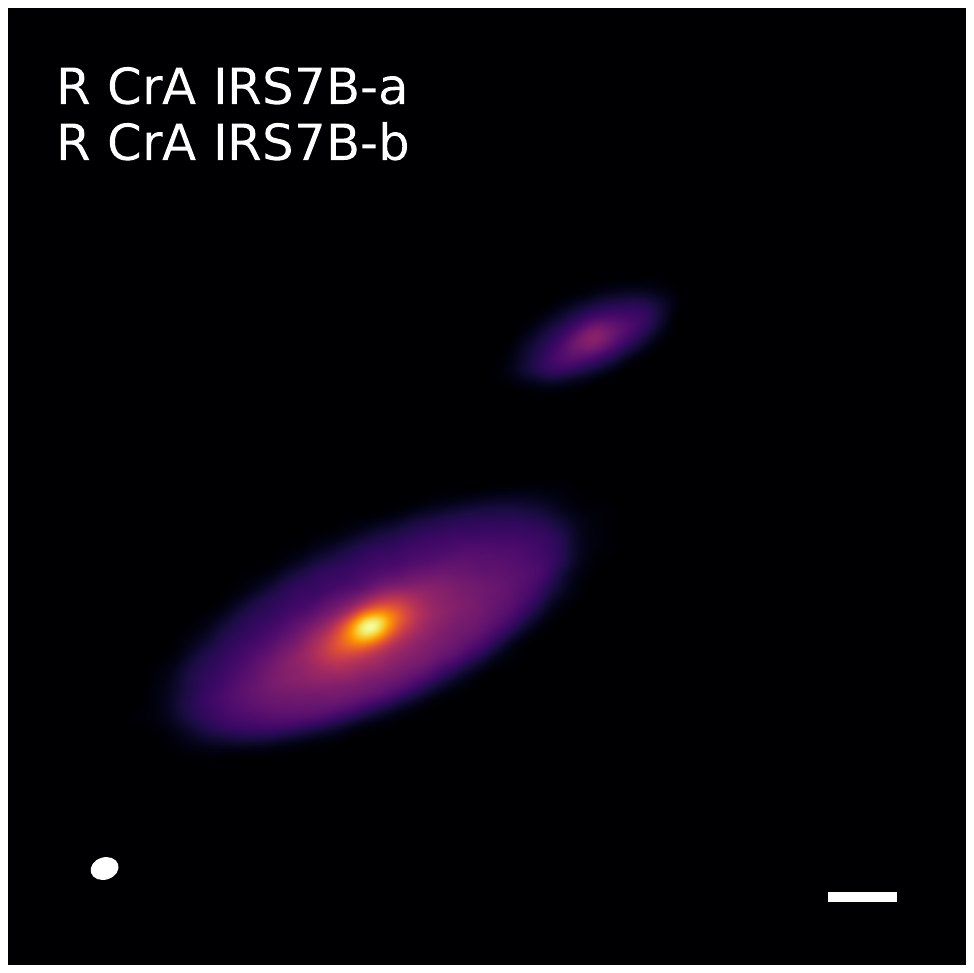}\includegraphics[width=0.5\linewidth]{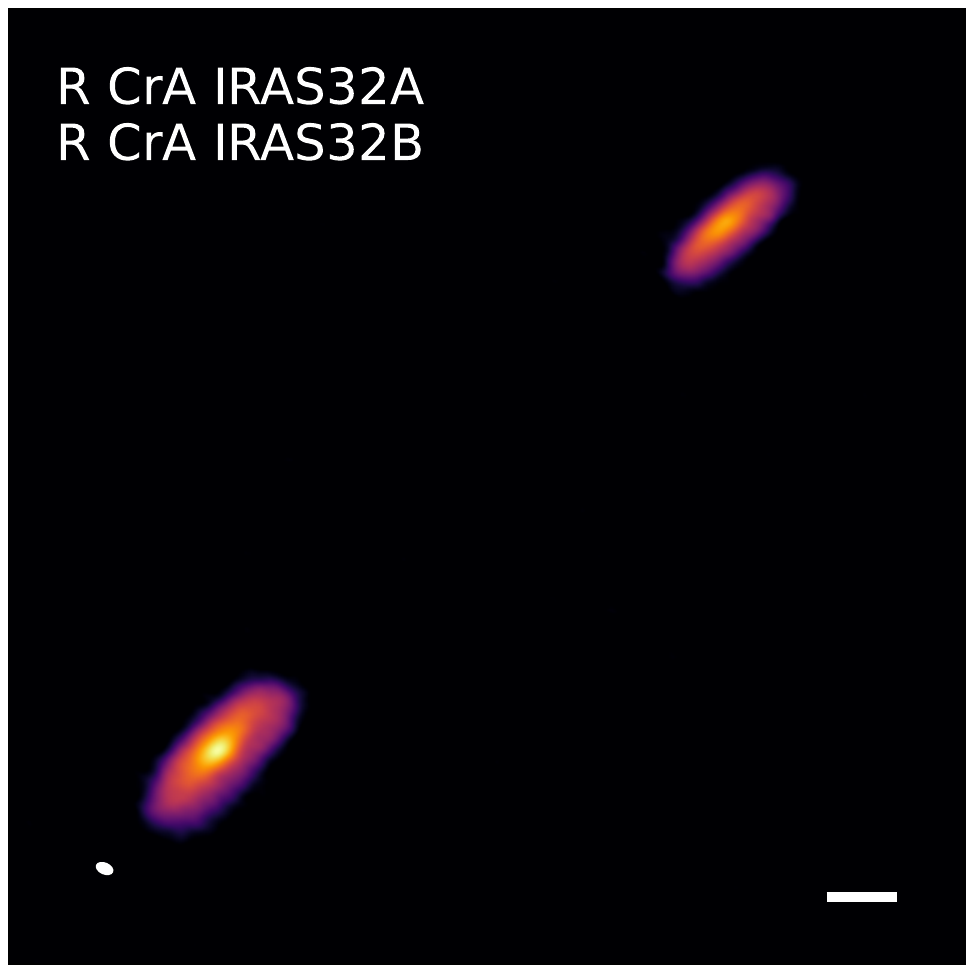}
    \caption{The best fit model image for the Binaries in our study. }
    \label{fig:Binary}
\end{figure*}

\begin{deluxetable*}{l|cccc}
\tablecaption{Best-Fit Parameters and Uncertainties for the Gaussian fit. \label{Table2}}
\tablehead{
\colhead{Source} & \colhead{$x_0$ (\arcsec)} & \colhead{$y_0$ (\arcsec)} & \colhead{$\log r_w$ (\arcsec)} & \colhead{$i$ (deg)} \\
& \colhead{$p.a$ (deg)} & \colhead{$\log(I_{\nu,o})$} & \colhead{$\log r_{En}$ (\arcsec)} & \colhead{$\log(I_{\nu,Eo})$}
}
\startdata
BHR71IRS2 & $-5.1\times10^{-1}\pm7.6\times10^{-5}$ & $-7.2\times10^{-1}\pm7.2\times10^{-5}$ & $-1.3\pm1.1\times10^{-3}$ & $3.6\times10^{1}\pm3.6\times10^{-1}$ \\
 & $1.8\times10^{2}\pm6.5\times10^{-1}$ & $-1.6\pm1.0\times10^{-6}$ & $1.2\times10^{-2}\pm3.1\times10^{-3}$ & $-1.6\pm2.6\times10^{-4}$ \\[0.5ex]
B335 & $5.4\times10^{-3}\pm3.1\times10^{-5}$ & $-8.2\times10^{-3}\pm3.1\times10^{-5}$ & $-1.6\pm8.4\times10^{-4}$ & $4.1\times10^{1}\pm2.1\times10^{-1}$ \\
 & $8.0\times10^{1}\pm2.6\times10^{-1}$ & $-1.6\pm9.6\times10^{-4}$ & $-9.4\times10^{-1}\pm2.6\times10^{-3}$ & $-1.6\pm1.7\times10^{-3}$ \\[0.5ex]
IRAS16253 & $2.1\times10^{-1}\pm2.0\times10^{-4}$ & $-9.3\times10^{-1}\pm1.4\times10^{-4}$ & $-1.3\pm2.7\times10^{-3}$ & $6.8\times10^{1}\pm3.5\times10^{-1}$ \\
 & $2.4\times10^{1}\pm3.5\times10^{-1}$ & $-1.9\pm8.6\times10^{-4}$ & $8.2\times10^{-2}\pm1.9\times10^{-2}$ & $-2.3\pm2.3\times10^{-2}$ \\[0.5ex]
IRAS 16544-1604 & $4.1\times10^{-2}\pm7.3\times10^{-5}$ & $-1.2\times10^{-1}\pm6.8\times10^{-5}$ & $-1.1\pm4.8\times10^{-4}$ & $7.4\times10^{1}\pm4.4\times10^{-2}$ \\
 & $1.3\times10^{2}\pm4.8\times10^{-2}$ & $-1.3\pm2.6\times10^{-4}$ & $2.7\times10^{-1}\pm1.0\times10^{-2}$ & $-1.8\pm1.4\times10^{-2}$ \\[0.5ex]
GSS30IRS3 & $-7.2\times10^{-2}\pm7.6\times10^{-5}$ & $-4.0\times10^{-1}\pm4.3\times10^{-5}$ & $-6.0\times10^{-1}\pm2.0\times10^{-4}$ & $7.3\times10^{1}\pm2.3\times10^{-2}$ \\
 & $1.9\times10^{1}\pm1.6\times10^{-2}$ & $-9.0\times10^{-1}\pm3.6\times10^{-4}$ & $-4.7\times10^{-1}\pm3.5\times10^{-3}$ & $-2.0\pm4.2\times10^{-3}$ \\[0.5ex]
IRAS15398 & $-9.5\times10^{-2}\pm9.9\times10^{-5}$ & $-1.6\times10^{-1}\pm8.9\times10^{-5}$ & $-1.8\pm4.9\times10^{-3}$ & $4.1\times10^{1}\pm1.3$ \\
 & $3.1\times10^{1}\pm2.0$ & $-2.0\pm9.2\times10^{-4}$ & $-1.0\times10^{-1}\pm3.0\times10^{-3}$ & $-2.0\pm9.2\times10^{-4}$ \\[0.5ex]
IRS5N & $2.4\times10^{-1}\pm9.2\times10^{-5}$ & $-6.9\times10^{-1}\pm4.8\times10^{-5}$ & $-7.9\times10^{-1}\pm3.2\times10^{-4}$ & $6.7\times10^{1}\pm4.1\times10^{-2}$ \\
 & $1.7\times10^{2}\pm3.3\times10^{-2}$ & $-1.0\pm8.1\times10^{-4}$ & $-7.9\times10^{-1}\pm6.8\times10^{-4}$ & $-2.2\pm1.2\times10^{-2}$ \\[0.5ex]
IRAS04166 & $7.8\times10^{-2}\pm3.1\times10^{-5}$ & $-1.8\times10^{-1}\pm3.2\times10^{-5}$ & $-1.2\pm3.0\times10^{-4}$ & $4.6\times10^{1}\pm6.8\times10^{-2}$ \\
 & $3.2\times10^{1}\pm9.9\times10^{-2}$ & $-1.1\pm1.8\times10^{-4}$ & $-7.7\times10^{-2}\pm5.9\times10^{-3}$ & $-1.7\pm6.9\times10^{-3}$ \\[0.5ex]
IRAS32 & $6.2\times10^{-1}\pm3.9\times10^{-5}$ & $-1.5\pm3.7\times10^{-5}$ & $-1.0\pm2.6\times10^{-4}$ & $6.8\times10^{1}\pm2.8\times10^{-2}$ \\
 & $4.6\times10^{1}\pm3.0\times10^{-2}$ & $-1.1\pm1.9\times10^{-4}$ & $-2.0\times10^{-1}\pm8.3\times10^{-3}$ & $-2.1\pm1.2\times10^{-2}$ \\[0.5ex]
IRAS32-b & $-3.3\times10^{-1}\pm5.5\times10^{-5}$ & $-4.9\times10^{-1}\pm5.0\times10^{-5}$ & $-1.1\pm4.2\times10^{-4}$ & $7.1\times10^{1}\pm4.8\times10^{-2}$ \\
 & $4.1\times10^{1}\pm4.5\times10^{-2}$ & $-1.4\pm2.7\times10^{-4}$ & $2.8\times10^{-1}\pm5.8\times10^{-3}$ & $-1.4\pm7.6\times10^{-3}$ \\[0.5ex]
BHR71IRS1 & $-2.1\pm2.4\times10^{-5}$ & $-1.5\times10^{-1}\pm2.4\times10^{-5}$ & $-9.4\times10^{-1}\pm1.4\times10^{-4}$ & $4.0\times10^{1}\pm3.1\times10^{-2}$ \\
 & $9.7\pm4.5\times10^{-2}$ & $-4.5\times10^{-1}\pm1.8\times10^{-4}$ & $-3.8\times10^{-1}\pm2.9\times10^{-3}$ & $-1.1\pm5.9\times10^{-3}$ \\[0.5ex]
Ced110IRS4 & $-2.3\times10^{-1}\pm7.9\times10^{-5}$ & $-6.9\times10^{-1}\pm4.3\times10^{-5}$ & $-7.6\times10^{-1}\pm3.3\times10^{-4}$ & $7.6\times10^{1}\pm2.2\times10^{-2}$ \\
 & $1.4\times10^{1}\pm2.1\times10^{-2}$ & $-1.1\pm4.4\times10^{-4}$ & $-4.9\times10^{-1}\pm2.7\times10^{-3}$ & $-1.9\pm3.8\times10^{-3}$ \\[0.5ex]
Ced110IRS4-b & $1.1\pm9.2\times10^{-4}$ & $-5.5\times10^{-1}\pm6.5\times10^{-4}$ & $-1.3\pm2.2\times10^{-3}$ & $7.8\times10^{1}\pm1.7$ \\
 & $1.8\times10^{2}\pm9.6\times10^{-1}$ & $-2.8\pm3.7\times10^{-3}$ & $1.6\pm7.3\times10^{-1}$ & $-2.9\pm5.6\times10^{-2}$ \\[0.5ex]
IRS7B & $-3.4\times10^{-3}\pm4.1\times10^{-5}$ & $-2.6\times10^{-1}\pm2.7\times10^{-5}$ & $-6.9\times10^{-1}\pm1.2\times10^{-4}$ & $6.9\times10^{1}\pm1.1\times10^{-2}$ \\
 & $2.5\times10^{1}\pm1.2\times10^{-2}$ & $-4.3\times10^{-1}\pm8.7\times10^{-5}$ & $7.7\times10^{-1}\pm1.3\times10^{-3}$ & $-4.4\times10^{-1}\pm7.3\times10^{-4}$ \\[0.5ex]
IRS7B-b & $-4.1\times10^{-1}\pm2.0\times10^{-4}$ & $2.9\times10^{-1}\pm1.4\times10^{-4}$ & $-8.2\times10^{-1}\pm1.0\times10^{-6}$ & $6.6\times10^{1}\pm2.9\times10^{-4}$ \\
 & $2.4\times10^{1}\pm1.4\times10^{-2}$ & $-1.4\pm6.0\times10^{-6}$ & $-8.2\times10^{-1}\pm1.4\times10^{-5}$ & $-2.1\pm2.9\times10^{-3}$ \\[0.5ex]
IRAS04169 & $3.6\times10^{-1}\pm3.4\times10^{-5}$ & $-2.8\times10^{-1}\pm4.1\times10^{-5}$ & $-1.1\pm2.3\times10^{-4}$ & $4.3\times10^{1}\pm5.2\times10^{-2}$ \\
 & $5.0\times10^{1}\pm8.3\times10^{-2}$ & $-1.0\pm1.3\times10^{-4}$ & $8.1\times10^{-1}\pm5.5\times10^{-2}$ & $-1.4\pm2.2\times10^{-1}$ \\[0.5ex]
TMC1A & $-3.9\times10^{-3}\pm8.2\times10^{-5}$ & $-7.0\times10^{-3}\pm6.1\times10^{-5}$ & $-9.7\times10^{-1}\pm4.8\times10^{-4}$ & $5.4\times10^{1}\pm6.0\times10^{-2}$ \\
 & $1.7\times10^{2}\pm7.4\times10^{-2}$ & $-7.8\times10^{-1}\pm6.3\times10^{-4}$ & $-4.6\times10^{-1}\pm3.9\times10^{-3}$ & $-1.2\pm3.9\times10^{-3}$ \\[0.5ex]
OphIRS43 & $2.8\pm1.2\times10^{-4}$ & $1.5\pm1.1\times10^{-4}$ & $-1.4\pm5.8\times10^{-3}$ & $7.8\times10^{1}\pm3.5\times10^{-1}$ \\
 & $4.7\times10^{1}\pm2.8\times10^{-1}$ & $-2.0\pm7.3\times10^{-4}$ & $-4.9\times10^{-2}\pm1.0\times10^{-2}$ & $-2.2\pm1.3\times10^{-2}$ \\[0.5ex]
OphIRS43 VLA2 & $2.9\pm6.8\times10^{-4}$ & $9.0\times10^{-1}\pm6.2\times10^{-4}$ & $-2.0\pm5.3\times10^{-2}$ & $7.4\times10^{1}\pm1.4\times10^{1}$ \\
 & $1.0\times10^{2}\pm1.0\times10^{1}$ & $-3.0\pm5.5\times10^{-3}$ & $3.0\times10^{-1}\pm3.4\times10^{-2}$ & $-3.0\pm1.5\times10^{-2}$ \\[0.5ex]
L1489IRS & $3.1\pm7.7\times10^{-5}$ & $-1.8\times10^{-1}\pm1.1\times10^{-4}$ & $-1.8\pm5.6\times10^{-3}$ & $4.9\times10^{1}\pm1.2$ \\
 & $8.6\times10^{1}\pm1.1$ & $-2.2\pm1.1\times10^{-3}$ & $-4.9\times10^{-1}\pm3.1\times10^{-3}$ & $-2.2\pm1.1\times10^{-3}$ \\[0.5ex]
OphIRS63 & $1.9\times10^{-3}\pm3.5\times10^{-5}$ & $-2.9\times10^{-3}\pm3.6\times10^{-5}$ & $-7.9\times10^{-1}\pm2.2\times10^{-4}$ & $5.9\times10^{1}\pm2.9\times10^{-2}$ \\
 & $6.0\times10^{1}\pm3.3\times10^{-2}$ & $-7.7\times10^{-1}\pm8.2\times10^{-5}$ & $-5.7\times10^{-1}\pm2.1\times10^{-4}$ & $-7.7\times10^{-1}\pm8.2\times10^{-5}$ \\
\enddata
\end{deluxetable*}

\begin{rotatetable*}
\begin{deluxetable*}{|l|ccccc}
\tabletypesize{\tiny}
\tablecaption{Best-Fit Parameters and Uncertainties for the PLCT model \label{tab:PLCT}}
\tablewidth{0pt}
\tablehead{
\colhead{Source} & \colhead{$x_0$(\arcsec)} & \colhead{$y_0$ (\arcsec)} & \colhead{$\log r_w$(\arcsec)} & \colhead{$i$ (deg)} & \colhead{$p.a.$ (deg)} \\
\colhead{} & \colhead{$\log(I_{\nu,0})$} & \colhead{$\gamma_1$} & \colhead{$\gamma_2$} & \colhead{$\log r_{En}$(\arcsec)} & \colhead{$\log(I_{\nu,E0})$}
}
\startdata
BHR71IRS2 & $-5.2\times10^{-1}\pm5.0\times10^{-5}$ & $-7.2\times10^{-1}\pm5.3\times10^{-5}$ & $-3.9\pm1.0\times10^{-1}$ & $4.5\times10^{1}\pm5.4\times10^{-1}$ & $1.8\times10^{2}\pm5.9\times10^{-1}$ \\
~ & $-1.7\pm2.7\times10^{-3}$ & $1.3\times10^{-1}\pm4.8\times10^{-2}$ & $1.7\pm5.6\times10^{-3}$ & $-2.8\times10^{-1}\pm6.1\times10^{-3}$ & $-1.7\pm2.7\times10^{-3}$ \\ \hline
B335 & $5.7\times10^{-3}\pm2.9\times10^{-5}$ & $-8.3\times10^{-3}\pm3.0\times10^{-5}$ & $-6.0\pm2.1\times10^{-2}$ & $3.4\times10^{1}\pm2.1\times10^{-1}$ & $7.7\times10^{1}\pm3.8\times10^{-1}$ \\
~ & $-1.4\pm1.1\times10^{-3}$ & $-3.0\pm1.9\times10^{-2}$ & $1.7\pm8.5\times10^{-4}$ & $-6.1\times10^{-1}\pm3.8\times10^{-3}$ & $-1.6\pm3.3\times10^{-3}$ \\ \hline
IRAS16253 & $2.1\times10^{-1}\pm1.9\times10^{-4}$ & $-9.3\times10^{-1}\pm1.4\times10^{-4}$ & $-9.6\times10^{-1}\pm1.6\times10^{-2}$ & $6.9\times10^{1}\pm3.7\times10^{-1}$ & $2.4\times10^{1}\pm3.6\times10^{-1}$ \\
~ & $-1.9\pm9.8\times10^{-4}$ & $8.4\times10^{-1}\pm2.8\times10^{-2}$ & $-5.8\times10^{-1}\pm2.6\times10^{-1}$ & $1.3\times10^{-1}\pm1.8\times10^{-2}$ & $-2.3\pm2.3\times10^{-2}$ \\ \hline
IRAS 16544-1604  & $4.1\times10^{-2}\pm7.1\times10^{-5}$ & $-1.2\times10^{-1}\pm6.5\times10^{-5}$ & $-8.1\times10^{-1}\pm1.9\times10^{-3}$ & $7.4\times10^{1}\pm4.4\times10^{-2}$ & $1.3\times10^{2}\pm4.9\times10^{-2}$ \\
~ & $-1.3\pm3.0\times10^{-4}$ & $2.2\times10^{-1}\pm6.9\times10^{-3}$ & $-1.4\pm4.9\times10^{-2}$ & $2.2\times10^{-1}\pm1.1\times10^{-2}$ & $-1.8\pm1.5\times10^{-2}$ \\ \hline
GSS30IRS3 & $-7.1\times10^{-2}\pm7.4\times10^{-5}$ & $-4.0\times10^{-1}\pm4.2\times10^{-5}$ & $-3.1\times10^{-1}\pm9.6\times10^{-4}$ & $7.2\times10^{1}\pm1.7\times10^{-2}$ & $2.0\times10^{1}\pm1.7\times10^{-2}$ \\
~ & $-8.8\times10^{-1}\pm2.4\times10^{-4}$ & $5.2\times10^{-1}\pm2.5\times10^{-3}$ & $-6.1\times10^{-1}\pm1.4\times10^{-2}$ & $-3.1\times10^{-1}\pm1.2\times10^{-3}$ & $-2.3\pm8.7\times10^{-3}$ \\ \hline
IRAS15398 & $-9.5\times10^{-2}\pm1.1\times10^{-4}$ & $-1.6\times10^{-1}\pm9.8\times10^{-5}$ & $1.6\times10^{-1}\pm2.2\times10^{-3}$ & $5.1\times10^{1}\pm5.3\times10^{-1}$ & $4.7\times10^{1}\pm1.1$ \\
~ & $-1.7\pm1.5\times10^{-3}$ & $1.5\pm1.7\times10^{-3}$ & $-4.9\times10^{1}\pm1.7$ & $1.6\times10^{-1}\pm2.2\times10^{-3}$ & $-1.7\pm1.5\times10^{-3}$ \\ \hline
IRS5N & $2.4\times10^{-1}\pm7.9\times10^{-5}$ & $-6.9\times10^{-1}\pm4.2\times10^{-5}$ & $-4.4\times10^{-1}\pm3.7\times10^{-4}$ & $6.5\times10^{1}\pm2.9\times10^{-2}$ & $1.7\times10^{2}\pm3.4\times10^{-2}$ \\
~ & $-9.9\times10^{-1}\pm2.7\times10^{-4}$ & $7.0\times10^{-1}\pm1.2\times10^{-3}$ & $-4.1\pm5.3\times10^{-2}$ & $-4.4\times10^{-1}\pm7.4\times10^{-3}$ & $-3.0\pm2.7\times10^{-2}$ \\ \hline
IRAS04166 & $7.8\times10^{-2}\pm3.1\times10^{-5}$ & $-1.8\times10^{-1}\pm3.3\times10^{-5}$ & $-9.2\times10^{-1}\pm7.5\times10^{-4}$ & $4.6\times10^{1}\pm7.0\times10^{-2}$ & $3.2\times10^{1}\pm9.9\times10^{-2}$ \\
~ & $-1.2\pm2.1\times10^{-4}$ & $5.4\times10^{-1}\pm3.3\times10^{-3}$ & $-2.8\pm6.0\times10^{-2}$ & $-1.4\times10^{-1}\pm5.9\times10^{-3}$ & $-1.8\pm6.5\times10^{-3}$ \\ \hline
IRAS32 & $6.2\times10^{-1}\pm5.3\times10^{-5}$ & $-1.5\pm3.4\times10^{-5}$ & $-7.2\times10^{-1}\pm3.1\times10^{-4}$ & $6.8\times10^{1}\pm2.5\times10^{-2}$ & $4.6\times10^{1}\pm3.1\times10^{-2}$ \\
~ & $-1.1\pm3.1\times10^{-4}$ & $6.1\times10^{-1}\pm1.6\times10^{-3}$ & $-6.8\pm1.1\times10^{-1}$ & $-3.2\times10^{-1}\pm7.4\times10^{-3}$ & $-2.1\pm9.0\times10^{-3}$ \\ \hline
IRAS32-b & $-3.3\times10^{-1}\pm9.9\times10^{-5}$ & $-4.9\times10^{-1}\pm4.9\times10^{-5}$ & $-8.4\times10^{-1}\pm1.2\times10^{-3}$ & $7.0\times10^{1}\pm5.1\times10^{-2}$ & $4.1\times10^{1}\pm4.8\times10^{-2}$ \\
~ & $-1.4\pm3.3\times10^{-4}$ & $3.6\times10^{-1}\pm4.7\times10^{-3}$ & $-2.7\pm9.4\times10^{-2}$ & $2.3\times10^{-1}\pm5.4\times10^{-3}$ & $-1.4\pm6.8\times10^{-3}$ \\ \hline
BHR71IRS1 & $-2.1\pm2.3\times10^{-5}$ & $-1.5\times10^{-1}\pm2.5\times10^{-5}$ & $-7.0\times10^{-1}\pm5.7\times10^{-4}$ & $4.0\times10^{1}\pm3.2\times10^{-2}$ & $9.6\pm4.5\times10^{-2}$ \\
~ & $-4.7\times10^{-1}\pm4.0\times10^{-4}$ & $2.5\times10^{-1}\pm1.9\times10^{-3}$ & $-7.6\times10^{-1}\pm1.0\times10^{-2}$ & $-4.8\times10^{-1}\pm3.1\times10^{-3}$ & $-1.2\pm3.2\times10^{-3}$ \\ \hline
Ced110IRS4 & $-2.3\times10^{-1}\pm6.6\times10^{-5}$ & $-6.9\times10^{-1}\pm3.8\times10^{-5}$ & $-3.4\times10^{-1}\pm4.8\times10^{-4}$ & $7.6\times10^{1}\pm1.8\times10^{-2}$ & $1.4\times10^{1}\pm2.3\times10^{-2}$ \\
~ & $-1.0\pm3.4\times10^{-4}$ & $8.4\times10^{-1}\pm1.2\times10^{-3}$ & $-2.6\pm4.7\times10^{-2}$ & $-3.4\times10^{-1}\pm8.1\times10^{-4}$ & $-2.4\pm1.2\times10^{-2}$ \\ \hline
Ced110IRS4-b & $1.1\pm9.4\times10^{-4}$ & $-5.5\times10^{-1}\pm6.5\times10^{-4}$ & $-1.3\pm2.6\times10^{-3}$ & $7.8\times10^{1}\pm2.0$ & $1.8\times10^{2}\pm1.2$ \\
~ & $-2.8\pm5.9\times10^{-3}$ & $4.0\times10^{-3}\pm5.5\times10^{-3}$ & $5.6\times10^{-1}\pm3.0\times10^{-2}$ & $2.0\pm6.8\times10^{-1}$ & $-2.9\pm4.3\times10^{-2}$ \\ \hline
IRS7B & $-2.2\times10^{-3}\pm3.7\times10^{-5}$ & $-2.6\times10^{-1}\pm2.5\times10^{-5}$ & $-3.6\times10^{-1}\pm1.4\times10^{-4}$ & $6.8\times10^{1}\pm1.0\times10^{-2}$ & $2.5\times10^{1}\pm1.1\times10^{-2}$ \\
~ & $-4.4\times10^{-1}\pm8.9\times10^{-5}$ & $6.6\times10^{-1}\pm5.3\times10^{-4}$ & $-8.0\pm6.3\times10^{-2}$ & $7.4\times10^{-1}\pm1.2\times10^{-3}$ & $-4.4\times10^{-1}\pm5.2\times10^{-4}$ \\ \hline
IRS7B-b & $-4.1\times10^{-1}\pm1.2\times10^{-4}$ & $2.9\times10^{-1}\pm8.4\times10^{-5}$ & $-8.0\times10^{-1}\pm1.2\times10^{-3}$ & $6.6\times10^{1}\pm1.0\times10^{-1}$ & $2.4\times10^{1}\pm9.8\times10^{-2}$ \\
~ & $-1.5\pm8.5\times10^{-4}$ & $5.3\times10^{-1}\pm7.1\times10^{-3}$ & $-6.1\pm3.8\times10^{-1}$ & $-6.8\times10^{-1}\pm1.3\times10^{-2}$ & $-2.7\pm2.6\times10^{-2}$ \\ \hline
IRAS04169 & $3.6\times10^{-1}\pm3.4\times10^{-5}$ & $-2.8\times10^{-1}\pm4.0\times10^{-5}$ & $-8.1\times10^{-1}\pm6.2\times10^{-4}$ & $4.3\times10^{1}\pm5.5\times10^{-2}$ & $5.0\times10^{1}\pm9.7\times10^{-2}$ \\
~ & $-1.0\pm1.2\times10^{-4}$ & $1.6\times10^{-2}\pm4.5\times10^{-3}$ & $-3.6\pm5.4\times10^{-2}$ & $7.0\times10^{-1}\pm6.3\times10^{-2}$ & $-1.6\pm1.9\times10^{-1}$ \\ \hline
TMC1A & $-3.4\times10^{-3}\pm7.4\times10^{-5}$ & $-6.8\times10^{-3}\pm5.7\times10^{-5}$ & $-6.6\times10^{-1}\pm7.1\times10^{-4}$ & $5.4\times10^{1}\pm6.0\times10^{-2}$ & $1.7\times10^{2}\pm7.4\times10^{-2}$ \\
~ & $-7.9\times10^{-1}\pm8.1\times10^{-4}$ & $5.5\times10^{-1}\pm2.7\times10^{-3}$ & $-2.0\pm4.2\times10^{-2}$ & $-4.8\times10^{-1}\pm3.9\times10^{-3}$ & $-1.2\pm3.6\times10^{-3}$ \\ \hline
OphIRS43 & $2.8\pm1.2\times10^{-4}$ & $1.5\pm1.1\times10^{-4}$ & $-1.1\pm5.8\times10^{-3}$ & $7.8\times10^{1}\pm3.5\times10^{-1}$ & $4.7\times10^{1}\pm2.8\times10^{-1}$ \\
~ & $-2.0\pm7.3\times10^{-4}$ & $5.1\times10^{-1}\pm3.3\times10^{-2}$ & $-5.0\pm7.7\times10^{-1}$ & $-5.9\times10^{-2}\pm1.0\times10^{-2}$ & $-2.2\pm1.3\times10^{-2}$ \\ \hline
OphIRS43 VLA2 & \nodata & \nodata & \nodata & \nodata & \nodata \\
~ & \nodata & \nodata & \nodata & \nodata & \nodata \\ \hline
L1489IRS & $3.1\pm7.9\times10^{-5}$ & $-1.8\times10^{-1}\pm1.0\times10^{-4}$ & $-3.2\pm5.5\times10^{-1}$ & $8.4\times10^{1}\pm4.5\times10^{-1}$ & $1.6\times10^{2}\pm1.1$ \\
~ & $-1.5\pm1.7\times10^{-3}$ & $3.8\times10^{-1}\pm2.1\times10^{-2}$ & $3.7\pm1.0$ & $-2.3\pm1.1\times10^{-2}$ & $-2.3\pm1.8\times10^{-3}$ \\ \hline
OphIRS63 & $1.9\times10^{-3}\pm2.8\times10^{-5}$ & $-3.2\times10^{-3}\pm2.9\times10^{-5}$ & $-2.6\times10^{-1}\pm1.9\times10^{-4}$ & $5.0\times10^{1}\pm2.2\times10^{-2}$ & $6.0\times10^{1}\pm2.9\times10^{-2}$ \\
~ & $-4.4\times10^{-1}\pm1.0\times10^{-4}$ & $8.1\times10^{-1}\pm4.3\times10^{-4}$ & $-2.0\pm1.2\times10^{-2}$ & $2.1\pm5.6\times10^{-1}$ & $-1.7\pm7.3\times10^{-1}$ \\
\enddata
\tablecomments{Each source occupies two rows. The first row contains $x_0$, $y_0$, $\log r_w$, $i$, and $p.a.$. The second row contains $\log(I_{\nu,0})$, $\gamma_1$, $\gamma_2$, $\log r_{En}$, and $\log I_{\nu,E0} $.}
\end{deluxetable*}
\end{rotatetable*}

\begin{rotatetable*}
\begin{deluxetable*}{|l|ccccc}
\tabletypesize{\tiny}
\tablecaption{Best-Fit Parameters and Uncertainties for the PLCT with asymmetry model \label{tab:asy}}
\tablewidth{0pt}
\tablehead{
\colhead{\shortstack{Source \\ \phantom{.} \\ \phantom{.}}} & 
\colhead{\shortstack{$x_0$(\arcsec) \\ $\log(I_{\nu,0})$ \\ $F_{\nu,a_0}$}} & 
\colhead{\shortstack{$y_0$(\arcsec) \\ $\gamma_1$ \\ $\phi_c$ (deg)}} & 
\colhead{\shortstack{$\log r_w$(\arcsec) \\ $\gamma_2$ \\ $\phi_w$(deg)}} & 
\colhead{\shortstack{$i$ (deg) \\ $\log r_c$(\arcsec) \\ $\log r_{En}$ (\arcsec)}} & 
\colhead{\shortstack{$p.a.$ (deg) \\ $\log r_{w,a}$(\arcsec) \\ $\log I_{\nu,E0}$}}
}
\startdata
BHR71IRS2 & & & & & \\ \hline
B335 & $-8.1\times10^{-3}\pm1.6\times10^{-4}$ & $-2.1\times10^{-2}\pm7.1\times10^{-5}$ & $-2.2\pm2.4\times10^{-2}$ & $3.0\times10^{1}\pm1.9\times10^{-2}$ & $-1.3\times10^{2}\pm2.5\times10^{-1}$ \\
~ & $-1.4\pm2.0\times10^{-3}$ & $-8.4\times10^{-1}\pm3.7\times10^{-3}$ & $1.4\pm9.4\times10^{-3}$ & $-1.9\pm4.0\times10^{-3}$ & $-1.5\pm3.0\times10^{-3}$ \\
~ & $5.0\pm4.4\times10^{-2}$ & $-3.8\times10^{1}\pm1.2\times10^{-2}$ & $-7.8\times10^{1}\pm3.9\times10^{-1}$ & $-6.5\times10^{-1}\pm5.4\times10^{-3}$ & $-1.7\pm3.5\times10^{-3}$ \\ \hline
IRAS16253 & $2.1\times10^{-1}\pm3.2\times10^{-4}$ & $-9.2\times10^{-1}\pm4.1\times10^{-4}$ & $-8.4\times10^{-1}\pm8.5\times10^{-3}$ & $6.6\times10^{1}\pm5.0\times10^{-1}$ & $2.6\times10^{1}\pm6.5\times10^{-1}$ \\
~ & $-1.9\pm9.7\times10^{-4}$ & $9.2\times10^{-1}\pm1.6\times10^{-2}$ & $-2.0\pm5.1\times10^{-1}$ & $-8.6\times10^{-1}\pm2.0\times10^{-2}$ & $-1.1\pm3.9\times10^{-2}$ \\
~ & $-1.0\pm2.8\times10^{-3}$ & $6.2\times10^{1}\pm1.6$ & $7.7\times10^{1}\pm2.8$ & $1.4\times10^{-1}\pm1.7\times10^{-2}$ & $-2.2\pm2.2\times10^{-2}$ \\ \hline
IRAS 16544-1604 & $4.0\times10^{-2}\pm9.0\times10^{-5}$ & $-1.1\times10^{-1}\pm1.0\times10^{-4}$ & $-8.2\times10^{-1}\pm2.5\times10^{-3}$ & $7.2\times10^{1}\pm5.8\times10^{-2}$ & $1.3\times10^{2}\pm5.1\times10^{-2}$ \\
~ & $-1.3\pm2.9\times10^{-4}$ & $3.1\times10^{-1}\pm1.1\times10^{-2}$ & $-1.3\pm5.5\times10^{-2}$ & $-1.3\pm4.0\times10^{-2}$ & $-1.2\pm2.9\times10^{-2}$ \\
~ & $-5.4\times10^{-1}\pm7.5\times10^{-3}$ & $-8.2\times10^{1}\pm4.2\times10^{-1}$ & $5.2\times10^{1}\pm7.7\times10^{-1}$ & $2.3\times10^{-1}\pm1.1\times10^{-2}$ & $-1.8\pm1.5\times10^{-2}$ \\ \hline
GSS30IRS3 & $-7.8\times10^{-2}\pm1.3\times10^{-4}$ & $-4.1\times10^{-1}\pm8.4\times10^{-5}$ & $-3.2\times10^{-1}\pm1.5\times10^{-3}$ & $7.2\times10^{1}\pm1.9\times10^{-2}$ & $2.0\times10^{1}\pm1.6\times10^{-2}$ \\
~ & $-8.8\times10^{-1}\pm2.4\times10^{-4}$ & $8.5\times10^{-2}\pm5.7\times10^{-3}$ & $-9.0\times10^{-1}\pm2.2\times10^{-2}$ & $-2.9\pm6.1\times10^{-2}$ & $-7.0\times10^{-1}\pm2.2\times10^{-3}$ \\
~ & $2.3\pm1.9\times10^{-2}$ & $9.5\times10^{1}\pm1.4\times10^{-1}$ & $5.3\times10^{1}\pm2.5\times10^{-1}$ & $-3.2\times10^{-1}\pm1.6\times10^{-3}$ & $-2.2\pm7.3\times10^{-3}$ \\ \hline
IRAS15398 & $-9.2\times10^{-2}\pm1.2\times10^{-4}$ & $-1.6\times10^{-1}\pm1.1\times10^{-4}$ & $2.5\times10^{-1}\pm8.3\times10^{-3}$ & $2.4\times10^{1}\pm1.7$ & $3.4\times10^{1}\pm1.5$ \\
~ & $-1.5\pm1.2\times10^{-3}$ & $1.4\pm2.5\times10^{-3}$ & $-9.7\pm3.9\times10^{-1}$ & $-6.1\times10^{-1}\pm3.6\times10^{-3}$ & $-5.6\times10^{-1}\pm3.5\times10^{-3}$ \\
~ & $-1.0\pm5.4\times10^{-5}$ & $-1.8\times10^{2}\pm1.2\times10^{-1}$ & $1.8\times10^{2}\pm8.7\times10^{-3}$ & $2.6\times10^{-1}\pm1.7\times10^{-3}$ & $-1.5\pm1.2\times10^{-3}$ \\ \hline
IRS5N & $2.4\times10^{-1}\pm1.3\times10^{-4}$ & $-7.0\times10^{-1}\pm1.0\times10^{-4}$ & $-4.5\times10^{-1}\pm4.2\times10^{-4}$ & $6.5\times10^{1}\pm3.7\times10^{-2}$ & $1.7\times10^{2}\pm4.0\times10^{-2}$ \\
~ & $-9.9\times10^{-1}\pm2.1\times10^{-4}$ & $7.0\times10^{-1}\pm1.6\times10^{-3}$ & $-4.0\pm5.3\times10^{-2}$ & $-9.4\times10^{-1}\pm3.1\times10^{-2}$ & $-7.8\times10^{-1}\pm2.2\times10^{-2}$ \\
~ & $-2.1\times10^{-1}\pm3.3\times10^{-3}$ & $9.1\times10^{1}\pm6.9\times10^{-1}$ & $5.8\times10^{1}\pm7.3\times10^{-1}$ & $1.9\pm6.0\times10^{-1}$ & $-2.0\pm5.8\times10^{-1}$ \\ \hline
IRAS04166 & $7.6\times10^{-2}\pm5.5\times10^{-5}$ & $-1.8\times10^{-1}\pm7.4\times10^{-5}$ & $-9.6\times10^{-1}\pm2.0\times10^{-3}$ & $4.5\times10^{1}\pm8.2\times10^{-2}$ & $3.3\times10^{1}\pm1.1\times10^{-1}$ \\
~ & $-1.2\pm2.1\times10^{-4}$ & $4.5\times10^{-1}\pm9.4\times10^{-3}$ & $-1.6\pm5.5\times10^{-2}$ & $-1.3\pm5.5\times10^{-3}$ & $-1.6\pm1.4\times10^{-2}$ \\
~ & $-4.4\times10^{-1}\pm6.5\times10^{-3}$ & $-9.7\times10^{1}\pm5.0\times10^{-1}$ & $7.2\times10^{1}\pm7.9\times10^{-1}$ & $-1.3\times10^{-1}\pm5.8\times10^{-3}$ & $-1.8\pm6.4\times10^{-3}$ \\ \hline
IRAS32 & $6.2\times10^{-1}\pm8.3\times10^{-5}$ & $-1.5\pm8.7\times10^{-5}$ & $-7.0\times10^{-1}\pm3.6\times10^{-4}$ & $6.7\times10^{1}\pm3.7\times10^{-2}$ & $4.6\times10^{1}\pm3.0\times10^{-2}$ \\
~ & $-1.1\pm2.0\times10^{-4}$ & $4.8\times10^{-1}\pm2.5\times10^{-3}$ & $-1.0\times10^{1}\pm1.4\times10^{-2}$ & $-2.9\pm5.8\times10^{-2}$ & $-9.1\times10^{-1}\pm1.9\times10^{-3}$ \\
~ & $1.4\pm1.3\times10^{-2}$ & $9.0\times10^{1}\pm2.4\times10^{-1}$ & $8.4\times10^{1}\pm4.7\times10^{-1}$ & $-2.9\times10^{-1}\pm7.9\times10^{-3}$ & $-2.1\pm8.6\times10^{-3}$ \\ \hline
IRAS32-b & $-3.4\times10^{-1}\pm1.2\times10^{-4}$ & $-5.0\times10^{-1}\pm1.0\times10^{-4}$ & $-8.9\times10^{-1}\pm2.7\times10^{-3}$ & $7.0\times10^{1}\pm4.9\times10^{-2}$ & $4.2\times10^{1}\pm4.4\times10^{-2}$ \\
~ & $-1.4\pm2.8\times10^{-4}$ & $2.8\times10^{-1}\pm1.3\times10^{-2}$ & $-1.8\pm7.9\times10^{-2}$ & $-1.4\pm8.8\times10^{-2}$ & $-1.2\pm4.4\times10^{-2}$ \\
~ & $-6.5\times10^{-1}\pm8.2\times10^{-3}$ & $-8.9\times10^{1}\pm3.2\times10^{-1}$ & $8.4\times10^{1}\pm8.0\times10^{-1}$ & $2.4\times10^{-1}\pm5.1\times10^{-3}$ & $-1.4\pm6.7\times10^{-3}$ \\ \hline
BHR71IRS1 & $-2.1\pm9.5\times10^{-5}$ & $-1.2\times10^{-1}\pm1.3\times10^{-4}$ & $-7.2\times10^{-1}\pm7.9\times10^{-4}$ & $4.1\times10^{1}\pm7.6\times10^{-2}$ & $8.9\pm5.9\times10^{-2}$ \\
~ & $-4.6\times10^{-1}\pm4.4\times10^{-4}$ & $-3.1\times10^{-2}\pm3.8\times10^{-3}$ & $-7.1\times10^{-1}\pm9.8\times10^{-3}$ & $-8.5\times10^{-1}\pm4.7\times10^{-3}$ & $-7.0\times10^{-1}\pm6.6\times10^{-3}$ \\
~ & $-6.1\times10^{-1}\pm1.1\times10^{-3}$ & $8.0\times10^{1}\pm1.5\times10^{-1}$ & $1.2\times10^{2}\pm2.4\times10^{-1}$ & $-4.5\times10^{-1}\pm4.1\times10^{-3}$ & $-1.2\pm4.7\times10^{-3}$ \\ \hline
Ced 110 IRS4A & $-2.3\times10^{-1}\pm5.1\times10^{-5}$ & $-6.9\times10^{-1}\pm3.2\times10^{-5}$ & $-3.1\times10^{-1}\pm4.0\times10^{-4}$ & $7.6\times10^{1}\pm1.5\times10^{-2}$ & $1.4\times10^{1}\pm1.7\times10^{-2}$ \\
~ & $-1.0\pm2.6\times10^{-4}$ & $1.0\pm1.6\times10^{-3}$ & $-6.0\pm7.0\times10^{-3}$ & $-2.3\pm9.3\times10^{-2}$ & $-1.9\pm3.2\times10^{-2}$ \\
~ & $-8.5\times10^{-1}\pm9.7\times10^{-3}$ & $-1.1\times10^{1}\pm4.6$ & $1.8\times10^{2}\pm6.3\times10^{-1}$ & $-3.1\times10^{-1}\pm6.6\times10^{-4}$ & $-2.3\pm9.5\times10^{-3}$ \\ \hline
Ced 110 IRS4B & & & & & \\ \hline
IRS7B & $6.3\times10^{-3}\pm1.0\times10^{-4}$ & $-2.5\times10^{-1}\pm6.9\times10^{-5}$ & $-3.7\times10^{-1}\pm1.9\times10^{-4}$ & $6.7\times10^{1}\pm1.8\times10^{-2}$ & $2.5\times10^{1}\pm1.3\times10^{-2}$ \\
~ & $-4.4\times10^{-1}\pm9.7\times10^{-5}$ & $7.0\times10^{-1}\pm9.9\times10^{-4}$ & $-9.9\pm8.9\times10^{-2}$ & $-7.1\times10^{-3}\pm7.5\times10^{-3}$ & $-9.7\times10^{-3}\pm7.5\times10^{-3}$ \\
~ & $7.5\times10^{-1}\pm7.2\times10^{-3}$ & $-8.4\times10^{1}\pm1.8\times10^{-1}$ & $9.3\times10^{1}\pm2.5\times10^{-1}$ & $7.4\times10^{-1}\pm1.3\times10^{-3}$ & $-4.4\times10^{-1}\pm5.6\times10^{-4}$ \\ \hline
IRS7B-b & & & & & \\ \hline
IRAS04169 & $3.6\times10^{-1}\pm4.1\times10^{-5}$ & $-2.8\times10^{-1}\pm4.8\times10^{-5}$ & $-7.9\times10^{-1}\pm3.3\times10^{-4}$ & $4.3\times10^{1}\pm4.6\times10^{-2}$ & $5.0\times10^{1}\pm7.7\times10^{-2}$ \\
~ & $-1.0\pm1.2\times10^{-4}$ & $1.1\times10^{-1}\pm2.8\times10^{-3}$ & $-6.0\pm2.3\times10^{-2}$ & $-1.2\pm2.0\times10^{-3}$ & $-1.7\pm1.6\times10^{-2}$ \\
~ & $3.5\times10^{-1}\pm7.3\times10^{-3}$ & $-8.9\times10^{1}\pm9.2\times10^{-1}$ & $1.3\times10^{2}\pm1.2$ & $6.8\times10^{-1}\pm6.8\times10^{-2}$ & $-1.6\pm1.9\times10^{-1}$ \\ \hline
TMC1A & $-2.6\times10^{-3}\pm9.3\times10^{-5}$ & $-5.2\times10^{-3}\pm6.8\times10^{-5}$ & $-5.4\times10^{-1}\pm3.4\times10^{-3}$ & $5.4\times10^{1}\pm8.0\times10^{-2}$ & $1.7\times10^{2}\pm8.6\times10^{-2}$ \\
~ & $-7.3\times10^{-1}\pm1.3\times10^{-3}$ & $5.4\times10^{-1}\pm3.5\times10^{-3}$ & $-4.8\times10^{-1}\pm3.8\times10^{-2}$ & $-5.0\times10^{-1}\pm2.3\times10^{-3}$ & $-1.0\pm6.8\times10^{-3}$ \\
~ & $-8.1\times10^{-1}\pm4.7\times10^{-3}$ & $-6.3\times10^{1}\pm1.5$ & $1.8\times10^{2}\pm2.7\times10^{-1}$ & $-3.5\times10^{-1}\pm5.0\times10^{-3}$ & $-1.3\pm7.3\times10^{-3}$ \\ \hline
OphIRS43 & $2.8\pm1.3\times10^{-3}$ & $1.5\pm1.4\times10^{-3}$ & $-1.1\pm3.6\times10^{-3}$ & $7.8\times10^{1}\pm2.5\times10^{-1}$ & $4.6\times10^{1}\pm2.1\times10^{-1}$ \\
~ & $-2.0\pm6.2\times10^{-4}$ & $5.3\times10^{-1}\pm2.6\times10^{-2}$ & $-5.1\pm3.8\times10^{-1}$ & $-5.4\times10^{-2}\pm1.7\times10^{-1}$ & $-1.2\pm1.2\times10^{-1}$ \\
~ & $2.1\pm4.0\times10^{-1}$ & $4.0\times10^{1}\pm1.6\times10^{1}$ & $3.1\times10^{1}\pm2.0\times10^{1}$ & $1.3\times10^{-3}\pm3.2\times10^{-3}$ & $-2.1\pm4.9\times10^{-3}$ \\ \hline
OphIRS43 VLA2 & & & & & \\ \hline
L1489IRS & $3.1\pm1.0\times10^{-6}$ & $-1.8\times10^{-1}\pm4.0\times10^{-6}$ & $-3.2\pm3.4\times10^{-3}$ & $7.8\times10^{1}\pm2.0\times10^{-2}$ & $1.6\times10^{2}\pm5.4\times10^{-3}$ \\
~ & $-1.6\pm1.0\times10^{-5}$ & $1.3\pm1.0\times10^{-4}$ & $2.0\times6.3\times10^{-5}$ & $1.6\times10^{-1}\pm5.1\times10^{-4}$ & $-1.3\times10^{-1}\pm8.2\times10^{-4}$ \\
~ & $3.8\pm7.9\times10^{-4}$ & $5.4\times10^{1}\pm9.2\times10^{-2}$ & $1.8\times10^{2}\pm4.4\times10^{-3}$ & $6.8\times10^{-2}\pm2.0\times10^{-6}$ & $-1.6\pm2.3\times10^{-5}$ \\ \hline
OphIRS63 & $1.8\times10^{-3}\pm2.8\times10^{-5}$ & $-3.1\times10^{-3}\pm2.3\times10^{-5}$ & $-2.5\times10^{-1}\pm2.2\times10^{-4}$ & $4.9\times10^{1}\pm2.1\times10^{-2}$ & $6.0\times10^{1}\pm2.4\times10^{-2}$ \\
~ & $-4.6\times10^{-1}\pm2.0\times10^{-4}$ & $9.5\times10^{-1}\pm7.0\times10^{-4}$ & $-4.9\pm3.3\times10^{-2}$ & $-2.5\pm4.2\times10^{-1}$ & $-2.0\pm6.5\times10^{-2}$ \\
~ & $-8.6\times10^{-1}\pm1.8\times10^{-2}$ & $-9.9\times10^{-1}\pm3.6$ & $1.8\times10^{2}\pm2.5\times10^{-1}$ & $-2.5\times10^{-1}\pm2.2\times10^{-4}$ & $-1.9\pm7.0\times10^{-3}$ \\
\enddata
\tablecomments{Each source occupies three rows. Row 1: $x_0$, $y_0$, $\log r_w$, $i$, $p.a.$ Row 2: $\log(I_{\nu,0})$, $\gamma_1$, $\gamma_2$, $\log r_c$, $\log r_{w,a}$. Row 3: $F_{\nu,a_0}$, $\phi_c$, $\phi_w$, $\log r_{En}$, $\log I_{\nu,E0}$.}
\end{deluxetable*}
\end{rotatetable*}

\begin{deluxetable*}{l|c|c}[h]
\tablecaption{Parameters for the models for edge on sources \label{tab:edge_on}}
\tablehead{
 \colhead{Parameter} & \colhead{IRAS 04302+2247} & \colhead{L1527 IRS}
}
\startdata
\multicolumn{3}{c}{{Rectangle}} \\
\hline
$x_0$ (\arcsec) & $-2.0 \times 10^{-2} \pm 1.1 \times 10^{-4}$ & $-4.4 \times 10^{-1} \pm 2.1 \times 10^{-5}$ \\
$y_0$ (\arcsec) & $-1.7 \times 10^{-1} \pm 5.5 \times 10^{-4}$ & $-3.7 \times 10^{-1} \pm 4.9 \times 10^{-5}$ \\
log$x_w$ (\arcsec) & $8.3 \times 10^{-2} \pm 1.2 \times 10^{-5}$ & $-5.9 \times 10^{-1} \pm 2.1 \times 10^{-5}$ \\
log$y_w$ (\arcsec) & $-9.1 \times 10^{-1} \pm 4.7 \times 10^{-4}$ & $-1.2 \pm 4.1 \times 10^{-4}$ \\
$p.a.$ (deg) & $84.8 \pm 8.0 \times 10^{-4}$ & $-88.6 \pm 85.7$ \\
log $I_{\nu,0}$ & $-8.0 \times 10^{-1} \pm 3.5 \times 10^{-4}$ & $-8.7 \times 10^{-1} \pm 1.8 \times 10^{-4}$ \\
log$r_{En}$ (\arcsec) & $-5.0 \times 10^{-1} \pm 4.5 \times 10^{-3}$ & $-4.1 \times 10^{-1} \pm 7.5 \times 10^{-4}$ \\
log $I_{\nu, E_0}$ & $-1.9 \pm 5.3 \times 10^{-3}$ & $-1.3 \pm 7.6 \times 10^{-4}$ \\
\hline
\multicolumn{3}{c}{{Broken Power-Law Rectangle}} \\
\hline
$x_0$ (\arcsec) & $-2.0 \times 10^{-2} \pm 1.1 \times 10^{-4}$ & $-4.4 \times 10^{-1} \pm 2.1 \times 10^{-5}$ \\
$y_0$ (\arcsec) & $-1.7 \times 10^{-1} \pm 5.4 \times 10^{-4}$ & $-3.7 \times 10^{-1} \pm 5.5 \times 10^{-5}$ \\
log$x_w$ (\arcsec) & $1.2 \times 10^{-1} \pm 1.6 \times 10^{-3}$ & $-1.9 \times 10^{-1} \pm 1.7 \times 10^{-3}$ \\
log$y_w$ (\arcsec) & $-9.1 \times 10^{-1} \pm 4.4 \times 10^{-4}$ & $-1.2 \pm 3.3 \times 10^{-4}$ \\
$p.a.$ (deg) & $84.8 \pm 7.3 \times 10^{-3}$ & $91.8 \pm 88.9$ \\
log $I_{\nu,0}$ & $-7.9 \times 10^{-1} \pm 3.9 \times 10^{-4}$ & $-8.2 \times 10^{-1} \pm 2.0 \times 10^{-4}$ \\
$\gamma_{in}$ & $3.8 \times 10^{-3} \pm 5.7 \times 10^{-3}$ & $1.1 \times 10^{-2} \pm 7.4 \times 10^{-4}$ \\
$\gamma_{out}$ & $7.7 \times 10^{-1} \pm 5.9 \times 10^{-2}$ & $1.6 \pm 3.0 \times 10^{-3}$ \\
log$x_b$ & $1.1 \times 10^{-1} \pm 1.5 \times 10^{-2}$ & $-8.1 \times 10^{-1} \pm 5.3 \times 10^{-4}$ \\
$\Delta$ & $6.3 \times 10^{-1} \pm 5.9 \times 10^{-2}$ & $1.3 \times 10^{-2} \pm 3.4 \times 10^{-3}$ \\
log$r_{En}$ (\arcsec) & $-7.5 \times 10^{-2} \pm 6.2 \times 10^{-3}$ & $-4.7 \times 10^{-1} \pm 1.1 \times 10^{-3}$ \\
log $I_{\nu, E_0}$ & $-1.7 \pm 7.3 \times 10^{-3}$ & $-1.4 \pm 1.1 \times 10^{-3}$ \\
\hline
\multicolumn{3}{c}{{Flared Asymmetric BPLRT}} \\
\hline
$x_0$ (\arcsec) & $-2.3 \times 10^{-2} \pm 2.1 \times 10^{-4}$ & $-4.4 \times 10^{-1} \pm 2.2 \times 10^{-5}$ \\
$y_0$ (\arcsec) & $-1.3 \times 10^{-1} \pm 2.1 \times 10^{-3}$ & $-3.7 \times 10^{-1} \pm 1.1 \times 10^{-4}$ \\
log$x_w$ (\arcsec) & $7.5 \times 10^{-4} \pm 9.5 \times 10^{-4}$ & $-4.5 \times 10^{-1} \pm 1.3 \times 10^{-2}$ \\
log$y_w$ (\arcsec) & $-1.1 \pm 2.2 \times 10^{-3}$ & $-1.4 \pm 1.5 \times 10^{-3}$ \\
$p.a.$ (deg) & $84.7 \pm 9.2 \times 10^{-3}$ & $91.9 \pm 1.0 \times 10^{-2}$ \\
log$I_{\nu,0}$ & $-7.5 \times 10^{-1} \pm 4.3 \times 10^{-4}$ & $-7.7 \times 10^{-1} \pm 3.2 \times 10^{-4}$ \\
$\gamma_{in}$ & $1.8 \pm 1.6 \times 10^{-1}$ & $-4.2 \times 10^{-1} \pm 2.9 \times 10^{-2}$ \\
$\gamma_{out,1}$ & $-2.6 \times 10^{-1} \pm 3.3 \times 10^{-2}$ & $3.2 \pm 4.3 \times 10^{-2}$ \\
$\gamma_{out,2}$ & $-4.5 \times 10^{-1} \pm 3.3 \times 10^{-2}$ & $3.4 \pm 4.5 \times 10^{-2}$ \\
$\epsilon$ & $1.4 \times 10^{-1} \pm 5.0 \times 10^{-2}$ & $1.0 \times 10^{-2} \pm 3.6 \times 10^{-4}$ \\
log$x_b$ (\arcsec) & $-2.6 \times 10^{-1} \pm 2.1 \times 10^{-2}$ & $-7.1 \times 10^{-1} \pm 2.8 \times 10^{-3}$ \\
$\Delta$ & $2.4 \times 10^{-1} \pm 3.4 \times 10^{-2}$ & $3.7 \times 10^{-1} \pm 9.7 \times 10^{-3}$ \\
$\gamma_x$ & $2.8 \pm 1.2 \times 10^{-2}$ & $2.4 \times 10^{-1} \pm 1.6 \times 10^{-2}$ \\
$\gamma_y$ & $1.6 \pm 8.7 \times 10^{-3}$ & $2.0 \pm 8.4 \times 10^{-3}$ \\
$A$ & $4.1 \times 10^{-1} \pm 2.3 \times 10^{-2}$ & $2.1 \pm 9.8 \times 10^{-2}$ \\
$x_s$  (\arcsec) & $9.4 \times 10^{-1} \pm 6.4 \times 10^{-2}$ & $3.3 \times 10^{-1} \pm 1.0 \times 10^{-2}$ \\
$w$ & $6.5 \times 10^{-1} \pm 6.5 \times 10^{-2}$ & $1.7 \times 10^{-1} \pm 5.2 \times 10^{-3}$ \\
$log_{smooth}$ & $2.6 \pm 2.4 \times 10^{-2}$ & $1.5 \pm 1.2 \times 10^{-2}$ \\
log$r_{En}$ (\arcsec) & $1.0 \times 10^{-1} \pm 8.4 \times 10^{-3}$ & $-4.8 \times 10^{-1} \pm 2.2 \times 10^{-3}$ \\
log $I_{\nu, E_0}$ & $-1.8 \pm 1.2 \times 10^{-2}$ & $-1.7 \pm 2.2 \times 10^{-3}$ \\
\enddata
\tablecomments{For Rectangle models and Broken Power-Law Rectangle models, the truncation parameters $\gamma_x$ and = $\gamma_y = 4$ are fixed to be $4$.}
\end{deluxetable*}

\begin{deluxetable*}{lc}[h]
\tablecaption{Parameters for the best-fit models for edge-on disks \label{tab:edge_on_BF}}
\tablehead{
 \colhead{Parameter} & \colhead{}
}
\startdata
\multicolumn{2}{c}{{IRAS 04302+2247 }} \\
\hline
Model & Flared Asym. Minor-axis \\
$x_0$ (\arcsec) & $-3.0 \times 10^{-2} \pm 2.8 \times 10^{-4}$ \\
$y_0$ (\arcsec) & $-1.3 \times 10^{-1} \pm 2.2 \times 10^{-3}$ \\
log$x_w$ (\arcsec) & $-6.6 \times 10^{-2} \pm 1.1 \times 10^{-2}$ \\
log$y_w$ (\arcsec)  & $-1.1 \pm 3.7 \times 10^{-3}$ \\
$p.a.$ (deg) & $84.7 \pm 1.1 \times 10^{-2}$ \\
log$I_{\nu,0}$ & $-7.5 \times 10^{-1} \pm 4.6 \times 10^{-4}$ \\
$\gamma_{in}$ & $2.0 \pm 1.8 \times 10^{-2}$ \\
$\gamma_{out1}$ & $-1.0 \pm 9.7 \times 10^{-2}$ \\
$\gamma_{out2}$ & $-1.2 \pm 9.8 \times 10^{-2}$ \\
$\epsilon$ (\arcsec) & $9.8 \times 10^{-2} \pm 4.6 \times 10^{-2}$ \\
log$x_b$ (\arcsec) & $-2.3 \times 10^{-1} \pm 1.1 \times 10^{-2}$ \\
$\Delta$ & $2.9 \times 10^{-1} \pm 2.4 \times 10^{-2}$ \\
$\gamma_x$ & $2.4 \pm 5.1 \times 10^{-2}$ \\
$\gamma_y$ & $1.5 \pm 1.2 \times 10^{-2}$ \\
$A$ & $3.4 \times 10^{-1} \pm 2.1 \times 10^{-2}$ \\
$x_s$ (\arcsec) & $9.4 \times 10^{-1} \pm 7.2 \times 10^{-2}$ \\
w (\arcsec) & $7.4 \times 10^{-1} \pm 8.2 \times 10^{-2}$ \\
a & $4.1 \times 10^{-1} \pm 1.4 \times 10^{-2}$ \\
$x_g$ (\arcsec) & $6.0 \times 10^{-2} \pm 1.6 \times 10^{-2}$ \\
$y_g$ (\arcsec) & $6.9 \times 10^{-2} \pm 8.6 \times 10^{-4}$ \\
log$x_{wg}$ (\arcsec) & $-2.0 \times 10^{-1} \pm 5.0 \times 10^{-3}$ \\
log$y_{wg}$ (\arcsec) & $-1.5 \pm 2.1 \times 10^{-2}$ \\
log$\mathrm{smooth}$ & $2.6 \pm 1.2 \times 10^{-2}$ \\
log $r_{En}$ (\arcsec) & $1.3 \times 10^{-1} \pm 8.5 \times 10^{-3}$ \\
log $I_{\nu, E_0}$ & $-1.8 \pm 1.3 \times 10^{-2}$ \\
\hline
\multicolumn{2}{c}{{L1527 IRS}} \\
\hline
Model & (Sym-gapped Sigmoid) \\
$x_0$ (\arcsec) & $-4.4 \times 10^{-1} \pm 2.0 \times 10^{-5}$ \\
$y_0$ (\arcsec) & $-3.6 \times 10^{-1} \pm 7.1 \times 10^{-5}$ \\
log$x_w$ (\arcsec) & $-5.2 \times 10^{-1} \pm 1.2 \times 10^{-3}$ \\
log$y_w$(\arcsec)  & $-1.4 \pm 1.8 \times 10^{-3}$ \\
$p.a.$ (deg) & $91.9 \pm 9.8 \times 10^{-3}$ \\
log$I_{\nu,0}$ & $-7.8 \times 10^{-1} \pm 3.6 \times 10^{-4}$ \\
$\gamma_{in}$ & $-5.2 \times 10^{-1} \pm 2.3 \times 10^{-2}$ \\
$\gamma_{out1}$ & $3.7 \pm 4.7 \times 10^{-3}$ \\
$\gamma_{out2}$ & $4.0 \pm 2.8 \times 10^{-3}$ \\
$\epsilon$ & $1.6 \times 10^{-2} \pm 2.9 \times 10^{-3}$ \\
log$x_b$ (\arcsec) & $-5.2 \times 10^{-1} \pm 1.2 \times 10^{-3}$ \\
$\Delta$ & $5.2 \times 10^{-1} \pm 1.1 \times 10^{-2}$ \\
$\gamma_x$ & $1.4 \pm 2.3 \times 10^{-2}$ \\
$\gamma_y$ & $1.9 \pm 9.4 \times 10^{-3}$ \\
$A$ & $1.5 \pm 4.8 \times 10^{-2}$ \\
$x_s$ (\arcsec) & $2.5 \times 10^{-1} \pm 1.0 \times 10^{-2}$ \\
w (\arcsec) & $1.5 \times 10^{-1} \pm 7.0 \times 10^{-3}$ \\
log$\mathrm{smooth}$ & $1.6 \pm 1.0 \times 10^{-2}$ \\
$x_{gap}$ (\arcsec) & $3.0 \times 10^{-1} \pm 8.2 \times 10^{-4}$ \\
log $w_{gap}$ (\arcsec) & $-6.6 \times 10^{-1} \pm 7.4 \times 10^{-3}$ \\
log $\Delta_{gap}$ & $-1.8 \times 10^{-1} \pm 2.2 \times 10^{-3}$ \\
log $r_{En}$ (\arcsec) & $-4.6 \times 10^{-1} \pm 2.8 \times 10^{-3}$ \\
log $I_{\nu, E_0}$ & $-1.7 \pm 2.5 \times 10^{-3}$ \\
\enddata
\end{deluxetable*}

\newpage

\onecolumngrid

\section{Best-Fit Model Modifications}
\label{advance_models}
As introduced in Section~\ref{sec:best-fit}, the best-fit models for many sources involve additional components beyond the standard models. These extensions are motivated by residual features such as localized brightness enhancement that cannot be captured by the asymmetric profile implemented in the standard model, multiple gaps, large-scale envelopes, or broken radial profiles. To accommodate these features, we develop a set of model modifications, including both additive components and alterations to the base radial or azimuthal profile. Each modification is described below with its mathematical expression, defining parameters, and physical motivation.

\subsection{Azimuthally Asymmetric component}

The azimuthally asymmetric component can also be defined in terms of a sine function as:

\begin{equation}
 F_{\nu,a2,\phi}(r) = F_{\nu,a2_0} \exp{\left(-\frac{(r - r_{c,a})^4}{2 \, {r_{w,a}}^4}\right)} \, \sin{\left(\phi - \phi_{w,a}\right)}
 \label{eq:8}
\end{equation}
where again $r_{c,a}$ is the radial center of the brightness asymmetry, $r_{w,a}$ and $\phi_{w,a}$ are characteristic radius and azimuthal angle. 

\subsection{Azimuthally Symmetric Ring Component} \label{sec:sym_ring}
As described in Section~\ref{sec:Inclined math}, azimuthally asymmetric ring features are implemented in the PLCT model. In some cases, however, we find that observed residuals exhibit ring or gap structure without significant azimuthal variation. For these sources, we include an azimuthally symmetric ring component as a simplified version of the asymmetric profile. This component adopts the same radial brightness distribution as given by \ref{eq:exp_gap},
\begin{equation}
 I_{\nu, ring}(r) = I_{\nu,r_0} \exp{\left(-\frac{(r - r_{c,r})^4}{2 \, {r_{w,r}}^4}\right)}
\end{equation} \label{eq:exp_gap}
which is a radially localized Gaussian-shaped function, centered at $r_{c,r}$, with half-width $r_{w,r}$. This component has no azimuthal dependence and thus represents an symmetrical ring or gap in the disk. It is applied to the PLCT profile as a multiplicative perturbation of the form $(1+a\cdot f(r))$ to produce the disk intensity profile $I_{\nu, disk}(r)$.
\begin{gather}
 f_{ring}(r) = \exp{\left(-\frac{(r - r_{c,r})^4}{2 \, {r_{w,r}}^4}\right)} \label{eq:sym_ring}\\
 I_{\nu, disk}(r) = I_{\nu, PLCT}(r) \cdot (1+a\cdot f_{ring}(r)) \label{eq:PLCT_sym_ring}
\end{gather}

The symmetric ring component is included in the best-fit model under two main circumstances. First, it is used when the residuals reveal ring-like or gap-like structures that are approximately symmetric in azimuth. Second, in some sources where no clearly identifiable structure is visible in the residuals, the inclusion of a symmetric ring yields a significantly improved fit compared to azimuthally asymmetric models.

\subsection{Large scale misalignment} \label{sec:gauss_env}

For some sources, we observed the presence of a distinct larger structure that is misaligned or oriented differently relative to the central disk, suggesting either an outer disk or an extended circumstellar structure that has independent geometric parameters. To capture this feature, we introduced an independent Gaussian envelope component, which employs a similar Gaussian radial profile with same $x_0$ and $y_0$ center, but allows its own independent inclination and position angle. 

The brightness distribution is defined as,
\begin{equation}
 I_{\nu, E_2} (r_2) = I_{\nu,E_{0,2}} \exp{\left(-\frac{r_2^2}{2 \, {r_{En_2}}^2}\right)}\\ 
\end{equation}
with coordinates system of the envelope and $r_2$ are calculated by,
\begin{gather}
 xp_2 = (x-x_0) \cos(p.a._2) + (y-y_0) \sin(p.a._2)\\ 
 yp_2 = -(x-x_0) \sin(p.a._2) + (y-y_0) \cos(p.a._2)\\ 
 r_2 = \sqrt{{xp_2}^2 + \frac{{yp_2}^2}{\cos^2 i_2}}
\end{gather}
where $i_2$ and $p.a._2$ are the independent position angle and inclination of this envelope, respectively, and $r_{En_2}$ is the radius of this envelope. 
One or more Gaussian envelope components can be added to the PLCT profile as part of the disk intensity profile $I_{\nu,disk}$:
\begin{equation}
 I_{\nu,disk} = I_{\nu,PLCT}(r) + I_{\nu, E_2} (r_2) + I_{\nu, E_3} (r_3)
\end{equation}

Furthermore, when observed structural features are not associated with the central disk but rather belong to an outer disk, structural components can be directly combined into the disk intensity profile as a multiplicative perturbation of form $(1+a\cdot f(r))$, and adopt the same geometry parameters (such as $i_2$ and $p.a._2$) and coordinate systems (such as $xp_2$, $yp_2$ and $r_2$) as the corresponding outer disk. This approach allows us to describe multi-layered or misaligned disk configurations.

\subsection{Spiral Arm Component} \label{sec:spiral_arm}
In some cases, the residuals reveal a structure that is not confined to a fixed radius, but instead drifts in radius as a function of azimuthal angle. This feature cannot be adequately captured by standard azimuthally asymmetric profile which assumes a fixed radial location for the asymmetric component. To model such azimuthally drifting asymmetries, we adopt a spiral arm component, whose brightness evolves with both radius and azimuthal angle. This spiral arm intensity distribution is defined as the product of two components:
\begin{equation}
 I_{\nu, spiral}(r,\phi) = I_{\nu,r,sp}(r) f_{sp}(r,\phi) \label{eq:I_sp}
\end{equation}
where $I_{\nu,r,sp}(r)$ is an exponentially tapered power-law, controlling the radial brightness profile of the spiral arm with parameters independent of the primary PLCT profile:
\begin{equation}
I_{\nu,r,sp}(r) = I_{\nu,0}\left(\frac{r}{r_{w,sp}}\right)^{-\gamma_{1,sp}}\exp\left(-\frac{r}{r_{w,sp}}\right)^{2 - \gamma_{2,sp}} 
\end{equation}
The spiral shape is described by the $f_{sp}(r,\phi)$ component:
\begin{equation}
f_{sp}(r,\phi) = f_{az}\sum_{n}\exp\left[-\frac{(r - r_{sp,n}(\phi))^2}{2(f_{r} \cdot r_{sp,n}(\phi))^2}\right]
\end{equation}
where $r_{sp,n}(\phi)$ describes the radial location of the spiral arm for a given azimuthal angle $\phi$:
\begin{equation}
r_{sp,n}(\phi) = \frac{r_{w,sp}}{4}\exp\left[b_{pitch}\left(\phi + \phi_0 + 2\pi n\right)\right]
\end{equation}
where $b_{pitch}$ determines the tightness of the spiral, with larger absolute values corresponding to looser spirals, and the sign indicating the winding direction. $f_{az}$ adjusts the overall azimuthal intensity relative to the underlying radial profile, and $f_r$ adjusts the radial width of the spiral arms. The azimuthal offset angle $\phi_0$ sets the angular orientation of the spiral pattern. $n$ is a predefined set of integers, controlling how many $2\pi$ azimuthal periods are covered by the spiral pattern. The summation over all $n$ yields a spiral brightness enhancement that spans several rotations.

Since the spiral arm has its independent PLCT profile parameters, it is not a modulation of the primary PLCT profile of the disk but an independent additive component. Thus, the spiral arm component is directly added to the total intensity distribution to produce the disk intensity profile $I_{\nu, disk}(r,\phi)$:

\begin{equation}
I_{\nu,disk}(r,\phi) = I_{\nu,PLCT}(r) + I_{\nu,spiral}(r,\phi)
\end{equation}

\subsection{Broken Power-Law Profile} \label{sec:BPL}
When fitting several sources using the PLCT model, we observe compact and deep residuals near the central source, concentrated at small radii, which do not necessarily exhibit clear or structured morphology. Their presence sometimes influences the overall fit of the PLCT profile, pulling the profile to a steeper inner slope, which affects the model's ability to capture the overall disk profile. Ring and gap components added in the model can also be affected and become misaligned, compensating for these residuals rather than tracing the features in the outer disk. To address this issue, we introduce a broken power-law profile, which allows the inner and outer regions of the disk to follow different power-law indices. The broken-powerlaw profile is defined as:
\begin{equation}
I_{\nu,{bpl}}(r) = \left(\frac{r}{r_w}\right)^{-\gamma_{in}}
\left[ \frac{1}{2}\left(1 + \left(\frac{r}{r_w}\right)^{1/\Delta} \right) \right]^{-(\gamma_{out} - \gamma_{in})\Delta}
\end{equation}
where $\gamma_{in}$ and $\gamma_{out}$ correspond to the inner and outer power-law indices respectively. The transition between two power-laws occurs at the radius $r_w$, and the smoothness of this transition is controlled by $\Delta$, where a smaller $\Delta$ results in a sharper transition.
This broken power-law modifies the power-law core of the PLCT profile. To describe the steep drop on the edge of the outer disk plane, we apply an exponential taper to the broken power-law, similar to the PLCT profile:
\begin{equation}
I_{\nu,\mathrm{ETBPL}}(r) = I_{\nu,{bpl}}(r) \cdot \exp{\left[ -\left( \frac{r}{r_{exp}} \right)^{2 - \gamma_2} \right]}
\end{equation}
where $r_{exp}$ is the scaling radius of this exponential tail. This exponentially tapered broken power-law (ETBPL) profile is the form used in actual model fitting.  

\subsection{Weighted Combined Power-Law Profile} \label{sec:weighted_PL}
When fitting certain sources, we observe that residuals exhibit differences between the major and minor axis. In some cases, distinct features appear across each axis, whose shapes are not well matched by ring-like structures. These features often show approximate mirror symmetry along a single axis, but not full azimuthal symmetry. We interpret such patterns as arising from different radial profiles along the major and minor axes, so that the conventional axisymmetric brightness profile cannot capture the overall behavior. To address this issue, we introduced a weighted combination of two power-law profiles, where the relative contribution of each power-law component varies as a function of azimuthal angle. 

We define two azimuthally symmetric power-law brightness profiles, denoted by $f_1(r)$ and $f_2(r)$, which can be either simple power-laws or broken power-laws. Each power-law has a distinct set of parameters (e.g., $\gamma_{in,1}$, $\gamma_{out,1}$, $r_{w,1}$ for $f_1(r)$, and $\gamma_{in,2}$, $\gamma_{out,2}$, $r_{w,2}$ for $f_2(r)$). The final model is constructed as an azimuthally varied linear combination of the two profiles:
\begin{gather}
 w(\phi) = \frac{p_1 + p_2}{2} + \frac{p_1 - p_2}{2} \cos{(2\phi)}\\
 I_{\nu}(r,\phi) = [w(\phi) \cdot f_1(r) + \left[1 - w(\phi)\right] \cdot f_2(r)] \cdot \exp \left[-\left(\frac{r}{r_{exp}}\right)^{2-\gamma_2}\right]
\end{gather}
Here, $p_1$ and $p_2$ control the relative dominance of each profile along the major and minor axis respectively. $\phi = 0$ corresponds to the major axis, and $\phi = \frac{\pi}{2}$ corresponds to the minor axis. The weighting function $w(\phi)$ is defined such that along the major axis, the combination becomes:
\begin{equation}
 p_1 \cdot f_1(r) + (1-p_1) \cdot f_2(r)
\end{equation}
while along the minor axis, the combination becomes:
\begin{equation}
 p_2 \cdot f_1(r) + (1-p_2) \cdot f_2(r)
\end{equation}
Along other azimuthal directions, the combination smoothly transitions between these two configurations. The weighting function is normalized at each angle, so that the total contribution of $f_1(r)$ and $f_2(r)$ always sums to one. We assume that the behavior at the edge of the outer disk still obeys the PLCT profile, thus using the same universal exponential tail with $r_{exp}$ as the scaling radius.

This construction allows the model to adapt to brightness asymmetries aligned with the disk's geometric axes, improving the overall fitting without requiring explicit structural components.

\subsection{Asymmetric Radius Profile} \label{sec:asym_radius}
Previous asymmetric profiles have primarily been modeled by applying an additional localized modulation to an otherwise axisymmetric disk profile (Eq.~\ref{eq9}). However, some sources exhibit asymmetric morphologies where the stellar position is significantly offset from the geometric center of the disk. In such cases, localized features do not reproduce the observed emission, while a smooth, global asymmetry of the disk plane is required. To address this limitation, we introduce an alternative asymmetric modification in which the asymmetry is applied directly to the characteristic radius of the model, such that the radius is allowed to vary smoothly with azimuthal angle. In this formulation, the characteristic radius $r_w$ is replaced by an azimuthally dependent radius $r_w(\phi)$. Two functional forms are considered. The first is an exponential modulation:
\begin{equation}
 r_w(\phi) = r_{w,0} \exp\left[a \cos(\phi - \phi_0)\right],
\end{equation}
and the second is a linear modulation:
\begin{equation}
 r_w(\phi) = r_{w,0} \left[1 + a \frac{\cos(\phi - \phi_0) + 1}{2}\right],
\end{equation}
where $a$ controls the strength of the asymmetry and $\phi_0$ controls the azimuthal direction of the maximum radius. In practice, either form may be adopted depending on which provides a better fit to the data.
This asymmetric radius modification is not restricted to a specific profile, and can be applied to any disk model containing a characteristic radius parameter (such as Gaussian or PLCT). As an illustrative example, the application of a linear asymmetric radius modulation to a PLCT profile would be:
\begin{align}
 I_{\nu,PLCT,ar}(r,\phi) &= I_{\nu,0}\left(\frac{r}{r_w(\phi)}\right)^{-\gamma}
\exp\left[-\left(\frac{r}{r_w(\phi)}\right)^{2 - \gamma_2}\right] \\
 r_w(\phi) &= r_{w,0} \left[1 + a \frac{\cos(\phi - \phi_0) + 1}{2}\right]
\end{align}
When combined with asymmetric ring or gap profiles introduced elsewhere, this modulation enables a clearer separation between global disk asymmetry and localized non-axisymmetric substructures.

\section{Application of Structural Components in Best-Fit Models}
\subsection{BHR 71 IRS1} \label{BHR 71 IRS1}
BHR 71 IRS1's central peak exhibits a significant offset towards the southern side of the geometric disk center, such that the northern side shows a shallower radial profile that extends to larger radii, while the southern side has a steeper radial profile and is truncated at smaller radii. This contrast varies smoothly with azimuthal angle, indicating an overall disk-wide asymmetry. This is likely due to  optically thick emission and a flaring structure \citep{2024ApJ...974...21G}. In addition, a localized asymmetric structure is present on the northern side of the disk at intermediate radii.

When using a single PLCT profile with a conventional asymmetric profile (either 4th-order Gaussian or sinusoidal), we find that it is insufficient to reproduce these morphologies simultaneously. In such models, the model peak would be forced to deviate from the observed central peak, leaving strong residuals near the disk center. To address this, we adopted a PLCT profile with a linear asymmetric radius modification, combined with an additional 4th-order Gaussian asymmetric gap component. Thus, the final model has the formulation:
\begin{gather}
 I_{\nu, disk}(r,\phi) = I_{\nu, PLCT,ar}(r,\phi) \cdot [1 + a \cdot f_a(r,\phi)] \label{eq:BHR71IRS1_bestfit}
\end{gather}
where $I_{\nu, PLCT,ar}(r,\phi)$ employs a linear asymmetric radius $r_w(\phi)$ as defined in Appendix~\ref{sec:asym_radius}. $f_r(r,\phi)$ is the asymmetric component given by Eq.~\ref{eq7}.
This model can separately account for a smooth, global asymmetry of the disk plane and capture the localized non-axisymmetric structure observed on the northern side of the disk, resulting in a significantly improved fitting quality in BHR~71~IRS1. 

\subsection{L1489 IRS}
\label{L1489IRS}
When using the simple PLCT model fitting, we find that only the compact central source in L1489 IRS is well-fitted, while extended positive residuals remain at larger radii. Previous studies of L1489 IRS \citep{2023ApJ...951...11Y}
have identified a complex disk geometry composed of an inner disk, an intermediate disk, and an outer disk, with misaligned inclinations and position angles. This warped multi-disk structure matches the large-scale residuals of our fitting. Such structure also provides a natural explanation for the asymmetric ring-like residuals located approximately within the radial extent of the intermediate disk, where warped structure leads to uneven illumination between two sides of the disk plane, resulting in a localized brightness enhancement on one side.

To capture these features, the best-fit model for L1489 IRS includes a compact PLCT core representing the inner disk and the central source, and two extended Gaussian components (Appendix~\ref{sec:gauss_env}) corresponding to the intermediate and outer disks. Both extended disks are modeled with a distinct inclination and position angle, while the intermediate disk is further modulated by an asymmetric ring structure to fit the localized asymmetric brightness enhancement. The final model thus takes the form:
\begin{gather}
 I_{\nu,disk}(r,\phi) = I_{\nu,PLCT}(r) + I_{\nu, E_2} (r_2) \cdot [1+a \cdot f_a(r_2,\phi_2)] + I_{\nu, E_3} (r_3) \label{eq:L1489IRS_bestfit}\\
 f_a(r_2,\phi_2) = \exp{\left(-\frac{(r_2 - r_{c,a,2})^4}{2 \, {r_{w,a,2}}^4}\right)} \, \exp{\left(-\frac{(\phi_2 - \phi_{c,a,2})^4}{2 \, {\phi_{w,a,2}}^4}\right)}
\end{gather}
where each component is computed in its own coordinate system, and the asymmetric ring $f_a(r_2,\phi_2)$ is defined in the frame of the intermediate disk, sharing its global Gaussian radial profile. This construction is motivated by the residuals morphology, as well as physically by the illumination asymmetry arising from the warped structure.

This best-fit model effectively captures the observed multi-disk morphology and asymmetry in L1489 IRS, and yields residuals that are nearly featureless across the entire disk extent.

\subsection{{IRS 7B-a}}
{IRS 7B-a} exhibits a localized asymmetric structure in the disk emission. When using a PLCT profile combined with a conventional asymmetric profile, we find that the residuals display a broad, coherent pattern similar to a wide spiral-like structure. Motivated by the shape of the residuals, we introduce an additional spiral arm component (Appendix~\ref{sec:spiral_arm}). Based on subsequent residual analysis, we retain the asymmetric component in addition to the spiral arm. The final model is therefore given by:
\begin{gather}
 I_{\nu,disk}(r,\phi) = I_{\nu,PLCT}(r)\cdot\left[1 + a \cdot f_a(r,\phi)\right] + I_{\nu,spiral}(r,\phi) \label{eq:IRS7B_bestfit}
\end{gather}
where the asymmetric component $f_a(r,\phi)$ follows Eq.\ref{eq7}, and the spiral arm component $I_{\nu,spiral}(r,\phi)$ follows the formulation defined in Appendix~\ref{sec:spiral_arm}. The spiral arm component shares the same normalization $I_{\nu,0}$ as the underlying PLCT profile.

Compared to a conventional PLCT + asymmetric profile model, the final combined model yields a substantially improved fitting quality. Nevertheless, the residual map still exhibits global negative residuals. We note that although this spiral-arm model is the best-fit model of IRS~7B, the spiral arm component primarily serves as a flexible description of a complex
asymmetric structure with significant and rapidly varying azimuthal dependence. Therefore, it is not interpreted as evidence for a physical spiral arm in the disk, but rather as an effective parameterization to reproduce the observed emission morphology.

\subsection{Oph IRS 63}
\label{IRS63}
Oph IRS 63 is one of the best-studied eDisk sources and multiple annular substructures have been observed in the dust disk. When fitting Oph IRS 63 using the PLCT model, we observe annular structured residuals broadly consistent with the structures identified in previous studies. 
Based on this result, we adopted a model composed of a PLCT profile with two symmetric ring/gap components, which aims to recover the annular substructures identified by earlier observations. However, when applying this model, we observe residuals exhibit prominent negative features near the central source, primarily concentrated along the major axis within $\sim 0.1$\arcsec. These residuals do not exhibit typical gap morphology, but rather manifest as localized dips on either side of the source. One of the two Gaussian ring components in the model is instead drawn inward to $\sim 0.05$\arcsec to compensate for these central residuals instead of matching with the known gap locations.

To address this issue, we introduce two key modifications. First, we replace the power-law core in the PLCT profile with an broken power-law profile tapered by the same exponential tail (ETBPL, see Appendix~\ref{sec:BPL}) to better accommodate complex features in the inner disk region. Second, to reproduce the azimuthally non-uniform structure of the central negative residuals, we adopt a weighted combination of two broken power-law profiles with azimuthal dependence weights (Appendix~\ref{sec:weighted_PL}). This allows the radial profile to differ along the major and minor axes, enabling a steeper inner slope on the major axis while maintaining a smoother profile on the minor axis. These modifications allows better fits of the inner structure without distorting the overall disk shape, and help fix the structure misplacement. Two symmetric Gaussian ring/gap components are then added to reproduce the annular substructures. 
Such exponentially tapered weighted broken power-law (ETWBPL) is defined as:
\begin{gather}
 w(\phi) = \frac{p_1 + p_2}{2} + \frac{p_1 - p_2}{2} \cos{(2\phi)} \label{eq:ETWBPL_w(phi)}\\
 \mathrm{BPL}_1(r) = \left(\frac{r}{r_{w,1}}\right)^{-\gamma_{\mathrm{in},1}} \left[ \frac{1}{2}\left(1 + \left(\frac{r}{r_{w,1}}\right)^{1/\Delta_1} \right) \right]^{-(\gamma_{\mathrm{out},1} - \gamma_{\mathrm{in},1}) \Delta_1} \label{eq:ETWBPL_BPL1)} \\
 \mathrm{BPL}_2(r) = \left(\frac{r}{r_{w,2}}\right)^{-\gamma_{\mathrm{in},2}} \left[ \frac{1}{2}\left(1 + \left(\frac{r}{r_{w,2}}\right)^{1/\Delta_2} \right) \right]^{-(\gamma_{\mathrm{out},2} - \gamma_{\mathrm{in},2}) \Delta_2} \label{eq:ETWBPL_BPL2)} \\
 I_{\nu, \mathrm{ETWBPL}}(r,\phi) = I_{\nu,0}\left[ w(\phi) \cdot \mathrm{BPL}_1(r) + \left(1 - w(\phi)\right) \cdot \mathrm{BPL}_2(r) \right] \cdot \exp \left[-\left(\frac{r}{r_{\mathrm{exp}}}\right)^{2-\gamma_2} \right] \label{eq:ETWBPL_I)}\\
\end{gather} 
Based on this model, two symmetric ring/gap components are implemented as:
\begin{gather}
 f_{r,1}(r) = a \cdot \exp\left[ -\frac{(r - r_{c,3})^4}{2 r_{w,3}^4} \right] \\
 f_{r,2}(r) = b \cdot \exp\left[ -\frac{(r - r_{c,4})^4}{2 r_{w,4}^4} \right] \\
 I_{\nu}(r,\phi) = I_{\nu,\mathrm{ETWBPL}}(r,\phi) \cdot \left[1 + f_{r,1}(r) + f_{r,2}(r)\right] \label{eq:OphIRS63_bestfit}
\end{gather}

This best-fit model captures two negative-amplitude Gaussian gap components, located at angular radii of $0.280$\arcsec and $0.139$\arcsec with corresponding widths of $0.073$\arcsec and $0.034$\arcsec, respectively. These values are derived from the best-fit parameters $r_{c,3}$, $r_{c,4}$ and $r_{w,3}$, $r_{w,4}$. These locations are consistent with the values reported in \citet{2020Natur.586..228S}, confirming that our model effectively recovers the known substructures in Oph IRS 63.

\subsection{GSS 30 IRS3}
\label{GSS30}
GSS 30 IRS3 is a high-inclination source with a disk inclination of $\sim66^\circ$. The disk exhibits a pronounced overall asymmetry along the minor axis, with central peak clearly shifted toward northern side. This feature has been reported in \cite{2024A&A...690A..46S} and attributed to optically thick emission and disk flaring.

We first attempted to reproduce this morphology using a PLCT profile with an asymmetric radius modification (Appendix~\ref{sec:asym_radius}). However, due to the high inclination of the disk, which places it close to an edge-on configuration, together with the minor-axis asymmetry introduced by flaring, we found that an azimuthally smooth radius modification is insufficient to describe the observed profile. To address this, we adopted a weighted combined power-law profile (Appendix~\ref{sec:weighted_PL}), and applied the asymmetric radius modification only to the power-law component that dominates along the minor axis. This allows the model to capture the strong minor-axis asymmetry induced by both high inclination and disk flaring, while preventing these effects from propagating onto the major axis, and maintaining a continuous, smooth azimuthal transition.

In addition, we identified a clear change in the radial power-law behavior in the inner disk region. This truncation in the profile motivates the introduction of broken power-law profiles. Following the notation in Appendix~\ref{IRS63} for Oph~IRS~63, the formulation of the model is:
\begin{gather}
 I_{\nu, \mathrm{ETWBPL}}(r,\phi) = I_{\nu,0}\left[ w(\phi) \cdot \mathrm{BPL}_1(r) + \left(1 - w(\phi)\right) \cdot \mathrm{BPL}_2(r,\phi) \right] \cdot \exp \left[-\left(\frac{r}{r_{\mathrm{exp}}}\right)^{2-\gamma_2} \right]
\end{gather}
where the azimuthal weighting function $w(\phi)$ and the broken power-law profiles $\mathrm{BPL}_1$ and $\mathrm{BPL}_2$ follow Eqs.~\ref{eq:ETWBPL_w(phi)},~\ref{eq:ETWBPL_BPL1)},~\ref{eq:ETWBPL_BPL2)}. The key difference is that $\mathrm{BPL}_2(r,\phi)$ is modulated by an linear asymmetric characteristic radius:
\begin{equation}
 r_{w,2}(\phi) = r_{w,2} \left[1 + a_r \frac{\cos(\phi - \phi_0) + 1}{2}\right],
\end{equation}
Based on the residuals of this combined model, {we further identified a symmetric} ring-like structure in the residuals, indicating the presence of a potential substructure. We therefore introduced an additional symmetric ring/gap component (Appendix~\ref{sec:sym_ring}), such that the final model is given by:
\begin{gather}
 I_{\nu,disk}(r,\phi) = I_{\nu,\mathrm{ETWBPL}}(r,\phi) \cdot \left[1 + f_{r}(r)\right] \label{eq:GSS30IRS3_bestfit}
\end{gather}
where $f_{r}(r)$ is an symmetric ring/gap component as defined by Eq.~\ref{eq:sym_ring}.

With this model, we are able to better reproduce the radial structure of the disk. The symmetric component is preferred to be a negative gap, located at $0.259''$ ($\sim$35.8 au) with a width of $0.059''$ ($\sim$8.2 au). The gap introduces two apparent local bumps on either side, which are consistent with the bumps along the major axis at $\sim0.18''$ ($\sim$25 au) and $\sim0.36''$ ($\sim$50 au) as reported in \cite{2024A&A...690A..46S}, indicating the potential presence of substructures. 

\subsection{Oph IRS 43 VLA1}
Oph~IRS~43~VLA1 is intrinsically compact, but is surrounded by a large-scale envelope emission in the continuum. When using a single Gaussian or PLCT profile, the model tends to partially fit the extended envelope emission, preventing an accurate characterization of the disk itself.

To decouple the compact disk emission from the large-scale envelope, we adopt a model combining a PLCT with an additional Gaussian envelope component (Appendix~\ref{sec:gauss_env}), with independent position angles. The model formulation is therefore given by:
\begin{gather}
 I_{\nu,disk}(r)=I_{\nu,\mathrm{PLCT}}(r)+I_{\nu,E_2}(r_2) \label{eq:OphIRS43_bestfit}
\end{gather}
With this model, the dust disk emission is well isolated from the surrounding envelope emission. However, the residuals of Oph~IRS~43~VLA1 exhibit two symmetric circular positive structure located on opposite sides of the disk, offset from the principal axes. These features appear as localized bipolar-like emission enhancements in the observation. We explored modeling this structure using a two-spiral-arms component. While this approach yields a better fit, the spiral parameters do not converge and the resulting spiral morphology is not physically reasonable. Therefore, we retain the PLCT + Gaussian envelope as the best-fit model for Oph~IRS~43~VLA1.

\bibliography{edisk}
\end{document}